\documentclass[11pt,a4paper]{article}
\usepackage{amsmath}
\usepackage{amssymb,graphicx,amsthm,dsfont}
\usepackage[pagebackref=true]{hyperref}
\usepackage[capitalize]{cleveref}
\usepackage{mathtools}
\usepackage{booktabs}
\usepackage{tikz}

\newcommand{\swap}{{\mathrm{KT}}}

\usepackage{booktabs}
\usepackage{makecell}

\usepackage{booktabs}
\usepackage{multirow}
\usepackage[square,numbers]{natbib}
\usepackage{cleveref}
\usepackage{enumitem}
\usepackage{nicefrac}
\usepackage{siunitx}
\usepackage{pgfplots}
\usepackage{wrapfig}
\usepackage{authblk}
\usepackage{fullpage}

\newtheorem{definition}{Definition}

\pgfplotsset{compat=newest}

\makeatletter
\renewcommand{\paragraph}{%
	\@startsection{paragraph}{4}%
	{\z@}{1ex \@plus 1ex \@minus .2ex}{-1em}%
	{\normalfont\normalsize\itshape}%
}
\makeatother

\makeatletter
\def\moverlay{\mathpalette\mov@rlay}
\def\mov@rlay#1#2{\leavevmode\vtop{%
		\baselineskip\z@skip \lineskiplimit-\maxdimen
		\ialign{\hfil$\m@th#1##$\hfil\cr#2\crcr}}}
\newcommand{\charfusion}[3][\mathord]{
	#1{\ifx#1\mathop\vphantom{#2}\fi
		\mathpalette\mov@rlay{#2\cr#3}
	}
	\ifx#1\mathop\expandafter\displaylimits\fi}
\makeatother

\newcommand{\cupdot}{\charfusion[\mathbin]{\cup}{\cdot}}

\usepackage{caption}
\usepackage{subcaption}

 \title{Collecting, Classifying, Analyzing, and Using Real-World Elections}
\author {
	Niclas Boehmer,
	Nathan Schaar
}
\affil{ 
	Technische Universit{\"a}t Berlin,  Faculty IV, Algorithmics and Computational Complexity, Berlin, Germany\\
	niclas.boehmer@tu-berlin.de
}

\begin{document}
\maketitle

\begin{abstract}
	We present a collection of $7582$ real-world elections divided into $25$ datasets from various sources ranging from sports competitions over music charts to survey- and indicator-based rankings.
	We provide evidence that the collected elections complement already publicly available data from the PrefLib database, which is currently the biggest and most prominent source containing $701$ real-world elections from $36$ datasets \cite{DBLP:conf/aldt/MatteiW13}. 
	Using the map of elections framework \cite{DBLP:conf/atal/SzufaFSST20}, we divide the datasets into three categories and conduct an analysis of the nature of our elections. 
	To evaluate the practical applicability of previous theoretical research on (parameterized) algorithms and to gain further insights into the collected elections, we analyze different structural properties of our elections including the level of agreement between voters and election's distances from restricted domains such as single-peakedness. 
	Lastly, we use our diverse set of collected elections to shed some further light on several traditional questions from social choice, for instance, on the number of occurrences of the Condorcet paradox and on the consensus among different voting rules. 
\end{abstract}

\section{Introduction}
The area of computational social choice is concerned with the algorithmic and axiomatic analysis of collective decision-making problems, where given a set of agents with preferences over some alternatives the task is to select a ``compromise'' alternative \cite{DBLP:reference/choice/2016}. 
One important part of computational social choice is the study of algorithmic aspects of election-related problems such as the computation and manipulation of voting rules \cite{DBLP:reference/choice/Zwicker16,DBLP:reference/choice/LangX16,DBLP:reference/choice/FaliszewskiR16,DBLP:reference/choice/ConitzerW16,DBLP:reference/choice/CaragiannisHH16}.
While in the early years of the field the main focus lay on the study of the theoretical worst-case computational complexity of these problems, in recent years the focus has at least partially shifted towards the practical applicability of theoretical research  (see e.g., \cite{DBLP:journals/heuristics/FaliszewskiSST18,DBLP:conf/aaai/WangSSZJX19,DBLP:journals/jair/KellerHH19,DBLP:conf/atal/SzufaFSST20,DBLP:conf/aaai/GoldsmithLMP14,DBLP:conf/ijcai/BoehmerBFN21}). 
Two classical social choice questions which have been extensively studied from an empirical perspective are the number of occurrences of voting paradoxes \cite{chamberlin1984social,felsenthal1993empirical,DBLP:conf/atal/BrandtGS16,DBLP:journals/scw/PlassmannT14} and the consensus among voting rules \cite{chamberlin1984social,felsenthal1993empirical,regenwetter2006behavioral,regenwetter2007unexpected,DBLP:conf/aldt/Mattei11,darmann2019evaluative,popov2014consensus}. 
Nevertheless, there are still many subareas that lack empirical research. 
For instance, there are numerous theoretical papers designing parameterized algorithms for elections that are close to being single-peaked (see e.g. \cite{
	DBLP:journals/jcss/YangG17,DBLP:conf/atal/YangG15,DBLP:conf/ijcai/CornazGS13,DBLP:conf/atal/YangG14a,DBLP:journals/ai/FaliszewskiHH14,DBLP:conf/aaai/MenonL16,DBLP:journals/tcs/SkowronYFE15}).\footnote{An election is single-peaked if there exists a societal order of the candidates and each voter ranks candidates that are closer to its top-choice according to the societal order above those which are further away.} 
To the best of our knowledge, only \citet{DBLP:conf/ijcai/SuiFB13} measured the distance of real-world elections from being single-peaked and detected  that most elections are far away. Thus, the practical applicability of the developed algorithms is largely unclear.  

Given that some collective decision-making problems can have a crucial impact on people's lives, the general rarity of experimental works can be seen as slightly worrisome. 
While papers proposing a collective decision-making mechanism often conduct an analysis of the mechanism's axiomatic and computational properties, a rigorous testing on real-world data is often missing.
However, such an analysis would be highly beneficial, as axiomatic and computational complexity guarantees are often outperformed in practice \cite{DBLP:conf/atal/BaumeisterHR20,DBLP:journals/jair/Walsh11,DBLP:conf/aaai/ConitzerS06,DBLP:conf/ecai/Walsh10,DBLP:conf/aaai/FaliszewskiLPT18,DBLP:journals/ai/SkowronFS15,DBLP:conf/ijcai/BredereckF0N19}.
Being able to conduct meaningful tests is thus crucial to better understand the practical aspects of collective decision-making mechanisms and to ultimately select the one that behaves ``fairest'' in practice.

One reason for the general rarity of experimental works in voting validating the applicability of theoretical research  might be the lack of data. 
To tackle this issue, in 2013, \citet{trendspreflib,DBLP:conf/aldt/MatteiW13} started the very useful PrefLib platform, a database for real-world election data. 
Many community members have contributed to this popular platform over the past years and at the moment it contains $701$ real-world elections dived into $36$ datasets (see \citet[Table 5]{DBLP:journals/corr/abs-2105-07815} for a recent overview of the datasets).
Many elections from PrefLib are based on humans expressing opinions over alternatives, e.g., over candidates in an election, over movies, or types of sushi. 
However, due to this nature of these elections, most of them either have few candidates or voters express only partial preferences which can include many ties. 
In fact, as observed by \citet[Table 5]{DBLP:journals/corr/abs-2105-07815}, there are only $165$ elections from $8$ sources on PrefLib with $10$ or more candidates where votes include not too many ties. 
The goal of this paper is to contribute to the rise of experimental works in computational social choice by executing the following four steps: 

\paragraph{Step 1: Collecting Data.} In \Cref{sec:collect}, we present our collection of $7582$ real-world elections divided into $25$ datasets.
We preprocess the data by deleting candidates and voters until each voter ranks all candidates. 
Subsequently, to be able to better compare the properties of our elections, for each dataset we create $500$ elections containing $30$ voters over $15$ candidates.     
Our real-world elections differ from most of the already publicly available ones in three aspects: 
First, they contain virtually no ties and are of various sizes (the average number of candidates varies from around $20$ to above $800$, while the average number of voters ranges from around $12$ to over $1400$).
Moreover, even after deleting voters and candidates until all voters rank all candidates, most elections are still of at least medium size.
As a majority of algorithms are designed for such so-called \emph{complete} elections, this is a very important step to ensure the usefulness of our data for experimental works. 
In the past, elections have been often completed by appending missing candidates in random order or based on the preference of other voters \cite{DBLP:conf/aaai/Doucette14,DBLP:journals/corr/abs-2105-07815}. 
Our approach offers the clear advantage that preferences in the final election are not distorted in any way: 
Each pairwise ordering of a voter represents its true opinion.
Second, unlike a majority of elections on PrefLib, our datasets are not based on humans explicitly expressing preferences over alternatives.
Admittedly, this might be considered as a drawback of our data as political elections are still often thought of as the prime application of social choice theory.
Nevertheless, we want to remark that voting is also relevant and already used in many other contexts, e.g., in multi-agent systems, or in sports, when aggregating the results of multiple competitions into a final ranking.  
Third, around half of our datasets arise from time-based preferences, i.e., capture in one form or another the changing preferences of agents over time. 
Time-based elections might not directly match ones intuition for an election; however, preferences obtained at different points in time are also frequently collected in an election (for instance, when deciding on the overall winner of multiple competitions). 
Notably, while there are already some theoretical works dealing with such time-evolving preferences  \cite{DBLP:conf/atal/BoehmerN21,DBLP:conf/aaai/Lackner20,DBLP:conf/aaai/ParkesP13,DBLP:journals/corr/abs-2005-02300}, as pointed out by \citet{trendspreflib}, there are only very few such elections currently publicly available.

\paragraph{Step 2: Classifying Data.} In \Cref{sec:map}, we apply the map of elections framework of \citet{DBLP:conf/atal/SzufaFSST20} and \citet{DBLP:conf/ijcai/BoehmerBFNS21} to visualize the collected elections as points on a map. 
Using this, we detect that one of our datasets seems to fall into a so-far vacant part of the ``space of elections''.
Moreover, based on their positions on the map, we propose a classification of our datasets into three categories and observe in the subsequent experiments  that datasets from one category typically have similar properties.
This suggests that if one wants to run experiments on our data, it should be sufficient to use few datasets from each of the three categories.

\paragraph{Step 3: Analyzing Data.} In \Cref{sec:sim,sec:rest}, we analyze various structural properties of the collected elections. 
This analysis serves three purposes: First, we aim for a better understanding of the collected elections.
Second, we want to gain some insights into the relationship between the different properties.
Third, we try to contribute to putting the research on parameterized algorithms for voting-related problems on an empirical basis by measuring already used parameters.
Unfortunately, we find that most of them are typically quite large and thus that most algorithms developed for these parameters are probably not really practically usable on our data.
Briefly put, in  \Cref{sec:sim} we analyze the degree of similarity between voters in an election, while in \Cref{sec:rest} we check which of our elections are (close to) a restricted domain.

\paragraph{Step 4: Using Data.} In \Cref{sec:util}, we use our collected elections to address some classical and already empirically researched questions from social choice, such as the frequency of Condorcet winners and the consensus among voting rules. 
While we partly confirm previous findings, for instance, that most elections have a Condorcet winner and that voting rules often return the same winner, we find contradicting evidence for others and also identify some datasets showing a distinct behavior. 
This indicates that our datasets are quite different from each other with some of them showing rarely observable and non-standard behavior, making them collectively well-suited for experimental~research. 

Our original and preprocessed election data is available at \url{github.com/n-boehmer/Collecting-Classifying-Analyzing-and-Using-Real-World-Elections}.
Most (sub)sections start with a list of main take-away messages summarizing the contribution of this (sub)section.

\section{Preliminaries}
For a set $S$ and an integer $k\in \mathbb{N}$, we denote as ${{S}\choose{k}}$ the set of all $k$-element subsets of $S$. \smallskip

\noindent \textbf{Preference Orders.} For a set $C$ of candidates, let $\mathcal{L}(C)$ denote the set of all total orders over $C$. 
We refer to the elements of $\mathcal{L}(C)$ as preference orders, votes, or voters.\smallskip

\noindent \textbf{Elections.} An election $E$ is defined by a set of $C=\{c_1,\dots , c_m\}$ of $m$ candidates and a collection $V=(v_1,\dots, v_n)$ of $n$ voters with $v_i\in \mathcal{L}(C)$ for each $i\in [n]$.
For a voter $v\in V$ and two candidates $a,b\in C$, we write $a\succ_v b$ to denote that $v$ prefers $a$ to $b$.
We say that voter $v\in V$ ranks candidate $c\in C$ in position $i\in [m]$ if $v$ prefers exactly $i-1$ candidates from $C\setminus \{c\}$ to $c$.
We refer to the candidate which a voter ranks in the first position as its top-choice.
\smallskip

\noindent \textbf{Kendall Tau Distance.}
For two votes $v,v'\in \mathcal{L}(C)$, their Kendall tau distance $\swap(v,v')$ is defined as the number of candidate pairs on which orderings $v$ and $v'$ disagree: $|\{c,c'\in {{C}\choose{2}}\mid (c\succ_v c' \wedge c' \succ_{v'} c) \vee (c\succ_{v'} c' \wedge c' \succ_{v} c)\}|$. 
Alternatively, $\swap(v,v')$ can be interpreted as the minimum number of swaps of adjacent candidates that need to be performed to transform $v$ into $v'$. \smallskip

\noindent \textbf{Restricted Domains.}
We define three different restricted domains here.
In single-peaked elections, there is an order of the candidates and each voter prefers candidates that are closer to its top-choice with respect to the order to those further away:
\begin{definition}[\cite{black1948rationale}]
	An election $E=(C,V)$ is \emph{single-peaked} if there is a linear order $\rhd$  over $C$, sometimes called the societal order, such that for each three candidates $a,b,c\in C$ with $a\rhd b \rhd c$, for each $v\in V$, if $a\succ_v b$ then  $b\succ_v c$. 
\end{definition}
In single-crossing elections, there is an order of voters such that going through the voters according to the order, the ordering of each pair of candidates changes at most once. 
\begin{definition}[\cite{mirrlees1971exploration,roberts1977voting}]
	An election $E=(C,V)$ is \emph{single-crossing} if there is a linear order $\rhd$  over $V$ such that for each two candidates $c,c'\in C$, there do not exist three votes $v,v',v''\in V$ with $v\rhd v' \rhd v''$ such that $c\succ_v c'$, $c' \succ_{v'} c$, and $c\succ_{v''} c'$.
\end{definition}
Lastly, we define group-separable elections: 
\begin{definition}[\cite{inada1969simple,inada1964note}]
	An election $E=(C,V)$ is \emph{group-separable} if each subset $A\subseteq C$ of candidates  with $|A|\geq 2$ can be partitioned into two sets $A'$ and $A''$ such that each voter $v\in V$ prefers either all candidates from $A'$ to all candidates from $A''$ or the other way around.
\end{definition}

\noindent \textbf{Pearson Correlation Coefficient (PCC).} The Pearson 
correlation coefficient is a measure for the linear correlation between two 
quantities $x$ and $y$, where $1$ means that $x$ and $y$ are perfectly positively
linearly correlated, i.e., it always holds $y=mx+b$ for some $b$ and $m> 0$,
$0$ indicates no linear correlation, and $-1$ describes a perfect negative correlation ($m<0$). 
A Pearson Correlation Coefficient between $0.4$ and $0.69$  indicates a moderate correlation,  a value between $0.7$ and $0.89$ indicates a strong correlation, and a value between $0.9$ and $1$ indicates a very strong 
correlation \citep{schober2018correlation}.

\section{Collecting Real-World Elections} \label{sec:collect}
In this section, we describe the different election datasets that we collected (\Cref{sub:raw}) and explain how we obtained the elections we use in our experiments from them (\Cref{sub:normalized}).

\subsection{Raw Election Data} \label{sub:raw}
In the following, we list the different data sources that we used to create our elections, ranging from results of sports competitions over music charts and expert assessments to survey- or indicator-based rankings.\footnote{Notably, there are virtually no ties in the data (if there happens to be a tie, we break it arbitrarily).} 
For each data source, we describe how we created elections from the data; for some sources, we created two types of elections. 

From a methodological perspective, our elections are of one of two types: 
We say that an election is \emph{time-based} if each vote corresponds to an evaluation of the candidates at different points in time.
In contrast to this, we call an election \emph{criterion-based} if each vote corresponds to some, in principle, independent criterion judging the candidates at the same point in time.
In \Cref{tab:basicStats}, we indicate for each dataset the type, the number of contained elections, and their average size before and after the preprocessing (as described in \Cref{sub:normalized}).
We also collected further datasets which we do not include in our analysis for the sake of conciseness (see \Cref{se:moredata} for descriptions).

\paragraph{Boxing/Tennis (World) Rankings.} The boxing data (collected by \citet{boxing}) contains the Ultimate Fighting Championship rankings of the top 16 fighters in twelve different weight classes in different weeks between February 2013 and August 2021. The tennis data (collected by \citet{tennis}) contains weekly rankings of the top 100 male tennis players published by the ATP between January 1990 and September 2019. For each year (and weight class), we created a \emph{tennis top 100 (boxing top 16)} election where each player (fighter) is a candidate and each vote corresponds to the ranking of the players (fighters) in one week.

\paragraph{American Football.} The American football data (collected by \citet{college}) contains weekly power rankings of college football teams from different media outlets  for each season between 1997 to 2021.
We created two different types of elections with teams as candidates:
First, for each season and each media outlet, we created a \emph{football season} election where each vote corresponds to the power ranking of the teams in one week according to the media outlet.
Second, for each week in one of the seasons, we created a \emph{football week} election where each vote corresponds to the power ranking of the teams in this week according to one of the media outlets.

\paragraph{Formula 1.} The Formula 1 data (collected by \citet{Formula1}) contains the finishing times of each driver in each lap of a race between 1950 and 2020. From this we created two types of elections with drivers as candidates: First, for each year, we created a \emph{Formula 1 season} election where each vote corresponds to a race in this year and ranks the drivers by their finishing time in this race.\footnote{Notably, Formula 1 season elections from 1961 to 2008 are also available on preflib.com.} Second, for each race, we created a \emph{Formula 1 race} election where each vote corresponds to a lap in the race and ranks the drivers by the time they spend in this lap. 

\paragraph{Spotify.} For each day between the 1st of January 2017 and 9th January 2018, the Spotify data (collected by \citet{spotify}) contains a daily ranking of the 200 most listened songs in one of 53 countries.
We created two types of elections with songs as candidates:
First, for each month and each country, we created a \emph{spotify month} election where each vote corresponds to the ranking of the songs on one day of the month in the country. 
Second, for each day, we created a \emph{spotify day} election where each vote corresponds to the ranking of the songs on this day in one of the $53$ countries.

\begin{table}
	\begin{center}
		\begin{tabular}{ccccccccc}
				\toprule
				name & 
				type& 
				\multicolumn{3}{c}{raw} & \phantom{a} &
				\multicolumn{3}{c}{relevant complete}\\ \cmidrule{3-5}
				\cmidrule{7-9}
				&& \#Elec. & \makecell{Avg.\\ \#Voters} & \makecell{Avg.\\ \#Cand.}   & & \makecell{\#Elec.} & \makecell{Avg.\\ \#Voters} &  \makecell{Avg.\\ \#Cand. }    \\ \midrule \midrule
				%------
				boxing top 16 & time & 99 & 31.9 & 19.76 && 31 & 17.45 & 15.32 \\
				football season & time & 2746 & 12.28 & 152.36 && 2422 & 12.6 & 156.71 \\
				Formula 1 race & time & 454 & 61.3 & 20.46 && 396 & 47.2 & 17.93 \\
				Formula 1 season & time & 71 & 14.58 & 43.97 && 42 & 13.38 & 21.57 \\
				spotify month & time & 645 & 29.78 & 306.64 && 632 & 29.91 & 109.28 \\
				tennis top 100 & time & 29 & 50.48 & 140 && 29 & 49.9 & 62.31 \\
				Tour de France & time & 97 & 21.14 & 175.69 && 95 & 19.7 & 82.64 \\
				city ranking & crit. & 1 & 12 & 216 && 1 & 12 & 216 \\
				country ranking & crit. & 12 & 17.25 & 119.17 && 12 & 14.25 & 95.58 \\
				football week & crit. & 415 & 83.28 & 219.67 && 415 & 77.35 & 98.45 \\
				spotify day & crit. & 362 & 53.06 & 247.74 && 375 & 49.06 & 20.73 \\
				university ranking & crit. & 4 & 18.5  &  832.5 && 4  & 18.5  & 123.25 \\
				\bottomrule
		\end{tabular}
		\caption{Information about our election datasets.} \label{tab:basicStats}
	\end{center}
\end{table}

\paragraph{Tour de France.} For each edition of the Tour de France between 1903 and 2021, the data contains the completion times of all riders for each stage. The dataset was crawled by us from the website \url{procyclingstats.com}.
For each edition, we created one \emph{Tour de France} election in which the riders are the candidates and each vote corresponds to a stage  and ranks the riders by their completion time.

\paragraph{City Rankings.} The city data (collected by \citet{city}) contains twelve quantitative indicators for the life quality in $216$ different cities determined by movehub.com.  We created a single \emph{city ranking} election where each city is a candidate and each vote corresponds to the ranking of the cities with respect to one of the indicators.\footnote{Note that there is also a different dataset based on indicator-based rankings over cities available on preflib.com.}   

\paragraph{Country Rankings.} For each year between $2005$ and $2016$, the country ranking data (based on the popular world happiness report and collected by \citet{country}) contains different quantitative indicators for the happiness of citizens from over $100$ countries. 
For each year, we created a \emph{country ranking} election where the countries are the candidates and each vote ranks them according to one indicator.

\paragraph{University Rankings.} For each year between 2012 and 2015, the university ranking data (collected by \citet{university}) contains rankings of universities according to different criteria provided by three systems. 
For each year, we created a \emph{university ranking} election where the universities are the candidates and each vote ranks them according to one criterion used by one of the three~systems.

\subsection{From Raw to Normalized Elections} \label{sub:normalized}
In our experiments, we do not use the raw elections created as described in \Cref{sub:raw} but instead apply some preprocessing.
As a first step, by deleting voters and candidates, we converted each created election into a \emph{complete} election, i.e., an election where every voter ranks all candidates.
In general, it is not clear which candidates and voters to delete from an incomplete election to obtain a complete election that is as large as possible.
Interestingly, this problem corresponds to finding a maximum edge biclique in a bipartite graph \cite{DBLP:journals/tcs/LonardiSY06,DBLP:conf/sdm/ShahamYL16}, where we are given a bipartite graph $G=(V\cupdot U, E)$ and the task is to find a subset of vertices $V'\subseteq V$, $U'\subseteq U$ such that the graph restricted to vertices from $V'\cup U'$ is complete and $|V'|\cdot |U'|$ is maximized.\footnote{To convert our problem to the problem of finding a biclique, for an elections, $E=(C,V)$, 
	we create a vertex for each vote $v\in V$ and candidate $c\in C$ and connect two vertices $v\in V$ and $c\in C$ if $c$ appears in $v$. Each biclique then corresponds to a complete subelection of $E$ and the other way round.}
As this problem is NP-hard \cite{DBLP:journals/tcs/LonardiSY06}, we employ an excellent heuristic by \citet{DBLP:conf/sdm/ShahamYL16}.
An alternative way of completing the data used in the literature is to fill incomplete votes randomly or based on the preferences of other voters. 
However, our approach has the advantage that each vote is fully based on a real-world vote and not perturbed in any way. 

As in our experiments we are interested in elections with at least $15$ candidates\footnote{We chose this number to be as large as possible while still being able to include most of our elections.}, we call each election with $15$ or more candidates (and an arbitrary number of voters) \emph{relevant}. 
We display information about the number and size of the relevant complete elections from each dataset in \Cref{tab:basicStats}.
Comparing the information from \Cref{tab:basicStats} for relevant complete elections to the information for raw elections, it becomes clear that our raw datasets have a varied level of incompleteness.
Examples of raw datasets with a high level of incompleteness are university rankings and football week (which is to be expected because we combined the data of different systems here, each, by design, ranking a different number and subset of candidates) and the two spotify datasets (which is also to be expected because the 200 most listened daily songs are not the same in each country and are also not the same on different days). 
However, even after converting the elections into complete ones, most of them are still quite large with some of them having over $100$ candidates. 
Notably, none of the elections from PrefLib with over $5$ voters identified by \citet{DBLP:conf/ijcai/BoehmerBFNS21} have over $50$ candidates.

As the last step, similarly as done by \citet{DBLP:conf/ijcai/BoehmerBFNS21}, to be able to meaningfully compare the results of our experiments within datasets and between datasets, we created \emph{normalized} elections. 
For each dataset, we created $500$ elections with $15$ candidates and $30$ voters as follows. 
To create an election $E=(C,V)$, we uniformly at random selected one relevant complete election $F=(D,W)$ from the respective dataset. 
Subsequently, we sampled a subset of $15$ candidates $C$ uniformly at random from $D$. 
After that, to create $V$, we sampled $30$ times a vote uniformly at random from $W$ with replacement.
This means that a vote from $W$ can occur potentially multiple times in $V$ and that different normalized elections might be based on $F$.
In all our experiments presented in the following sections (with the exception of \Cref{sub:sim_tb}) we only use normalized elections and will no longer explicitly specify this.
We refer to the dataset containing all elections from all datasets as the \emph{aggregated} dataset.

\section{Drawing a Map of Our Elections} \label{sec:map} \label{sub:map}
\begin{itemize}
	\item Our datasets can be roughly partitioned into three groups and  capture a large part of the space of elections, some of which are not captured by elections from Preflib \cite{DBLP:conf/aldt/MatteiW13,DBLP:conf/ijcai/BoehmerBFNS21}.  
	\item Elections from one dataset and ``methodologically'' similarly generated elections are quite close to each other.
\end{itemize}

To get a feeling for the type of our elections and to be able to better relate the datasets to each other, we apply the ``map of elections'' framework.
In this framework, which has been developed by \citet{DBLP:conf/atal/SzufaFSST20} and \citet{DBLP:conf/ijcai/BoehmerBFNS21}, we take a set of elections and compute for each pair their so-called ``positionwise'' distance.\footnote{The positionwise distance is based on the notion of frequency matrices. In the frequency matrix of an election, each column corresponds to a candidate and each row to a position and an entry captures the fraction of voters ranking the respective candidate in the respective position. The distance between two elections then corresponds to the summed earth mover's distance between the columns of their frequency matrices with columns being rearranged to minimize this distance (see \cite{DBLP:conf/atal/SzufaFSST20,DBLP:conf/ijcai/BoehmerBFNS21} for details).\label{fote:pos}}
Afterward, using the embedding algorithm from \citet{fruchterman1991graph}, we draw a map of our elections where each election is represented by a dot with the Euclidean distance between two dots being as similar as possible to the distance between the respective two elections. 
Note that the position of an election on the map thus naturally depends on the set of depicted elections. 

To give a meaning to the absolute position of an election on the map, \citet{DBLP:conf/ijcai/BoehmerBFNS21} introduced what they call a compass consisting of four types of ``extreme'' elections capturing different kinds of (dis)agreement between voters and their convex combinations: 
\begin{description}
	\item[Identity] All voters have the same preference order. 
	\item[Uniformity] Each possible preference order appears exactly once. 
	\item[Antagonism] Half of the voters rank the candidates in the same order, while the other half ranks them in the opposite order.
	\item[Stratification] There is a partitioning of the candidates into two sets $A$ and $B$ of equal size and all possible preference orders where all candidates from $A$ are ranked before those from $B$ appear once. 
\end{description}
\begin{figure*}
	\centering
	\begin{subfigure}[b]{0.49\textwidth}
		\centering
		\includegraphics[width=\textwidth]{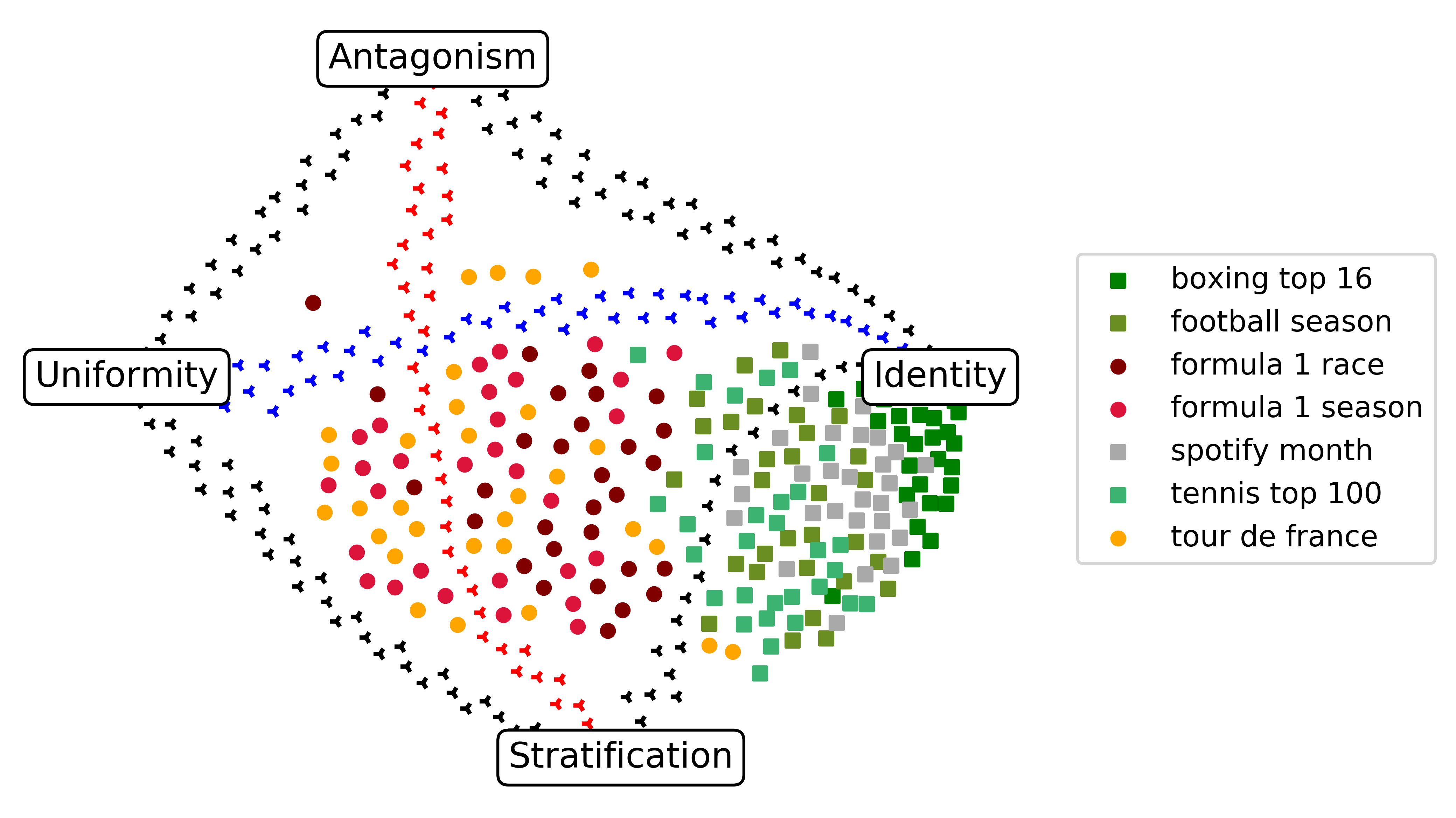}
		\caption{Time-based elections}\label{fig:maptime}
	\end{subfigure}
	\hfill
	\begin{subfigure}[b]{0.49\textwidth}
		\centering
		\includegraphics[width=\textwidth]{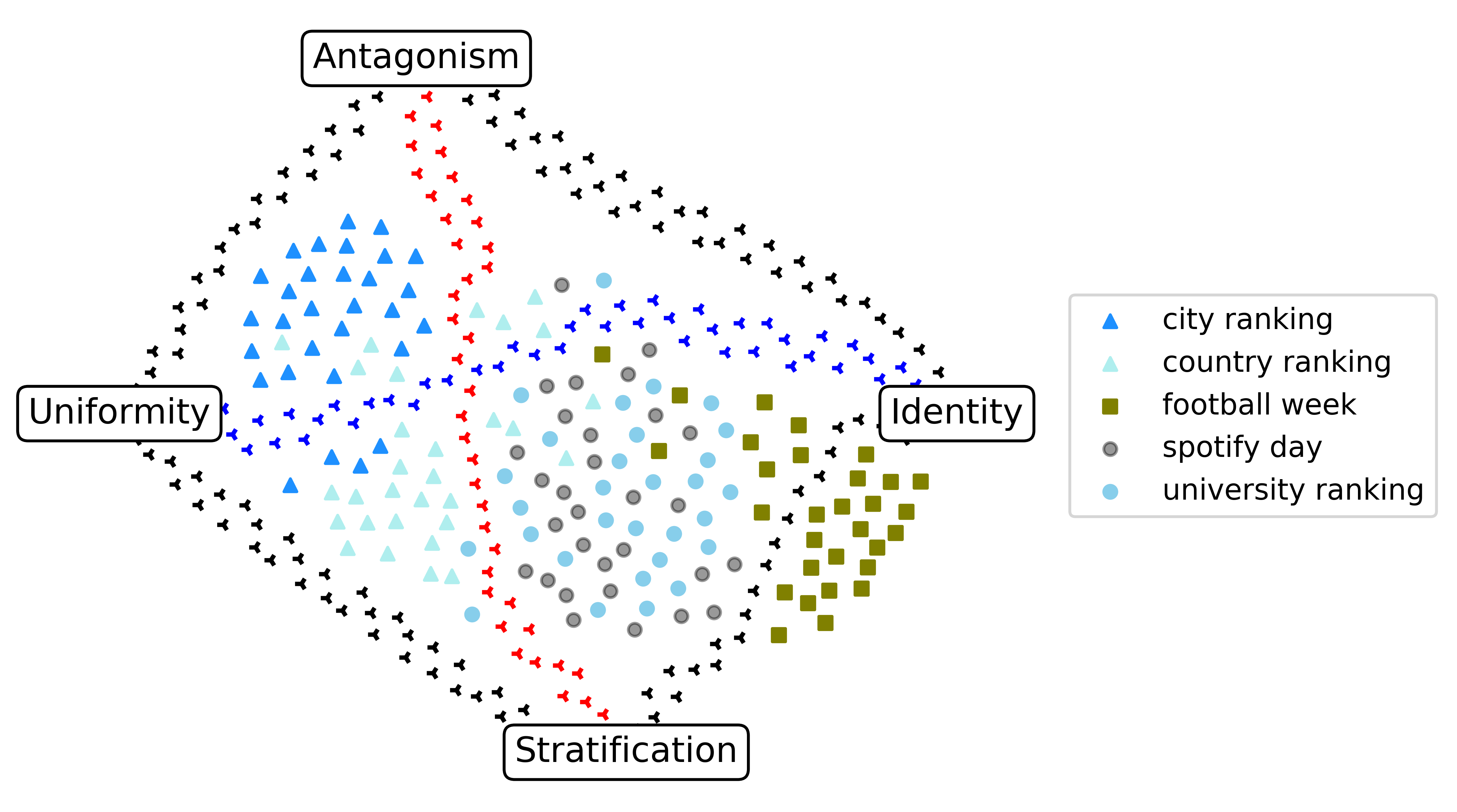}
		\caption{Criterion-based elections}\label{fig:mapcrit}
	\end{subfigure}
	\caption{Visualization of our elections as map of elections.} \label{fig:map}
\end{figure*}

\begin{figure}
	\centering
	\includegraphics[width=0.5\textwidth]{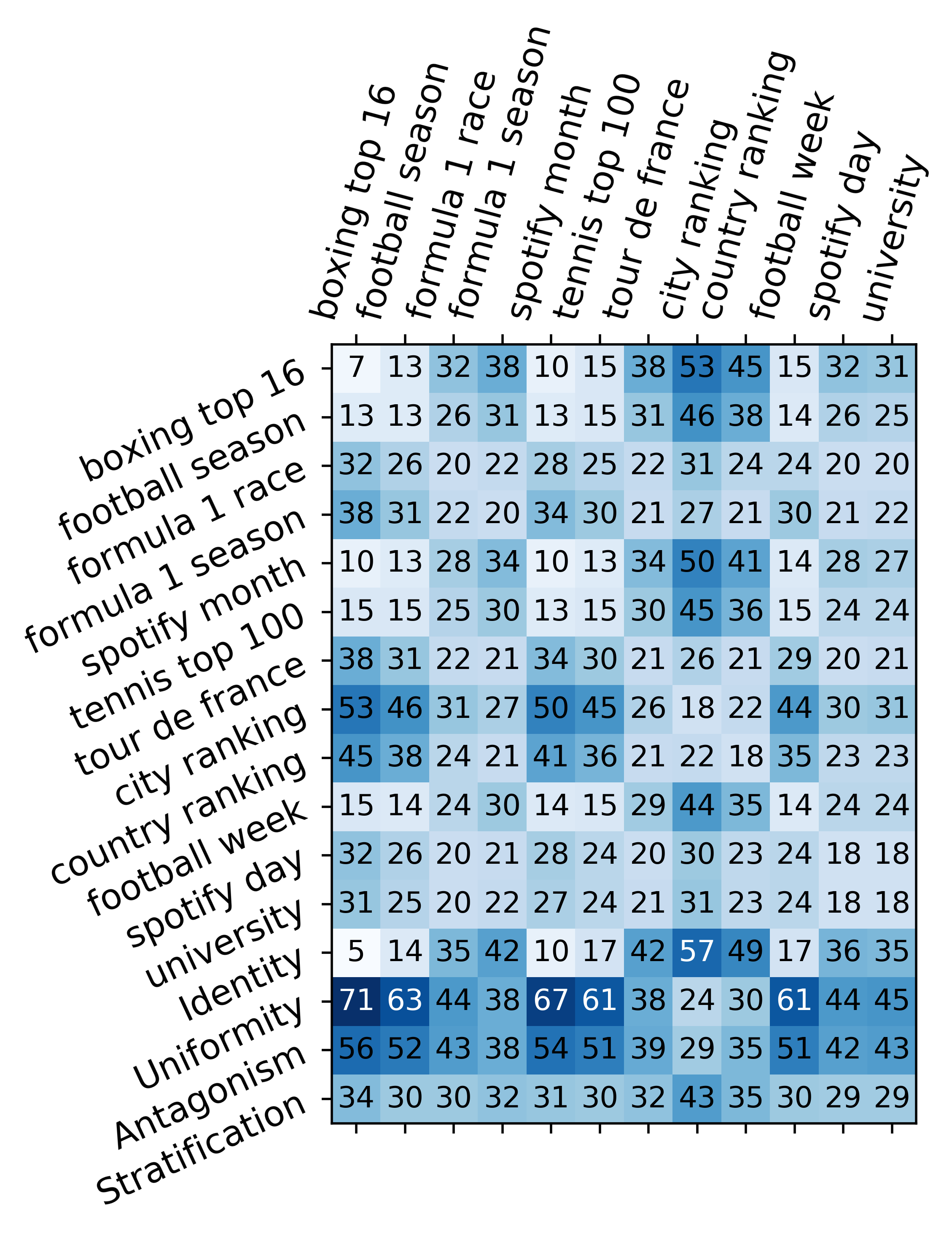}
	\caption{Average positionwise distance of election pairs from different datasets. The maximum distance between two election is $74.67$.}
	\label{fig:positionwisedistance}
\end{figure}

\paragraph{Setup.}
We created two maps of elections (\Cref{fig:map}) where each election is represented by a point whose shape and color indicate the dataset to which it belongs. 
To make the created maps not too crowded, we created a separate map for time-based (\Cref{fig:maptime}) and criterion-based (\Cref{fig:mapcrit}) elections. 
For each map, we included $30$ elections sampled uniformly at random from each normalized dataset and the compass elections introduced by \citet{DBLP:conf/ijcai/BoehmerBFNS21} together with their convex combinations appearing as ``paths''.   
Moreover, in \Cref{fig:positionwisedistance}, we also depict for each pair $A$ and $B$ of datasets, the average positionwise distance of elections from $A$ to elections from $B$ (note that in case we have $A=B$, we take the average over all pairs of elections from $A$). 
In the last four columns, we show the datasets' average distance from the compass elections.

\paragraph{Classifying Datasets.}
Examining \Cref{fig:map,fig:positionwisedistance}, it is possible to divide our datasets into three groups: 
The first group of datasets (boxing top 16, football season, spotify month, tennis top 100, and football week) drawn as squares all contain elections somewhat close to identity (see also \Cref{fig:positionwisedistance}). 
Notably, except for football week\footnote{Recalling that in football week elections the strength of college football teams at one point are judged by different systems (votes), it is also quite intuitive that these elections are close to identity, as one could argue that there exists a ``ground truth''.}, these are all time-based datasets. 
For all of them except spotify month, the ranking at a certain point in time also depends on previous information on candidates that also already influenced previous votes, so in some sense, votes are ``by design'' not independent here.\footnote{For spotify month this is not really the case ``by design''. However, also here similar effects are present. E.g. users often listen to playlists that only change slowly over time, implying that what users listened to on one day in some sense ``predicts'' what they will listen to on the next day.}
In contrast to this, in time-based elections from the other datasets (Formula 1 race, Formula 1 season, and Tour de France), which are further away from identity, one vote only depends on the performance of a candidate at some point in time (and not on previous performances)

The second group of datasets (Formula 1 race, Formula 1 season, Tour de France, spotify day, and university rankings)  drawn as circles constitute the ``middle'' part of our maps: This is also reflected in them being roughly at the same distance from identity and uniformity (while all are clearly closer to stratification than to antagonism; see also \Cref{fig:positionwisedistance}). 
What is particularly striking here is that despite the fact that these elections are seemingly not all simply close to a canonical extreme election like identity, there are surprising similarities between the datasets: 
In particular, university, Formula 1 race, and spotify day elections all fall in exactly the same area of the space of elections (the average distance of two elections from one of these datasets is very close to the average distance of two elections picked from two different of these datasets). 
The same also holds for Tour de France and Formula 1 season elections. 
Remarkably, Tour de France and Formula 1 season elections are also by design of a very similar nature in the sense that in both datasets players compete in a similar task on different days.
The similarity of these datasets indicates that whether players drive in cars or ride bicycles seems to be not so crucial for the resulting election (similar observations apply to boxing top 16 and tennis top 100, and city rankings and country~rankings).

The third group of datasets consists of city and country rankings and is drawn as triangles. 
Both are clearly different from the rest as they are significantly closer to uniformity than identity. 
Remarkably, the city ranking dataset is the only one of our datasets and the first known dataset which is significantly closer to antagonism (distance $29$) than stratification (distance $43$). 
Considering the underlying data which provides ratings of cities according to different indicators, the ``closeness'' to antagonism is quite plausible, as some of the studied indicators seem to capture in some sense contradicting objectives, e.g., big cities where inhabitants typically have access to a variety of healthcare facilities (being one of the indicators) are typically also quite polluted (being another indicator).

\paragraph{Captured Part of the Space of Elections.} It seems that our datasets contain elections of a different nature than those available on PrefLib: \citet[Figure 2b]{DBLP:conf/ijcai/BoehmerBFNS21} drew a map of elections including representatives of all PrefLib datasets with at least $10$ candidates, $10$ votes, and not too many ties. 
They observed that most elections are closer to uniformity than identity and closer to stratification than antagonism, thereby ending up in the bottom left quadrant of the map. 
In contrast to this, our elections are mostly located in the bottom right quadrant.
Nevertheless, we can confirm the observation of \citet{DBLP:conf/ijcai/BoehmerBFNS21} that real-world elections typically end up closer to stratification than antagonism (we also do not provide any elections that are in the top right quadrant).

\paragraph{Homogenicity of Elections from one Dataset.}
\Cref{fig:map,fig:positionwisedistance} also shed some light on the ``diversity'' of elections from one dataset: 
Looking at the maps, we can observe that while there is some mixing between elections from different datasets, elections from one dataset typically fall into the same area of the map. 
This indicates a certain kind of shared structure. 
However, this degree of homogenicity of a dataset, which can be quantified as the average distance of election pairs from this dataset (see ``diagonal'' entries from \Cref{fig:positionwisedistance}) depends on the dataset:
At the one extreme are boxing top 16, spotify month, and football season with an average distance of $7$, $10$, and $13$, respectively, and on the other extreme are Tour de France, Formula 1 race and Formula 1 season with an average distance of $21$, $20$, and $20$, respectively.

\section{Similarity Measures and their Correlation}  \label{sec:sim}
In addition to our analysis from the previous section based on the map of elections, in this section, we focus on one structural property of our elections, i.e., the similarity of different votes in one election. 
In \Cref{sub:sim}, we analyze how similar different votes from one election are using four different metrics and also inspect the metric's correlation. 
In \Cref{sub:sim_parts}, we analyze whether the top part, middle part or bottom part of different votes are more similar to each other.
Lastly, in \Cref{sub:sim_tb}, we restrict our focus to time-based elections and analyze the similarity of successive votes in those elections.

\subsection{Similarity Measures and their Correlation} \label{sub:sim}
\begin{itemize}
	\item The NP-hard to compute Kemeny score is always highly correlated with the average KT-distance of all vote~pairs and the EMD-positionwise distance from Identity.
	\item All datasets are quite homogenous with respect to the similarity of votes in elections.
	\item For most of our elections the values of all similarity measures are not small. 
\end{itemize}
In this subsection, we compare four measures capturing different facets of similarity. 
Similarity measures are also a potentially attractive parameter to develop parameterized algorithms because they can be understood as a ``distance from triviality'' parameterization, as most computational problems are easy if all votes are the same (see, e.g.,~\cite{DBLP:journals/tcs/BetzlerFGNR09}).

\paragraph{Setup.}
We consider four similarity measures:
\begin{description}
	\item[Maximum KT-distance] The maximum KT-distance among all pairs of votes: $\max_{v,v'\in {{V}\choose{2}}} \swap(v,v')$. 
	\item[Average KT-distance] The average KT-distance among all pairs of votes: $\nicefrac{\sum_{v,v'\in {{V}\choose{2}}} \swap(v,v')}{|{{V}\choose{2}}|}$.
	\item[Disagreeing pairs] The number of candidate pairs for which not all votes agree on their ordering: $|\big\{\{c,c'\}\in {{C}\choose{2}} \mid \exists v,v'\in V: c\succ_v c' \wedge c'\succ_{v'} c \big\}|$.
	\item[Kemeny score] The minimum summed KT-distance of a central order to all votes: $\min_{v^*\in \mathcal{L}(C)} \sum_{v\in V} \swap(v,v^*)$.\footnote{To compute the Kemeny score, we used code from \citet{DBLP:journals/aamas/BetzlerBN14}.}
\end{description}  
Note that the number of disagreeing pairs is always at least as large as the maximum KT-distance, which in turn is at least as large as the average KT-distance (all three values range from $0$ to $|{{C}\choose{2}}|$ so from $0$ to $105$ in our case).

\begin{figure}
	\centering
	\includegraphics[width=0.6\textwidth]{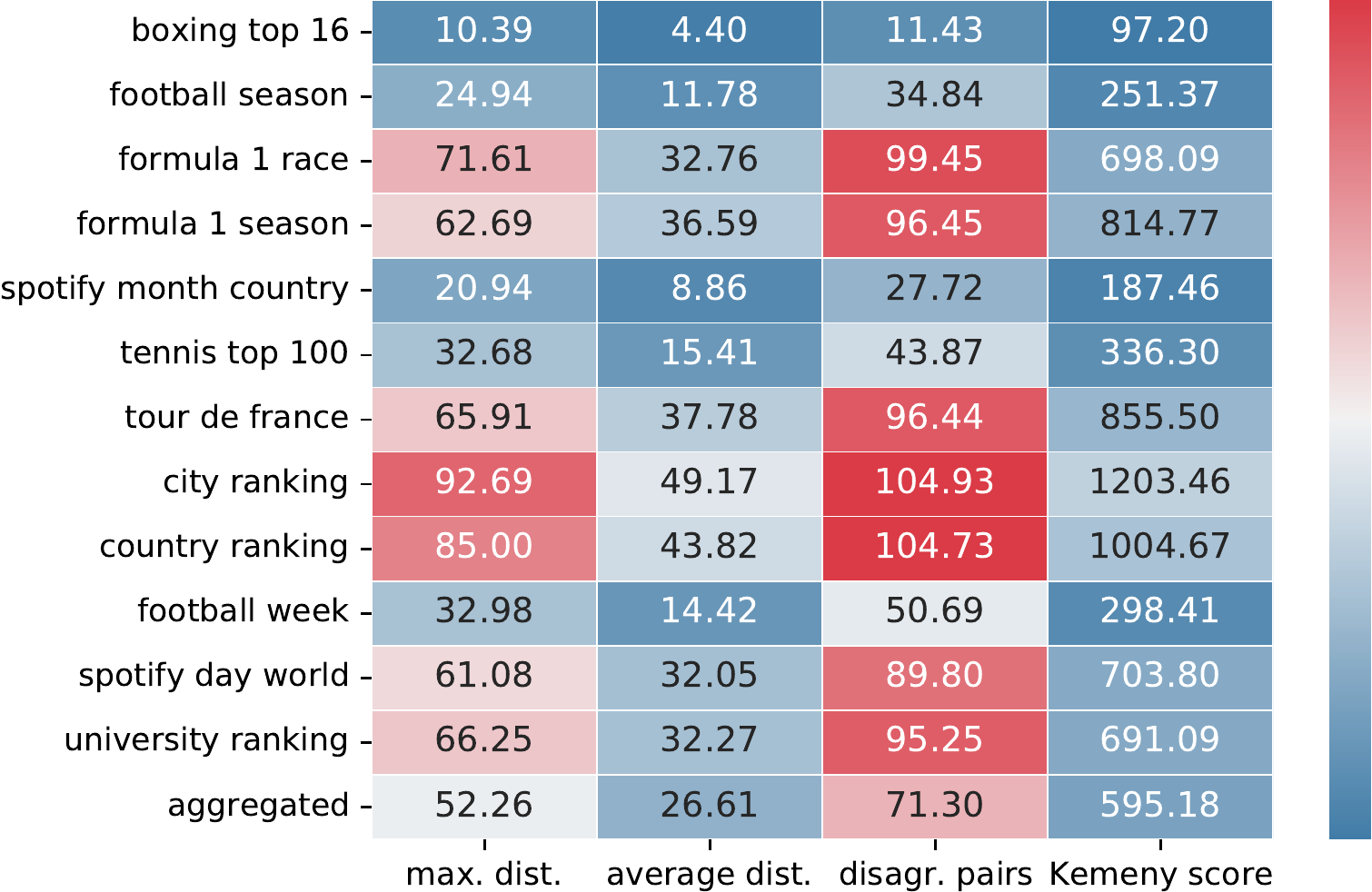}
	\caption{Average values for four different similarity measures. Colors encode the values normalized by the theoretically possible maximum value.}
	\label{fig:average-simmeasures}
\end{figure}

\paragraph{Values of Similarity Measures.}
In \Cref{fig:average-simmeasures}, for all four measures,  we depict for each dataset the value of the similarity measure averaged over all $500$ elections from the dataset.
Concerning the results on the aggregated dataset, what stands out is that the maximum KT-distance and the number of disagreeing pairs is quite high and in particular much higher than the average KT-distance (and comparing normalized values also than the Kemeny score). 
However, this is also quite intuitive in the sense that both the maximum distance and the number of disagreeing pairs might in the end only depend on two voters and are thus very sensitive to ``outliers'' (as soon as there are two voters with reversed preferences orders in an election, both values are at the maximum). 
Considering the results on the different datasets, especially the average number of disagreeing pairs clearly divides them (in line with our groups proposed in \Cref{sub:map}): 
Unsurprisingly, the datasets close to identity have a ``low'' average number of disagreeing pairs (always below 50).
The number is the lowest for boxing top 16 and spotify month with $11.43$ and $27.72$, respectively. This is quite remarkable as it means that \emph{all} voters agree on the ordering of $89.1\%$ and $73.6\%$ of all candidate pairs, respectively. 
For the ``middle'' datasets, the average number of disagreeing pairs is much higher and lies between $89.8$ and $99.45$ (this means that the voters only agree on the ordering of between $5.4\%$ and $14.4\%$ of all candidate pairs). For the two ``outliers'', city and country ranking, the average number of disagreeing pairs is very close to the maximum possible value of $105$ with $104.93$, respectively, $104.73$. 
As already discussed in \Cref{sub:map} one reason for this might be that in the two ``outlier'' datasets votes correspond to sometimes contradicting and opposing indicators, which can lead to two close-to-reversed votes.

Interestingly, for all considered datasets, a majority of elections from the dataset have close similarity scores. 
That is, for all four measures, the median and average value are nearly identical and the first and the third quantile differ only by around $10\%$ from the median.

\paragraph{Similarity Measures for Parameterized Algorithms.}
\citet{DBLP:journals/tcs/BetzlerFGNR09} developed different parameterized algorithms for computing the central order minimizing the Kemeny score: One algorithm running in $\mathcal{O}^*(2^m)$, where $m$ is the number of candidates. Another algorithm running in $\mathcal{O}^*(1.53^k)$ where $k$ is the Kemeny score, and an algorithm running in $\mathcal{O}^*(16^d)$ where $d$ is the average KT-distance (they also considered the maximum KT-distance between two votes as a parameter for a related problem).
Considering the average values on the aggregated dataset, the exponential part of the running time of these algorithms evaluate as follows. $2^m$ is $32768$, $1.53^k$ is \num{7.79e109}, and $16^d$ is \num{1.1e32}.   
Even on boxing top 16, where votes are most similar to each other, the number of candidates still leads to the best results ($2^m$ is $32768$, $1.53^k$ is \num{8.2e17}, $16^d$ is $198668$), partly questioning the practical usefulness of the algorithms for the two similarity parameterizations. 
Overall, it seems that the number of candidates is nearly always the best of our parameters to use. 
Considering the different similarity measures, the average KT-distance is clearly the smallest, which is also theoretically guaranteed; however, the gap to the other parameters might be seen as unexpectedly large.

\begin{figure}
	\centering
	\includegraphics[width=0.6\textwidth]{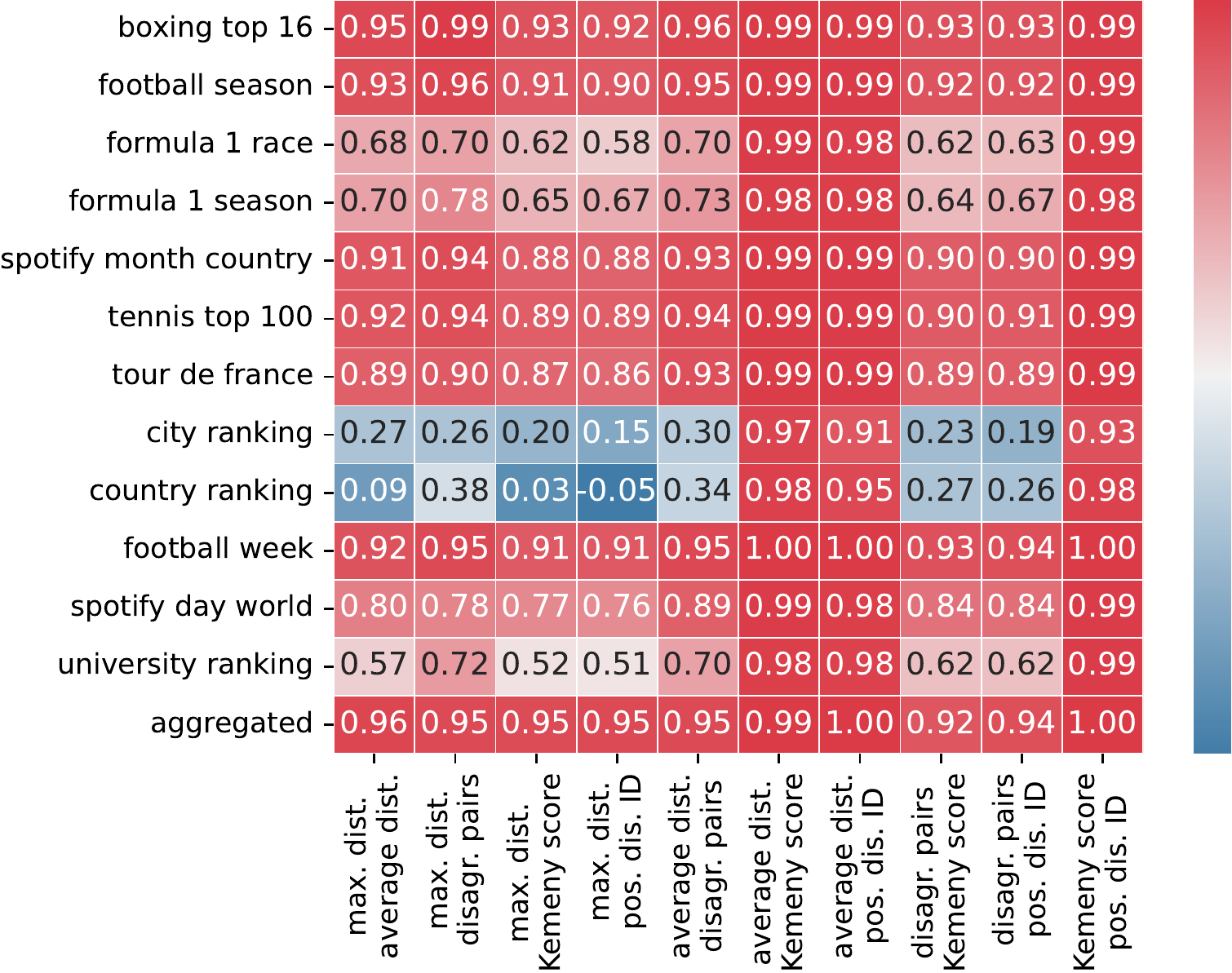}
	\caption{Correlation of similarity measures. Colors represent the respective value (blue means $0$; read means $1$).}
	\label{fig:correlation-simmeasures}
\end{figure} 
\paragraph{Correlation of Similarity Measures.}\label{sub:simcorr}
We also computed for each pair of metrics their Person correlation coefficient (see \Cref{fig:correlation-simmeasures}).
Here we also considered election's EMD-positionwise distance  (see \Cref{fote:pos}) from the Identity election, where all voters have the same preferences. 
Recall that this distance was also essential for our previously proposed classification of datasets.
For all our datasets, the correlation between the average KT-distance and the Kemeny score is between $0.97$ and $1$ (the detected linear relationship here is that the Kemeny score of an election is $23.8$ times the average KT-distance minus $37.8$). 
Moreover, the EMD-positionwise distance from Identity is similarly strongly correlated to the Kemeny score with a PCC value of $1.00$ on the aggregate dataset and PCC values ranging from $0.93$ to $1$ on the level of individual datasets (the detected linear relationship here is that the Kemeny score of an election is $21.6$ times the EMD-positionwise distance plus $53.8$).
This indicates that in most practical applications where one is interested in the  NP-hard to compute Kemeny score, it is sufficient to simply compute the average KT-distance or the EMD-positionwise distance from Identity.
Accordingly, the correlation between the EMD-positionwise distance and the average KT-distance is also similarly high.
For the other pairs of metrics, the correlation on the aggregate dataset is again very high ranging from $0.93$ to $0.96$, which is quite surprising given the different nature of the measures (also in terms of how sensitive they are to outliers). 
However, here there are some differences on the dataset level:
The two examples where the correlation is the lowest are city and country elections, where all pairs of metrics except the three pairs discussed above have a linear correlation between only $0.03$ and $0.34$.  
This, again, might be due to the fact that in these datasets votes are sometimes reverses of each other. 
This severely affect the maximum KT-distance and the number of disagreeing pairs yet only partly increase the average KT-distance, Kemeny score, and EMD-positionwise distance from Identity. 
However, it remains unclear why also the correlation between the number of disagreeing pairs and the maximum KT-distance is quite low here.

\subsection{Similarity in Different Parts of Votes} \label{sub:sim_parts}
\begin{itemize}
	\item Voters typically agree more on which candidates should be considered as high-quality or low-quality candidates than who should be considered as medium-quality candidates.
	\item Voters tend to rank candidates at the top more consistently in the same ordering than candidates at the bottom.
\end{itemize}

Having analyzed the general similarity of votes in one election, we now ask whether certain parts of votes exhibit a higher similarity than others.
For this, we divide each vote (consisting of $15$ candidates) into three parts each containing $8$ candidates: the \emph{top} part capturing positions one to eight, the \emph{middle} part capturing positions five to twelve, and the \emph{bottom} part capturing positions eight to fifteen (note that the different parts partly overlap).

\paragraph{Setup.}
For each election and each of the three parts, we computed two similarity measures:
\begin{description}
	\item[Pairwise intersection] We compute for each pair of votes the number of candidates that appear in both votes in the considered part. The pairwise intersection is this value averaged over all pairs of votes in the election.
	\item[Total intersection] Number of candidates that appear in all votes in the considered part.
\end{description}

\begin{figure}
	\centering
	\begin{subfigure}[b]{0.37\textwidth}
		\centering
		\includegraphics[width=\textwidth]{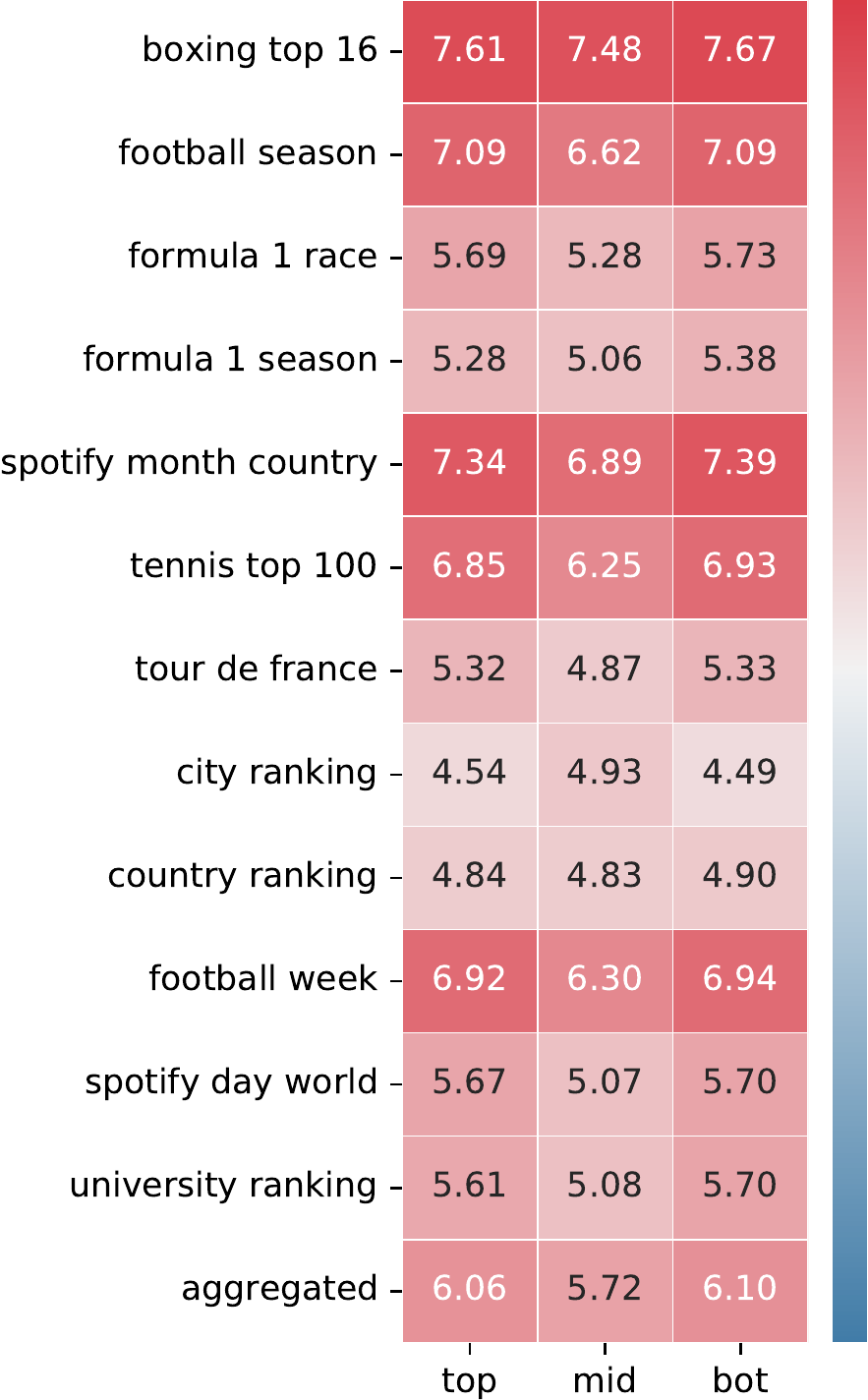}
		\caption{Pairwise intersection.}\label{fig:sect_params_big_pair} 
	\end{subfigure}
	\qquad \qquad \qquad
	\begin{subfigure}[b]{0.37\textwidth}
		\centering
		\includegraphics[width=\textwidth]{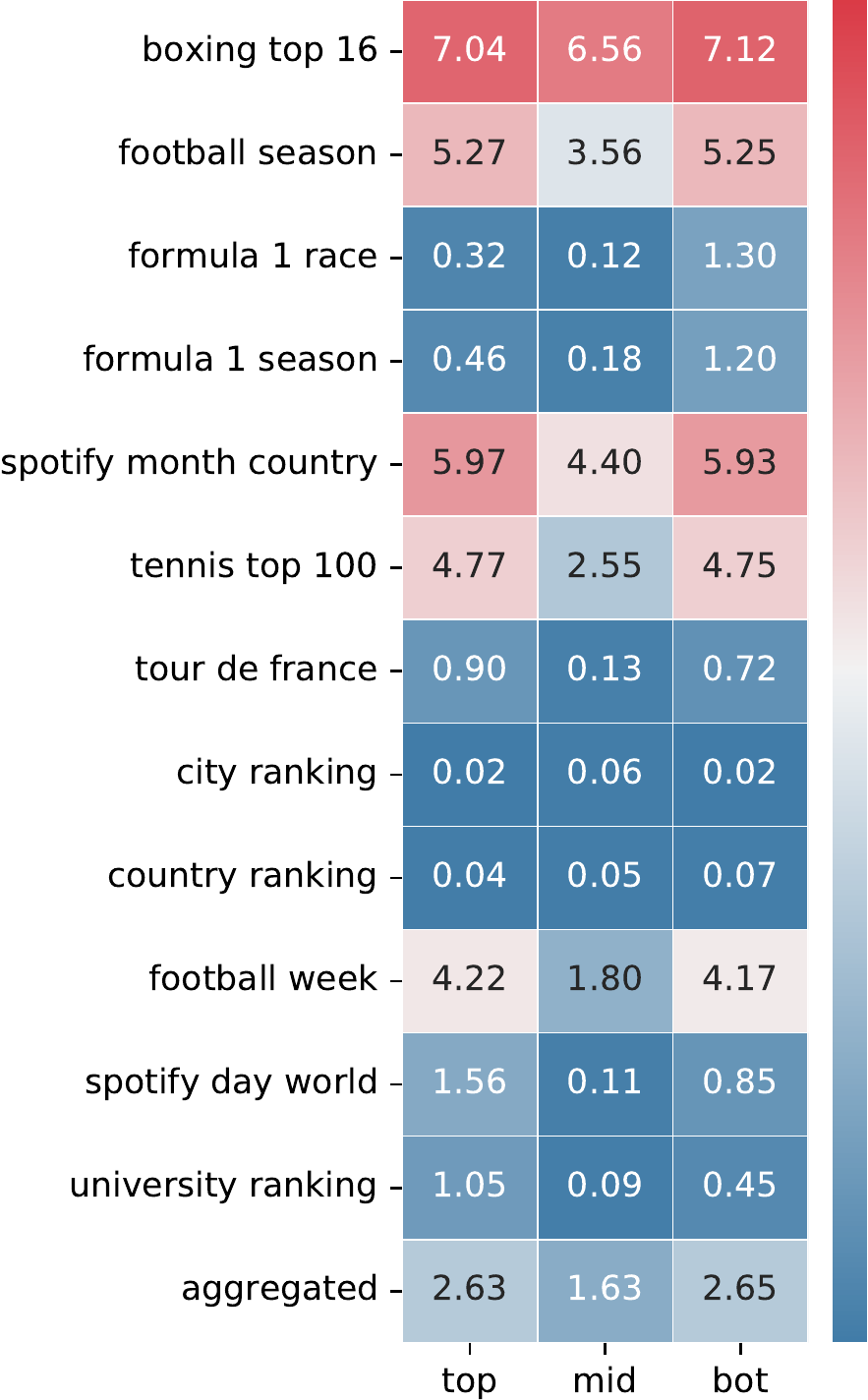}
		\caption{Total intersection.}\label{fig:sect_params_big_tot} 
	\end{subfigure}
	\caption{For each dataset, for two different similarity measures, average values for the three parts of the votes.} 
	\label{fig:sect_params_big} 
\end{figure}

\paragraph{Results.}
Averaged over all elections, the pairwise intersection in the top, middle, and bottom part is $6.06$, $5.27$, and $6.1$, respectively, while the total intersection is $2.63$, $1.63$, and $2.65$, respectively (see   \Cref{fig:sect_params_big} for separate results for each dataset).  
Recalling that each constructed part consists of eight candidates it is remarkable that each pair of votes roughly agrees on six candidates in each part.
This indicates a high pairwise consensus of voters on the general quality of candidates in all of our elections. 
In contrast to this, the number of candidates appearing in all votes in the same part is quite low, indicating a lower overall consensus. 
However, datasets that are close to identity also exhibit a high overall consensus, i.e., the total intersection for the top and bottom part ranges from $4.17$ to $7.12$, whereas for the middle part the values are a bit lower ranging from $1.8$ to $6.56$.

Comparing the different parts of the votes to each other, for all datasets except city rankings, the average and total intersection for the top and bottom part is higher than for the middle part. 
This suggests that while voters have similar opinions about who should be considered as a high-quality or low-quality candidate, opinions are more diverse concerning medium-quality candidates.
The difference is particularly strong for football week elections where the total intersection for the top and bottom part is $4.2$, but only $1.8$ for the middle part. 
Recalling that football week elections are based on power rankings of football teams published by different media outlets in the same week this is also quite intuitive, as there are typically some clearly strong and clearly weak teams that appear in the top, respectively, the bottom of each ranking, while the middle part is more subjective. 

The general trend that voters in most of our elections agree more which candidates should be ranked on the top or the bottom than in the middle also provides a possible justification why a great majority of our elections is closer to stratification than to antagonism (as discussed in \Cref{sec:map}). 
Under antagonism, half of the voters rank the candidates in one order and the other half of the voters rank them in the opposite order.
Consequently, all voters agree on which are the medium-quality candidates while the electorate is divided concerning who are the best and worst candidates. 
This is exactly the opposite of what we observe in our data. 

For the top and bottom part of elections from datasets close to identity, the total intersection value is large enough to also meaningfully analyze whether voters opinions about the \emph{ordering} of candidates are more similar at the top or at the bottom:
For this we computed for each election the average KT-distance between two votes restricted to the candidates appearing in all votes in this part. 
The results clearly indicate a stronger consensus at the top than at the bottom, in particular in boxing top 16 and tennis top 100 elections. 
For these two datasets this can be explained by the fact that in these elections it should be much more difficult to climb up on position when you are in the top part (you are one of the best players) compared to when you are in the bottom part (you are one of the worst ranked players).

\subsection{Similarity in Time-Based Elections} \label{sub:sim_tb}
\begin{itemize}
	\item For all time-based datasets except Formula~1 season, Formula~1 race and Tour de France, votes that appear closer to each other in time are more similar to each other.
	\item In time-based elections, candidates ranked in first and last position change less frequently than the candidates in other positions. 
\end{itemize}

For time-based elections, votes have a natural ordering which makes it possible to analyze the change of votes over time. 
In this section, we do not consider the normalized datasets (because they have no ordering); instead, for each raw election from a time-based dataset, we deleted all candidates that do not appear in all votes and subsequently discarded the election if less than $10$ candidates remain (it is important that we do not delete voters, as we also analyze the change between one voter and the following; we call two such voters \emph{successive}). 

\paragraph{Setup.}
For each election, we computed the following: 
\begin{description}
	\item[Average ordering change] The average number of times the pairwise ordering of two candidates is swapped over time. 
	\item[Maximum ordering change] The maximum number of times the pairwise ordering of any two candidates is swapped over time.
	\item[Average fluctuation] For each position, we compute the number of votes where the candidate on this position is different in the next vote. The average fluctuation is this value averaged over all positions.
\end{description}
Notably, the average ordering change value times the number of candidate pairs gives the summed KT-distance of all pairs of successive votes. 

However, evaluating  the values of these measures without additional information is very difficult, as the observed values might simply be due to the structure of the election and not due to the ordering of the votes (for instance, if two candidates have the same pairwise ordering in all but one vote then it is nearly irrelevant how the votes are ordered for our measures, while if they are ranked in the same order in half of the votes and in the opposite order in the other half, then the ordering of votes truly makes a difference). 
That is why, for each considered election, we created a copy where we shuffled all votes randomly and recomputed the above quantities. 
The average values for each dataset can be found in \Cref{tab:timestats}.

\begin{table}
	\begin{center}
		\begin{tabular}{cccccccccc}
			\toprule
			name & 
			\multicolumn{4}{c}{original} & \phantom{a} &
			\multicolumn{3}{c}{shuffled}\\ \cmidrule{2-5}
			\cmidrule{7-9}
			& \makecell{Avg.\\ ord. ch.} & \makecell{Max.\\ ord. ch.} & \makecell{Avg.\\ fluct.} &  \makecell{Cor.\\ KT+temp.} &  & \makecell{Avg.\\ ord. ch.} & \makecell{Max.\\ ord. ch.}  & \makecell{Avg.\\ fluct.}  \\ \midrule \midrule
			%------
			boxing top 16 & 0.82 & 2.82 & 2.7 & 0.7 && 1.97  & 10.08 & 14.98\\
			football season & 1.14 & 5.15 & 10.3 & 0.46 && 1.49 & 6.24 & 11.09\\
			Formula 1 race & 8.96 & 15.84 & 45.98 & 0.04 && 11.96  & 18.57 & 49.75\\
			Formula 1 season & 4.05 & 7.26 & 14.52& 0.03 && 4.08  & 7.17 & 14.53\\
			spotify month & 1.22 & 9.38  & 22.54& 0.86 && 1.92  & 11.64 & 25.64 \\
			tennis top 100 & 1.31 & 6.86 & 23.16& 0.93 && 4.43  & 16.9 & 42.79\\
			Tour de France & 4.63 & 9.27 & 19.47 & 0.05 && 4.84  & 9.44 & 19.53\\
			\bottomrule
		\end{tabular}
		\caption{For each dataset, average values of time-based similarity measures for original and shuffled elections and correlation between KT- and temporal distance.} \label{tab:timestats}
	\end{center}
\end{table} 

Lastly, for each dataset, we also computed the correlation between the KT- and temporal distance of two votes. 
For this for each pair of votes in some election from this dataset, we computed the KT-distance and the temporal distance (as the number of votes that come in between the two plus one). 
Afterwards we computed the PCC of these values. 
The results can be found in fifth column of \Cref{tab:timestats}.

\paragraph{Results.}
We first discuss the ordering change.
Comparing the results for the original and shuffled elections in \Cref{tab:timestats}, no clear difference is visible for Formula 1 season and Tour de France elections (remarkably, both datasets capture similar types of elections where players compete in multiple stages on different days against each other). 
This indicates that in elections from these datasets the time-based ordering of the vote does not have any direct consequences on the votes, i.e., performances of candidates in one stage have no clear connection to their performance in the next stage. 
For all other datasets, the time-based ordering induces a more ``structured'' election than a random ordering, indicating that successive voters have some form of relationship here. 
The most extreme example is tennis top 100, where following the true order of the votes the ordering of each candidate pair only changes $1.31$ times, on average, while following a random order it changes $4.43$ times.  
That votes have a strong temporal relationship in these elections can be explained by considering the design of the underlying ranking: 
Players gain points based on their performance in major tournaments and keep them for roughly one year. 
Thus, the number of points a player has typically does not drastically change between weeks (each corresponding to a vote) implying some form of temporal consistency. 
However, also in elections that are subjected to a higher degree of change such as Formula 1 race elections the temporal dimension of the data is visible: 
In the original elections, the ordering of the candidates (which in this context means which of the candidates needed less time for the respective lap in the underlying race) changes on average $8.96$ times, while if we shuffle votes (laps) then it changes on average $11.96$ times.

\begin{figure}
	\centering
	\begin{subfigure}[b]{0.35\textwidth}
		\centering
		\resizebox{\textwidth}{!}{\input{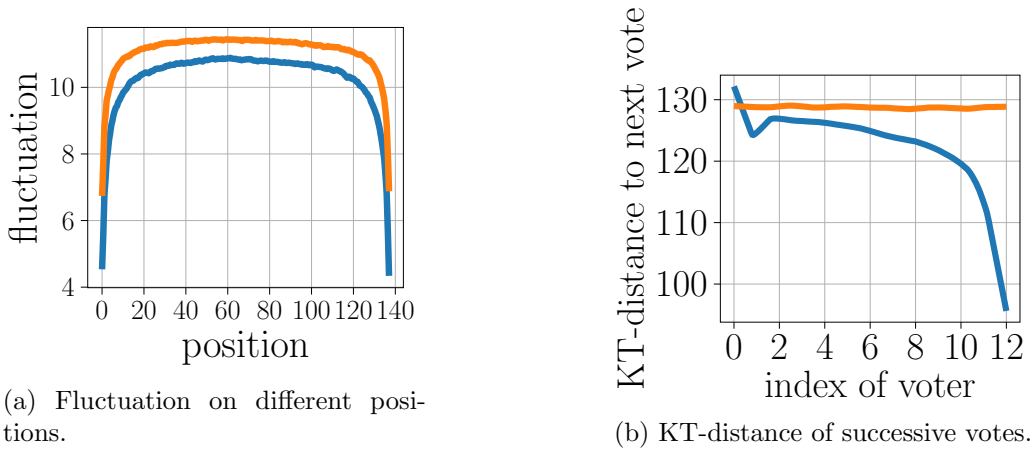}}
		\caption{Fluctuation on different positions.}\label{fig:time-based-example-position}
	\end{subfigure}
	\qquad \qquad \qquad
	\begin{subfigure}[b]{0.35\textwidth}
		\centering
		\resizebox{\textwidth}{!}{% This file was created with tikzplotlib v0.9.17.
\begin{tikzpicture}

\definecolor{color0}{rgb}{0.12156862745098,0.466666666666667,0.705882352941177}
\definecolor{color1}{rgb}{1,0.498039215686275,0.0549019607843137}

\begin{axis}[
tick align=outside,
tick pos=left,
x grid style={white!69.0196078431373!black},
xlabel={index of voter},
xmajorgrids,
xmin=-0.6, xmax=12.6,
xtick style={color=black},
y grid style={white!69.0196078431373!black},
ylabel={KT-distance to next vote},
ymajorgrids,
ymin=93.796273870517, ymax=133.998789007918,
ytick style={color=black},
every tick label/.append style={font=\Huge}, 
label style={font=\Huge}
]
\addplot [line width=4pt, color0]
table [row sep=\\] {%
0 132.171401977539\\0.756756782531738 124.648719787598\\0.792792797088623 124.36100769043\\0.804804801940918 124.299980163574\\0.852852821350098 124.26513671875\\0.864864826202393 124.278327941895\\0.924924850463867 124.406883239746\\1.04504501819611 124.786285400391\\1.11711716651917 125.033477783203\\1.32132136821747 125.791488647461\\1.44144141674042 126.254852294922\\1.54954957962036 126.660003662109\\1.59759759902954 126.833168029785\\1.60960960388184 126.849243164062\\1.71771776676178 126.930625915527\\1.84984982013702 126.95051574707\\2.04204201698303 126.890983581543\\2.21021032333374 126.825248718262\\2.28228235244751 126.786643981934\\2.43843841552734 126.697082519531\\2.59459447860718 126.622116088867\\2.69069075584412 126.589149475098\\2.79879879951477 126.564277648926\\3.01501512527466 126.526222229004\\3.23123121261597 126.463966369629\\3.36336326599121 126.433120727539\\3.44744753837585 126.415000915527\\3.61561560630798 126.393135070801\\3.73573565483093 126.357971191406\\3.78378367424011 126.338859558105\\4.03603601455688 126.228942871094\\4.31231212615967 126.083343505859\\4.48048067092896 125.978805541992\\4.54054069519043 125.945869445801\\4.90090084075928 125.755424499512\\5.2132134437561 125.57738494873\\5.52552556991577 125.377433776855\\5.57357358932495 125.33992767334\\5.62162160873413 125.297233581543\\6.04204225540161 124.875915527344\\6.63063049316406 124.229270935059\\6.70270252227783 124.162300109863\\6.84684705734253 124.041358947754\\6.93093109130859 123.967018127441\\7.03903913497925 123.874313354492\\7.21921920776367 123.733863830566\\7.49549531936646 123.552185058594\\7.65165185928345 123.439971923828\\7.72372388839722 123.380577087402\\8 123.206642150879\\8.3723726272583 122.745429992676\\8.42042064666748 122.686180114746\\8.60060024261475 122.440620422363\\8.84084129333496 122.068817138672\\9.02102088928223 121.760711669922\\9.47747707366943 120.850517272949\\9.60960960388184 120.570388793945\\9.88588619232178 119.880462646484\\10.0180177688599 119.519973754883\\10.2582578659058 118.749435424805\\10.2822818756104 118.670936584473\\10.3303298950195 118.449012756348\\10.4144144058228 118.037155151367\\10.5105104446411 117.474891662598\\10.6786785125732 116.329803466797\\10.8108110427856 115.23104095459\\10.9429426193237 113.895111083984\\11.0750751495361 112.448234558105\\11.1471471786499 111.559211730957\\11.2192192077637 110.345443725586\\12 95.6236572265625\\};
\addplot [line width=4pt, color1]
table [row sep=\\] {%
0 128.964599609375\\0.876876831054688 128.805221557617\\0.936936855316162 128.791290283203\\1.33333337306976 128.767379760742\\1.52552556991577 128.770278930664\\1.68168163299561 128.78971862793\\1.71771776676178 128.796127319336\\1.87387382984161 128.871887207031\\2.03003001213074 128.931182861328\\2.41441440582275 129.041839599609\\2.58258247375488 129.038757324219\\3.03903913497925 128.906768798828\\3.25525522232056 128.820251464844\\3.30330324172974 128.800598144531\\3.43543553352356 128.744338989258\\3.60360360145569 128.723571777344\\3.74774765968323 128.737091064453\\4 128.757186889648\\4.28828811645508 128.84765625\\4.39639616012573 128.85302734375\\4.58858871459961 128.876892089844\\4.69669675827026 128.895950317383\\4.80480480194092 128.915084838867\\5.02102088928223 128.905975341797\\5.2132134437561 128.872772216797\\5.4654655456543 128.82438659668\\5.62162160873413 128.805908203125\\6.01801824569702 128.723937988281\\6.42642641067505 128.687805175781\\6.48648643493652 128.691467285156\\6.55855846405029 128.698303222656\\6.70270252227783 128.688003540039\\6.85885906219482 128.663970947266\\7.01501512527466 128.605316162109\\7.23123121261597 128.558212280273\\7.41141128540039 128.529602050781\\7.55555534362793 128.493423461914\\7.68768787384033 128.4677734375\\7.71171188354492 128.464004516602\\7.7597599029541 128.476867675781\\8.14414405822754 128.585830688477\\8.28828811645508 128.62141418457\\8.34834861755371 128.635955810547\\8.40840816497803 128.654296875\\8.57657623291016 128.73957824707\\8.76876831054688 128.730438232422\\8.87687683105469 128.734512329102\\9.02102088928223 128.738967895508\\9.26126098632812 128.716735839844\\9.34534549713135 128.700103759766\\9.44144153594971 128.667755126953\\9.66967010498047 128.637741088867\\9.82582569122314 128.609161376953\\10.1621618270874 128.542877197266\\10.2942943572998 128.533752441406\\10.4144144058228 128.554306030273\\10.5345344543457 128.594314575195\\10.9189186096191 128.747161865234\\11.1351346969604 128.808731079102\\11.1951951980591 128.815567016602\\11.315315246582 128.813690185547\\12 128.856079101562\\};
\end{axis}

\end{tikzpicture}}
		\caption{KT-distance of successive votes.}\label{fig:time-based-example-votes}
	\end{subfigure}
	\caption{Time-based similarity in football season elections. Taking into account that elections have different sizes, we mapped the results for all elections on the average election size and afterwards averaged them to obtain the depicted graphs. In blue, the original data is shown; in orange, the shuffled data.} 
	\label{fig:three graphs}
\end{figure}

Next, for each dataset separately, we consider the PCC of the KT-distance and the temporal distance of each pair of votes in the election as an additional simple measure for the temporal correlation of the votes. 
Here, again, in the Formula 1 season and Tour de France data, no correlation is visible. 
Moreover, this metric also indicates that slightly contrary to our previous observations, there is nearly no linear correlation between the KT- and temporal distance of votes in the Formula 1 race data. 
In contrast to this, for boxing top 16, spotify month and tennis top 100, the correlation is strong. 
For tennis top 100, the correlation value is even $0.93$, indicating a close to perfectly linear correlation.

We also examined the average fluctuation for different positions.
Here, it is typically the case that the first and last positions exhibit less fluctuation, indicating that in time-based elections changes on the first and last position are rarer than changes on other positions. 
We show in \Cref{fig:time-based-example-position} an exemplarily plot for football season elections.
Moreover, we also analyzed the KT-distance between successive votes: 
While for most datasets no clear relationship between the index of votes in the election (i.e., whether it is the first, second, third, ... vote in the election) and its KT-distance to its successive vote is visible, football season elections form a clear exception:
In these elections, the average KT-distance between successive votes is usually around $125$ but drops to around $100$ for the last votes (see \Cref{fig:time-based-example-votes} for a visualization).
Recalling that each vote here represents the power ranking of football teams in different weeks of the season (with the first vote representing the first ranking and the last vote the final ranking) this is quite plausible because typically teams need some time within a season to show their ``true'' quality.

\section{Restricted Domains} \label{sec:rest} \label{sub:restricteddomain-freq} \label{sub:restricteddomain}

\begin{itemize}
	\item There are only few elections from a restricted domain and only some elections close to one.
	\item Elections that are close to one domain are typically also close to another.
	\item Elections from a restricted domain are typically quite degenerate. 
\end{itemize}

In this section, we analyze which of our elections are part of a restricted domain.
There are numerous papers analyzing the computational complexity of various problems on elections from different types of restricted domains (see e.g.,  \cite{DBLP:journals/tcs/SkowronYFE15,DBLP:journals/jair/BetzlerSU13,DBLP:journals/iandc/FaliszewskiHHR11,DBLP:conf/ijcai/FaliszewskiKO20,DBLP:conf/aaai/Walsh07,DBLP:journals/jair/BrandtBHH15,DBLP:journals/aamas/MagieraF17,DBLP:conf/atal/ElkindMOS16}  and  \citet{DBLP:conf/ijcai/ElkindLP16,trendsrestricted} for surveys). 
Possible motivations for these works are typically that restricted domains allow for nice combinatorial algorithms and the belief that they capture (close-to) realistic situations. 
We focus on the three arguably most popular restricted domains of single-peaked \cite{black1948rationale}, single-crossing \cite{mirrlees1971exploration,roberts1977voting}, and group-separable elections \cite{inada1969simple,inada1964note}.

We check here which of our elections fall into one of these domains and afterwards consider the candidate deletion and voter deletion distance of all elections from them.

\paragraph{Members in Restricted Domains.} Overall, only very few of our elections fall into a restricted domain. That is, for the $500$ boxing top $16$ elections, where votes are very similar to each other, the number of single-peaked/singe-crossing/group-separable elections is $77$/$138$/$101$. 
Moreover, we have one single-peaked election in the football season dataset and one in the spotify month dataset. So overall, only $1.3\%$, $2.3\%$, $1.6\%$ of our elections are single-peaked, single-crossing, and group-separable, respectively. 
Some other works have also analyzed the occurrences of elections from restricted domains and found even less evidence: 
\citet{regenwetter2007unexpected} analyzed  five-candidate American Psychological Association (APA) presidential elections and found no evidence of restricted domains. 
\citet{DBLP:conf/aldt/Mattei11} considered three- and four-candidate elections based on a Netflix price competition and found that $0.03\%$ of elections are single-peaked.

\paragraph{Distance to a Restricted Domain.} Given that only a few of our elections fall into a restricted domain, our goal now is to check whether more are at least close to one.
In particular, we consider the voter deletion and candidate deletion distance, i.e., the minimum number of voters/candidates that need to be deleted such that the resulting election falls into the restricted domain.
Notably, there are also many more distance measures (see, e.g., \cite{DBLP:journals/jair/ErdelyiLP17,DBLP:conf/ijcai/CornazGS13,DBLP:conf/sigecom/ElkindFS12}). 
Moreover, motivated by the many polynomial-time results on restricted domains, there are several papers developing parameterized algorithms for election-related problems for different distance measures to restricted domains (see \cite{DBLP:journals/ai/FaliszewskiHH14,DBLP:conf/aaai/MenonL16,DBLP:conf/aldt/MisraSV17} for algorithms parameterized by the voter and candidate deletion distance and \cite{DBLP:conf/ijcai/CornazGS13,DBLP:journals/jcss/YangG17,DBLP:conf/atal/YangG15,DBLP:conf/atal/YangG14a,DBLP:journals/tcs/SkowronYFE15} for examples for other distance measures). 

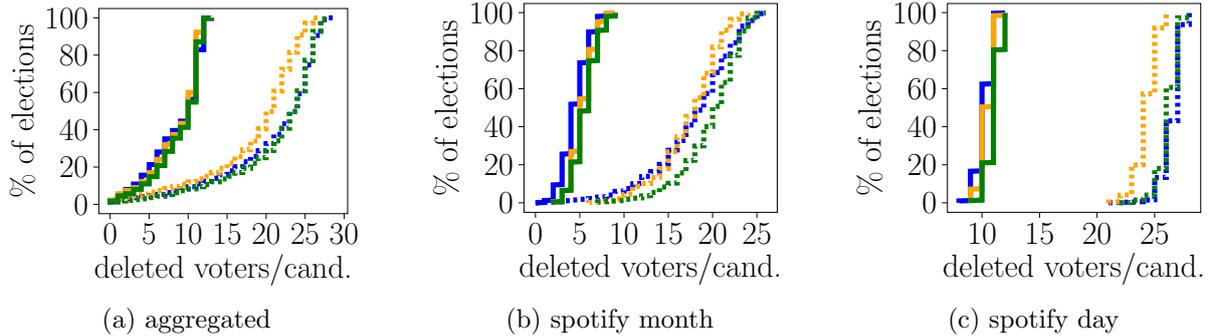
\begin{figure}
	\centering
	\begin{subfigure}[b]{0.3\textwidth}
		\centering
		\resizebox{\textwidth}{!}{% This file was created with tikzplotlib v0.9.17.
\begin{tikzpicture}

\definecolor{color0}{rgb}{1,0.647058823529412,0}

\begin{axis}[
legend cell align={left},
legend style={
  fill opacity=0.8,
  draw opacity=1,
  text opacity=1,
  at={(0.03,0.97)},
  anchor=north west,
  draw=white!80!black
},
tick align=outside,
tick pos=left,
x grid style={white!69.0196078431373!black},
xlabel={deleted voters/cand.},
xmin=-1.45, xmax=30.45,
xtick style={color=black},
y grid style={white!69.0196078431373!black},
ylabel={\% of elections},
every tick label/.append style={font=\Huge}, 
label style={font=\Huge},
ymin=-4.9996, ymax=104.9916,
ytick style={color=black}
]
\addplot [line width=4pt, blue]
table {%
0 0
0 1.29999995231628
1 1.31666672229767
1 4.34999990463257
2 4.36666679382324
2 7.93333339691162
3 7.94999980926514
3 10.8666667938232
4 10.8833332061768
4 15.6333332061768
5 15.6499996185303
5 21.3333339691162
6 21.3500003814697
6 28.0666675567627
7 28.0833339691162
7 35.1166648864746
8 35.1333351135254
8 39.5833320617676
9 39.5999984741211
9 44.5999984741211
10 44.6166648864746
10 56.2999992370605
11 56.3166656494141
11 82.9000015258789
12 82.9166641235352
12 99.5999984741211
13 99.6166687011719
13 99.9833297729492
};
%\addlegendentry{cand-sp}
\addplot [line width=4pt, blue, dashed]
table {%
0 0
0 2.33599996566772
1 2.35199999809265
1 2.6800000667572
2 2.69600009918213
2 3.04800009727478
3 3.06399989128113
3 3.54399991035461
4 3.55999994277954
4 4.08799982070923
5 4.10400009155273
5 4.74399995803833
6 4.76000022888184
6 5.3439998626709
7 5.3600001335144
7 6.30399990081787
8 6.32000017166138
8 7.12799978256226
9 7.14400005340576
9 8.22399997711182
10 8.23999977111816
10 9.40799999237061
11 9.42399978637695
11 11
12 11.0159997940063
12 12.3599996566772
13 12.3760004043579
13 13.9759998321533
14 13.9919996261597
14 15.8719997406006
15 15.8879995346069
15 17.9680004119873
16 17.9839992523193
16 20.1919994354248
17 20.2080001831055
17 22.8400001525879
18 22.8560009002686
18 25.6959991455078
19 25.7119998931885
19 28.7759990692139
20 28.7919998168945
20 32.6080017089844
21 32.6240005493164
21 37.568000793457
22 37.5839996337891
22 43.1599998474121
23 43.1759986877441
23 49.7760009765625
24 49.7919998168945
24 59.3440017700195
25 59.3600006103516
25 74.7119979858398
26 74.7279968261719
26 90.536003112793
27 90.552001953125
27 99.015998840332
28 99.0319976806641
28 99.984001159668
29 99.9919967651367
};
%\addlegendentry{cand-sp}
\addplot [line width=4pt, color0]
table {%
0 0
0 2.28333330154419
1 2.29999995231628
1 5.71666669845581
2 5.73333311080933
2 7.38333320617676
3 7.40000009536743
3 8.96666622161865
4 8.98333358764648
4 12.2666664123535
5 12.2833337783813
5 16.9833335876465
6 17
6 24.1166667938232
7 24.1333332061768
7 32.0499992370605
8 32.0666656494141
8 37.4166679382324
9 37.4333343505859
9 43.4166679382324
10 43.4333343505859
10 60.0166664123535
11 60.033332824707
11 92.283332824707
12 92.3000030517578
12 99.8666687011719
13 99.8833312988281
13 99.9833297729492
};
%\addlegendentry{cand-sc}
\addplot [line width=4pt, color0, dashed]
table {%
0 0
0 2.36800003051758
1 2.3840000629425
1 2.74399995803833
2 2.75999999046326
2 3.64800000190735
3 3.66400003433228
3 4.40000009536743
4 4.41599988937378
4 5.51200008392334
5 5.52799987792969
5 6.58400011062622
6 6.59999990463257
6 7.40799999237061
7 7.42399978637695
7 8.46399974822998
8 8.47999954223633
8 9.52799987792969
9 9.54399967193604
9 10.7440004348755
10 10.7600002288818
10 12.039999961853
11 12.0559997558594
11 13.6560001373291
12 13.6719999313354
12 15.0240001678467
13 15.039999961853
13 17.0639991760254
14 17.0799999237061
14 19.3519992828369
15 19.3680000305176
15 22.1040000915527
16 22.1200008392334
16 25.5279998779297
17 25.5440006256104
17 29.3199996948242
18 29.3360004425049
18 33.9360008239746
19 33.9519996643066
19 40.5519981384277
20 40.568000793457
20 49.7519989013672
21 49.7680015563965
21 60.6800003051758
22 60.6959991455078
22 72.3280029296875
23 72.3440017700195
23 81.5120010375977
24 81.5279998779297
24 90.8399963378906
25 90.8560028076172
25 97.1439971923828
26 97.1600036621094
26 99.9199981689453
27 99.9359970092773
27 99.9919967651367
};
%\addlegendentry{cand-sc}
\addplot [line width=4pt, green!50.1960784313725!black]
table {%
0 0
0 1.66666662693024
1 1.68333327770233
1 4.51666688919067
2 4.53333330154419
2 5.69999980926514
3 5.71666669845581
3 8.03333377838135
4 8.05000019073486
4 10.9666662216187
5 10.9833335876465
5 14.8333330154419
6 14.8500003814697
6 20.7666664123535
7 20.783332824707
7 28.3999996185303
8 28.4166660308838
8 35.466667175293
9 35.4833335876465
9 41.25
10 41.2666664123535
10 54.7999992370605
11 54.8166656494141
11 87.1500015258789
12 87.1666641235352
12 99.9166641235352
13 99.9333343505859
13 99.9833297729492
};
%\addlegendentry{cand-gs}
\addplot [line width=4pt, green!50.1960784313725!black, dashed]
table {%
0 0
0 1.67200005054474
1 1.68799996376038
1 1.81599998474121
2 1.83200001716614
2 2.11999988555908
3 2.13599991798401
3 2.51200008392334
4 2.52800011634827
4 2.94400000572205
5 2.96000003814697
5 3.59200000762939
6 3.60800004005432
6 4.32000017166138
7 4.33599996566772
7 5.25600004196167
8 5.27199983596802
8 6.23199987411499
9 6.2480001449585
9 7.31199979782104
10 7.32800006866455
10 8.27999973297119
11 8.29599952697754
11 9.43200016021729
12 9.44799995422363
12 10.6319999694824
13 10.6479997634888
13 12.0799999237061
14 12.0959997177124
14 13.5839996337891
15 13.6000003814697
15 15.0719995498657
16 15.0880002975464
16 17.0480003356934
17 17.0639991760254
17 19.3519992828369
18 19.3680000305176
18 21.9839992523193
19 22
19 25.0240001678467
20 25.0400009155273
20 28.8880004882812
21 28.9039993286133
21 34.0400009155273
22 34.0559997558594
22 40.8160018920898
23 40.8320007324219
23 49.6080017089844
24 49.6240005493164
24 60.8079986572266
25 60.8240013122559
25 77.2320022583008
26 77.2480010986328
26 92.7360000610352
27 92.7519989013672
27 99.447998046875
28 99.463996887207
28 99.9919967651367
};
%\addlegendentry{cand-gs}
\end{axis}

\end{tikzpicture}}
		\caption{aggregated}\label{fig:distanceToRestricted-ag}
	\end{subfigure}
	\hfill
	\begin{subfigure}[b]{0.3\textwidth}
		\centering
		\resizebox{\textwidth}{!}{% This file was created with tikzplotlib v0.9.17.
\begin{tikzpicture}

\definecolor{color0}{rgb}{1,0.647058823529412,0}

\begin{axis}[
legend cell align={left},
legend style={
  fill opacity=0.8,
  draw opacity=1,
  text opacity=1,
  at={(0.97,0.03)},
  anchor=south east,
  draw=white!80!black
},
tick align=outside,
tick pos=left,
every tick label/.append style={font=\Huge}, 
label style={font=\Huge},
x grid style={white!69.0196078431373!black},
xlabel={deleted voters/cand.},
xmin=-1.3, xmax=27.3,
xtick style={color=black},
y grid style={white!69.0196078431373!black},
ylabel={\% of elections},
ymin=-4.99, ymax=104.79,
ytick style={color=black}
]
\addplot [line width=4pt, blue]
table {%
0 0
1 0.200000047683716
1 1.20000004768372
2 1.39999997615814
2 9.39999961853027
3 9.60000038146973
3 25.6000003814697
4 25.7999992370605
4 51.7999992370605
5 52
5 73.5999984741211
6 73.8000030517578
6 89.8000030517578
7 90
7 98
8 98.1999969482422
8 99.4000015258789
9 99.5999984741211
9 99.8000030517578
};
%\addlegendentry{cand-sp}
\addplot [line width=4pt, blue, dashed]
table {%
0 0
3 0.600000023841858
3 1
4 1.20000004768372
4 1.39999997615814
5 1.60000002384186
5 1.79999995231628
6 2
6 2.20000004768372
7 2.40000009536743
7 3
8 3.20000004768372
8 4
9 4.19999980926514
9 6
10 6.19999980926514
10 7.40000009536743
11 7.59999990463257
11 10
12 10.1999998092651
12 13.8000001907349
13 14
13 16.3999996185303
14 16.6000003814697
14 20.2000007629395
15 20.3999996185303
15 27.6000003814697
16 27.7999992370605
16 32.5999984741211
17 32.7999992370605
17 40.7999992370605
18 41
18 48.5999984741211
19 48.7999992370605
19 57.4000015258789
20 57.5999984741211
20 68.4000015258789
21 68.5999984741211
21 77
22 77.1999969482422
22 84
23 84.1999969482422
23 92.8000030517578
24 93
24 95.8000030517578
25 96
25 99.4000015258789
26 99.5999984741211
26 99.8000030517578
};
%\addlegendentry{votr-sp}
\addplot [line width=4pt, color0]
table {%
2 0
2 0.600000023841858
3 0.799999952316284
3 6.59999990463257
4 6.80000019073486
4 27
5 27.2000007629395
5 54.5999984741211
6 54.7999992370605
6 80.5999984741211
7 80.8000030517578
7 95.1999969482422
8 95.4000015258789
8 99.5999984741211
9 99.8000030517578
};
%\addlegendentry{cand-sc}
\addplot [line width=4pt, color0, dashed]
table {%
6 0
6 0.200000047683716
7 0.399999976158142
7 0.600000023841858
8 0.799999952316284
8 1
9 1.20000004768372
9 2
10 2.20000004768372
10 4.40000009536743
11 4.59999990463257
11 8.60000038146973
12 8.80000019073486
12 10
13 10.1999998092651
13 13.1999998092651
14 13.3999996185303
14 17.7999992370605
15 18
15 26.7999992370605
16 27
16 34.7999992370605
17 35
17 45.4000015258789
18 45.5999984741211
18 55.2000007629395
19 55.4000015258789
19 68
20 68.1999969482422
20 81.8000030517578
21 82
21 90.5999984741211
22 90.8000030517578
22 96.4000015258789
23 96.5999984741211
23 99.1999969482422
24 99.4000015258789
24 99.8000030517578
};
%\addlegendentry{votr-sc}
\addplot [line width=4pt, green!50.1960784313725!black]
table {%
2 0
2 0.399999976158142
3 0.600000023841858
3 6.40000009536743
4 6.59999990463257
4 21.3999996185303
5 21.6000003814697
5 48.2000007629395
6 48.4000015258789
6 74.5999984741211
7 74.8000030517578
7 90.5999984741211
8 90.8000030517578
8 98
9 98.1999969482422
9 99.8000030517578
};
%\addlegendentry{cand-gs}
\addplot [line width=4pt, green!50.1960784313725!black, dashed]
table {%
7 0
7 0.399999976158142
8 0.600000023841858
8 0.799999952316284
10 1.20000004768372
10 1.79999995231628
11 2
11 2.40000009536743
12 2.59999990463257
12 3.59999990463257
13 3.79999995231628
13 5.40000009536743
14 5.59999990463257
14 8.19999980926514
15 8.39999961853027
15 10.1999998092651
16 10.3999996185303
16 15.6000003814697
17 15.8000001907349
17 22.2000007629395
18 22.3999996185303
18 29.7999992370605
19 30
19 40
20 40.2000007629395
20 50.5999984741211
21 50.7999992370605
21 63.2000007629395
22 63.4000015258789
22 77.4000015258789
23 77.5999984741211
23 90.1999969482422
24 90.4000015258789
24 97.1999969482422
25 97.4000015258789
25 99.5999984741211
26 99.8000030517578
};
%\addlegendentry{votr-gs}
\end{axis}

\end{tikzpicture}}
		\caption{spotify month}\label{fig:distanceToRestricted.spomon}
	\end{subfigure}
	\hfill
	\begin{subfigure}[b]{0.3\textwidth}
		\centering
		\resizebox{\textwidth}{!}{% This file was created with tikzplotlib v0.9.17.
\begin{tikzpicture}

\definecolor{color0}{rgb}{1,0.647058823529412,0}

\begin{axis}[
legend cell align={left},
legend style={
  fill opacity=0.8,
  draw opacity=1,
  text opacity=1,
  at={(0.5,0.09)},
  anchor=south,
  draw=white!80!black
},
tick align=outside,
tick pos=left,
every tick label/.append style={font=\Huge}, 
label style={font=\Huge},
x grid style={white!69.0196078431373!black},
xlabel={deleted voters/cand.},
xmin=7, xmax=29,
xtick style={color=black},
y grid style={white!69.0196078431373!black},
ylabel={\% of elections},
ymin=-4.99, ymax=104.79,
ytick style={color=black}
]
\addplot [line width=4pt, blue]
table {%
8 0
8 1
9 1.20000004768372
9 16.6000003814697
10 16.7999992370605
10 62.4000015258789
11 62.5999984741211
11 99.5999984741211
12 99.8000030517578
};
%\addlegendentry{cand-sp}
\addplot [line width=4pt, blue, dashed]
table {%
21 0
24 0.600000023841858
24 1.39999997615814
25 1.60000002384186
25 13.1999998092651
26 13.3999996185303
26 43
27 43.2000007629395
27 93.5999984741211
28 93.8000030517578
28 99.8000030517578
};
%\addlegendentry{votr-sp}
\addplot [line width=4pt, color0]
table {%
9 0
9 7.19999980926514
10 7.40000009536743
10 50.4000015258789
11 50.5999984741211
11 98.4000015258789
12 98.5999984741211
12 99.8000030517578
};
%\addlegendentry{cand-sc}
\addplot [line width=4pt, color0, dashed]
table {%
21 0
21 0.399999976158142
22 0.600000023841858
22 4
23 4.19999980926514
23 19.7999992370605
24 20
24 57.2000007629395
25 57.4000015258789
25 91.8000030517578
26 92
26 99.8000030517578
};
%\addlegendentry{votr-sc}
\addplot [line width=4pt, green!50.1960784313725!black]
table {%
9 0
9 1.20000004768372
10 1.39999997615814
10 21
11 21.2000007629395
11 80.4000015258789
12 80.5999984741211
12 99.8000030517578
};
%\addlegendentry{cand-gs}
\addplot [line width=4pt, green!50.1960784313725!black, dashed]
table {%
22 0
23 0.200000047683716
23 1.39999997615814
24 1.60000002384186
24 4
25 4.19999980926514
25 18
26 18.2000007629395
26 60.5999984741211
27 60.7999992370605
27 97.1999969482422
28 97.4000015258789
28 99.8000030517578
};
%\addlegendentry{votr-gs}
\end{axis}

\end{tikzpicture}}
		\caption{spotify day}\label{fig:distanceToRestricted.spodai}
	\end{subfigure}
	\caption{For different datasets, fraction of elections within a given candidate deletion (solid) or voter deletion distance (dashed) from single-peakedness (blue), single-crossingness (orange), and group-separability~(green).}
	\label{fig:distanceToRestricted}
\end{figure}

For each of our elections, we computed the voter and candidate deletion distance from single-peakedness, single-crossingness, and group-separability.\footnote{For single-peaked candidate deletion we used the polynomial-time algorithm from \citet{DBLP:journals/jair/ErdelyiLP17} and for single-crossing voter deletion the polynomial-time algorithm from \citet{DBLP:journals/mss/BredereckCW16}.
	For single-peaked voter deletion and single-crossing candidate deletion, we used the FPT algorithms based on conversions to hitting set by \citet{DBLP:conf/aaai/ElkindL14}. For the voter and candidate deletion distance to group separability, we again used fixed-parameter tractable algorithms of  \citet{DBLP:conf/aaai/ElkindL14}. For our implementation, we employed \citet{gurobi}.}   
In \Cref{fig:distanceToRestricted-ag}, we show the results on the aggregated dataset as a cumulative distribution function. 
For the candidate deletion distance, the picture is very similar for all three restricted domains:
There are around $15\%$ of elections within distance $4$, around $28\%$ within distance $6$, around $56\%$ within distance $10$, and around $99\%$ within distance $12$. 
Considering that we have seen in the previous part that there are considerably more single-crossing elections than single-peaked or group-separable elections, the similarity between the domains here is partly unexpected.

For voter deletion, there is some difference between the restricted domains: 
For all three restricted domains, $15\%$ of elections are within distance $14$ and are more or less uniformly distributed within this distance. 
For single-peakedness and group-separability, $25\%$ of all elections are within a distance of $18$, $50\%$ within a distance of $23$, and $99\%$ within a distance of $27$. 
For single-crossingness, distances are typically one smaller, as  $25\%$ of all elections fall within distance $17$ and $50\%$ within distance $20$. 
This slight difference might be because in contrast to the other two domains, for single-crossingness an ordering of the voters is needed which might be easier to construct if we can choose which voters to delete (however, for single-peakedness the same is true for candidate deletion, yet no such effect is visible). 
Comparing the normalized voter deletion distance to the normalized candidate deletion distance it seems that the latter is typically slightly smaller. 
Nevertheless, there is a strong linear correlation between the candidate deletion and voter deletion distance of an election.

Examining the results on the dataset level, there are significant differences: 
The general trend here is that the higher the average Kemeny score of a dataset is the further is the dataset on  average from a restricted domain. 
One dataset from our close to identity group which contains many elections with a low Kemeny score are spotify month election, and in \Cref{fig:distanceToRestricted.spomon} we depict the cumulative distribution for this dataset. 
Notably, for all three restricted domains, elections from this dataset have the second-lowest candidate deletion distance, i.e., $50\%$ of the spotify month elections have a candidate deletion distance of $5$ and smaller. 
In contrast to this, in \Cref{fig:distanceToRestricted.spodai} we show the plot for spotify day elections which belong to the middle datasets and have higher Kemeny scores.
Here for all elections, at least $9$ candidates or at least $21$ voters need to be deleted to make it fall into one of our three restricted domains, indicating that this dataset is far away from a restricted domain. 
Given that one can see the whole spotify data as one huge election, the opposite behavior of spotify day and spotify month elections highlights the natural fact that depending on which votes from a large election are taken into account very different elections arise. 
To sum up, we have found only little evidence of elections from restricted domains and also only a few elections at a small distance (recall that $5$ candidates are not really a small number here, as this corresponds to $33\%$ of candidates). 
Thus, both the voter and candidate deletion distance are probably too large on many real-world elections for the usage of parameterized algorithms.

To the best of our knowledge, the work of \citet{DBLP:conf/ijcai/SuiFB13} is the only other work that studies the distance of real-world elections from a restricted domain. 
In particular, they analyzed two large elections (both with $9$ candidates and over $\num{10000}$ votes) related to the 2002 Irish general election and found that for both more than $95\%$ of voters need to be deleted to make the election single-peaked. 
This indicates that our elections are comparably quite close to a restricted domain (which also might be due to the case that we have fewer voters). 
Moreover, they also considered the minimum number of axes such that each vote is single-peaked with respect to one axis and found that for one election more than $36$ and for the other more than $263$ axes are needed. 
On the positive side, they found that the voter deletion distance of their two elections to, so-called, two-dimensional single-peakedness is much smaller. 
Testing how far our elections are from multi-dimensional single-peakedness would be an interesting direction for future work.

\paragraph{Membership in Multiple Restricted Domains.}

\begin{figure*}
	\centering
	\begin{subfigure}[b]{0.135\textwidth}
		\centering
		\includegraphics[width=\textwidth]{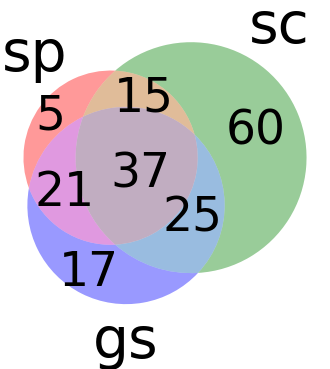}
		\caption{VD/CD 0} \label{fig:venn-original}
	\end{subfigure}
	\hfill
	\begin{subfigure}[b]{0.135\textwidth}
		\centering
		\includegraphics[width=\textwidth]{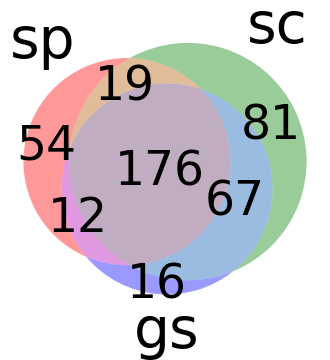}
		\caption{CD 1}
	\end{subfigure}
	\hfill
	\begin{subfigure}[b]{0.135\textwidth}
		\centering
		\includegraphics[width=\textwidth]{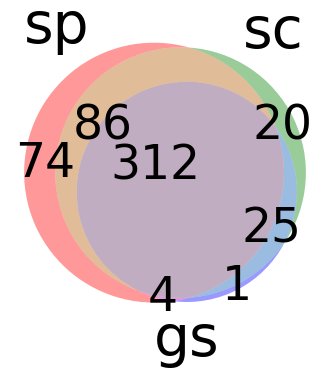}
		\caption{CD 2}
	\end{subfigure}
	\hfill
	\begin{subfigure}[b]{0.135\textwidth}
		\centering
		\includegraphics[width=\textwidth]{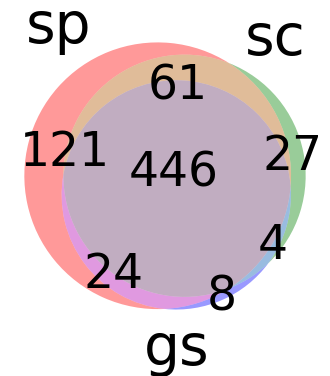}
		\caption{CD 3}
	\end{subfigure}
	\hfill
	\begin{subfigure}[b]{0.135\textwidth}
		\centering
		\includegraphics[width=\textwidth]{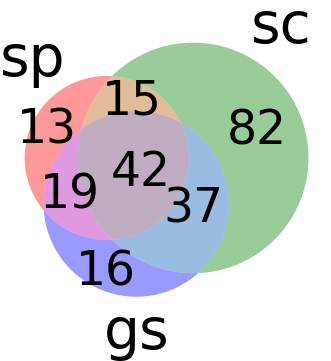}
		\caption{VD 2}
	\end{subfigure}
	\hfill
	\begin{subfigure}[b]{0.135\textwidth}
		\centering
		\includegraphics[width=\textwidth]{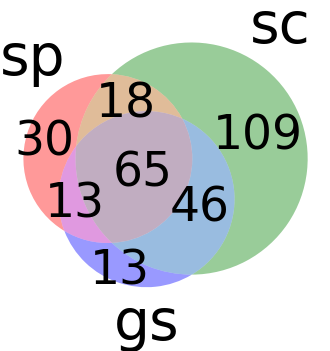}
		\caption{VD 4}
	\end{subfigure}
	\hfill
	\begin{subfigure}[b]{0.135\textwidth}
		\centering
		\includegraphics[width=\textwidth]{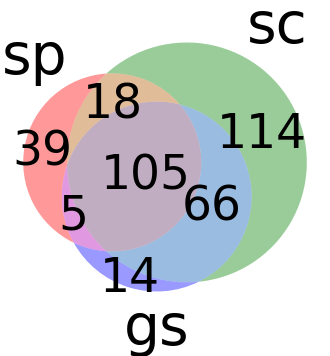}
		\caption{VD 6}
	\end{subfigure}
	\caption{For different voter deletion distances (VD) or candidate deletion distances (CD),	
		Venn diagrams of elections which are within the given distance from single-peakedness, single-crossingness, or group-separability. }
	\label{fig:venn}
\end{figure*}

Motivated by previous works by \citet{DBLP:journals/tcs/SkowronYFE15} and \citet{DBLP:journals/scw/ElkindFS20} on elections that are simultaneously single-peaked and single crossing and to better understand the relationship of the different restricted domains, we now analyze whether elections that are part of, or close to, one restricted domain are typically also part of or close to another. 
In \Cref{fig:venn-original}, we show a Venn diagram capturing all elections that are part of one of our three restricted domains: Each domain is represented by a circle and the overlap of different circles (and the numbers shown there) represents the intersection of domains. 
Notably, each restricted domain shares more elections with another domain than there are elections that are only part of this domain. 
In fact, there are almost as many elections that are part of all three domains ($37$) than elections that are single-peaked or group-separable but not single-crossing ($43$). This indicates that in practice there is a significant overlap of the domains and that algorithms for single-crossing elections can be applied to most elections falling in one of the three domains.

In \Cref{fig:venn}, we also display Venn diagrams for elections that are within a certain voter or candidate deletion distance from a restricted domain.
Elections that are close to one of the domains after deleting some candidates overlap even more than elections from restricted domains: Remarkably, considering elections within a candidate deletion distance of $3$, there are $3.5$ times more elections that are within this distance from all restricted domains, than those that are only close to one restricted domain. 
For the different voter deletion distance values the picture is roughly similar as in \Cref{fig:venn-original}.
So overall it seems that in real-world elections the different restricted domains and their closer environment heavily overlap and that it is, for instance, possible to apply algorithms for (close to) single-peaked or single-crossing elections to an overwhelming majority of (close to) group-separable elections.

Moreover, there is also a strong linear correlation between an election's distance from the different restricted domains: 
For each pair of our restricted domains, the Pearson correlation coefficient of the candidate deletion distances from the two is around $0.95$, while it is between $0.85$ and $0.9$ for the voter deletion distance. 

\paragraph{Further Considerations.}

In \Cref{sub:restricteddomain-deg}, we analyze the properties of elections that are (close to) being single-peaked or single-crossing and observe that they are typically quite degenerate, meaning that they have a low Kemeny score and that they fall into a small part of the space of all elections from the respective restricted domain. 
For single-peaked elections, we find that the top-choices of voters typically fall in the same area of the societal order.
Moreover, in \Cref{sub:general-restrictions}, we check which of our elections fulfills some more general restrictions.
Among others, we find that value-restricted elections \cite{sen1966possibility} occur quite frequently and that in the characterization of single-peaked, single-crossing, and group-separable elections via forbidden configurations one of the two configurations is redundant in practice.

\section{Case Study: How Different are Different Voting Rules?}  \label{sec:util}
In this section, we use our datasets to shed some further light on traditional questions from social choice.
While there is already quite some empirical research on the considered questions, nearly all of these works considered elections with $3$ to $5$ candidates coming from a single data source. 
Thus, our rich data allows us to take a broader look. 

One popular question arises around the notion of a Condorcet winner. 
A candidate $c$ is a strong (weak) Condorcet winner if for each other candidate $d$ more than (at least) half of the voters prefer $c$ to $d$. 
Previous research has found that strong Condorcet winners nearly always exist, i.e., the so-called Condorcet paradox occurs relatively rarely, and that the strong Condorcet efficiency, i.e., how often these rules select the strong Condorcet winner as a winner, of all rules is very high \cite{DBLP:conf/aldt/Mattei11,chamberlin1984social,darmann2019evaluative,popov2014consensus,DBLP:journals/scw/PlassmannT14}.
We investigate these issues in \Cref{sub:cond}.

In \Cref{sub:rules}, we analyze the level of agreement between different voting rules.
While from a theoretical and axiomatic perspective, voting rules significantly differ from each other, various authors provided evidence that most of them are very similar in practice \cite{DBLP:conf/aldt/Mattei11,chamberlin1984social,felsenthal1993empirical,regenwetter2006behavioral,regenwetter2007unexpected,mccabe2006exploratory,darmann2019evaluative,popov2014consensus}.

Overall, while parts of our results in this section are in line with previous studies, we also find evidence that suggests that the established consensus in the literature according to which in practice all voting rules are more or less the same should be relativized, as it seems to only apply if we have elections with a Condorcet winner and/or the number of voters divided by the number of candidates is large.

We start by defining all considered voting rules
(for each rule computing a score, all candidates with the highest score win.):
\begin{description}
	\item[Plurality] Each voter awards one point to its top-choice. 
	\item[Plurality with runoff] In the first round, each voter awards one point to its top-choice. If more than two candidates have the highest number of points, then delete all but them. Otherwise, we delete all candidates who do not have the highest or second-highest number of points. In the second round, each voter awards one point to the remaining candidate it ranks highest. 
	\item[Borda] For $i\in [m]$, each voter awards $m-i$ points to the candidate it ranks in the $i$th position. 
	\item[Copeland] A candidate $c\in C$ gets a point for each candidate $d\in C\setminus \{c\}$ for which a strict majority of voters prefers $c$ to $d$ and loses a point if a strict majority prefers $d$ to $c$.
	\item[Hare] In each round, each voter awards one point to its most preferred remaining candidate. After each round, a candidate with the lowest score gets deleted (ties are broken according to the preferences of the first voter). If all candidates in some round have the same number of points, then we return all of them as the winners.     
\end{description}

\begin{figure}
	\begin{subfigure}[t]{0.37\textwidth}
		\centering
		\includegraphics[width=\textwidth]{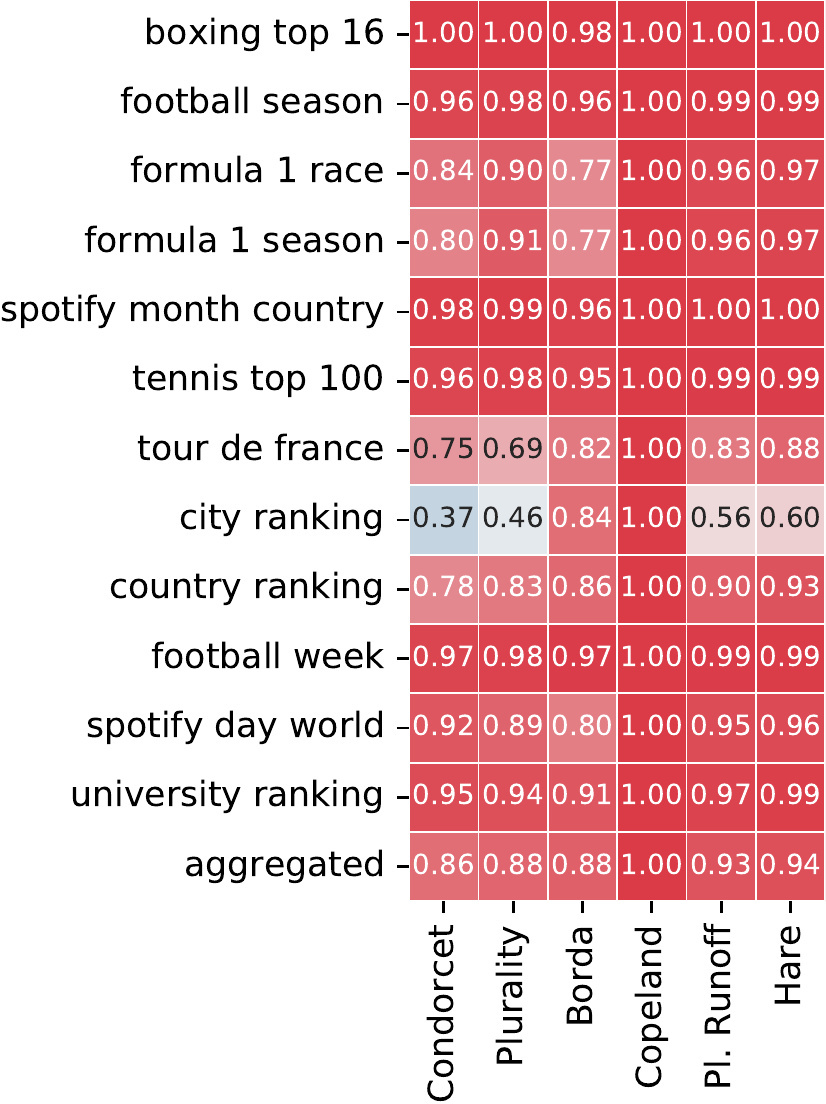}
		\caption{Strong Condorcet winner.} \label{fig:strongCond}
	\end{subfigure}
	\qquad \qquad \qquad
	\begin{subfigure}[t]{0.37\textwidth}
		\centering
		\includegraphics[width=\textwidth]{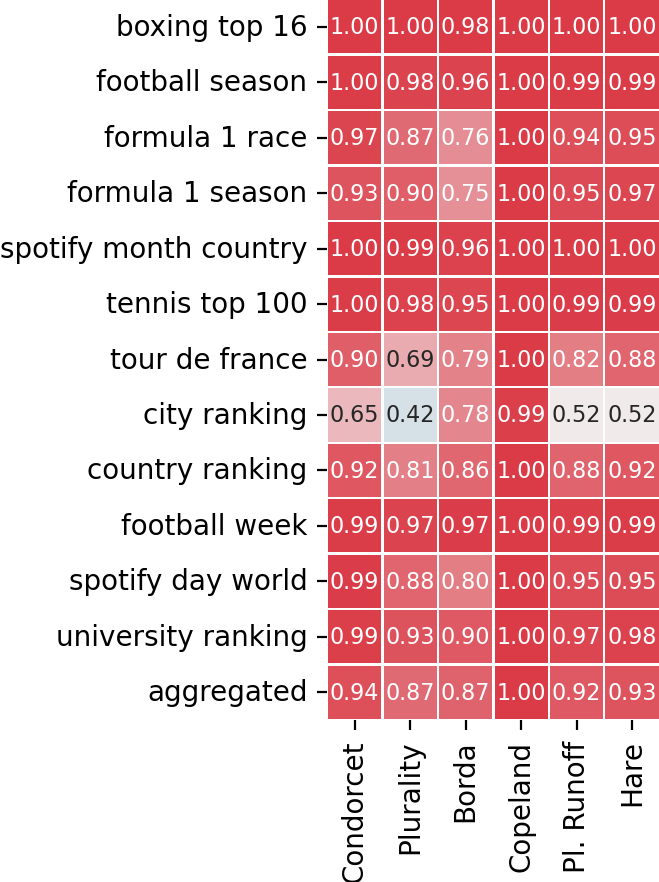}
		\caption{Weak Condorcet winner.}\label{fig:weakCond}
	\end{subfigure}
	\caption{In the first column, fraction of elections admitting a strong/weak Condorcet winner. In the other columns, strong/weak Condorcet efficiency of different voting rules.}
	\label{fig:Condorcet}
\end{figure}

\subsection{Condorcet Paradox and Condorcet Efficiency} \label{sub:cond}
\begin{itemize}
	\item Most of our elections ($86\%$) have a strong Condorcet winner and all voting rules  select them as a winner most of the times ($88\%$ or more). 
	\item In some of our datasets only few elections have a strong Condorcet winner and voting rules select it as a winner less frequently.  
\end{itemize}

In line with the literature, we first focus on strong Condorcet winners.
In \Cref{fig:strongCond}, in the first column, we depict for each of our datasets the fraction of elections admitting a strong Condorcet winner. 
While for all datasets from our first group of close to identity datasets around $96\%$ of elections admit a strong Condorcet winner, for the other datasets this fraction is (considerably) below $100\%$. 
The most extreme case are city ranking elections where only $37\%$ of the elections admit a strong Condorcet winner. 
Moreover, overall ``only'' $86\%$ of all our elections admit a strong Condorcet winner.
This is in contrast to previous works. For instance, \citet{popov2014consensus} reported that in one of their studied datasets $6.7\%$ of elections do not admit a strong Condorcet winner, while for all others this value is below $0.3\%$. 

Concerning the strong Condorcet efficiency of the different voting rules, results again significantly depend on the considered dataset. 
For close to identity datasets all voting rules have a very high Condorcet efficiency of $0.95$ and above (note that Copeland's voting rule is guaranteed to select a strong Condorcet winner if one exists). 
\citet{DBLP:conf/aldt/Mattei11} and \citet{popov2014consensus} also reported a Condorcet efficiency of $0.95$ and above for different rules. 
However, on our other datasets, the Condorcet efficiency can be much lower: For Plurality, Plurality with Runoff, and Hare, their Condorcet efficiency is the lowest on the city ranking dataset with $0.46$, $0.59$, and $0.6$, respectively. 
For Borda, the minimum Condorcet efficiency is $0.77$ on Formula 1 race and Formula 1 season elections. 
Interestingly, the other voting rules achieve a much higher efficiency on these two sets. 
Considering the results on the aggregated dataset, Hare and Plurality with Runoff have the highest Condorcet efficiency with $0.94$ and $0.93$ respectively, while Plurality and Borda both have a Condorcet efficiency of $0.88$. 
Given that Borda takes much more information into account than Plurality, it is slightly unexpected that both perform so similarly here.

In \Cref{fig:weakCond}, we depict the same statistics for the notion of weak Condorcet winners: 
A substantial fraction of our elections, i.e., $8\%$ of all elections, $15\%$ of tour de France, and $28\%$ of city ranking elections, admit a weak but no strong Condorcet winner. 
This is quite remarkable given that the distinction between a weak and a strong Condorcet winner almost appears like a tie-breaking issue.
The Condorcet efficiency of our rules slightly decreases when moving from strong to weak Condorcet winners. 
This is something to be expected because weak Condorcet winners, which are now also taken into account, have in general a slightly weaker standing in the election than strong Condorcet winners.

\subsection{Consensus among Voting Rules} \label{sub:rules}
\begin{itemize}
	\item Voting rules often agree on the returned winner because most elections have a Condorcet winner and voting rules often select them. 
	\item On elections without a Condorcet winner, Borda and Copeland, on the one hand, and Plurality, Plurality with Runoff and Hare, on the other hand, regularly agree on a winner.
	\item The rankings returned by different voting rules do not exhibit a strong correlation (and in some cases even none). 
\end{itemize}
In \Cref{sub:winnercons}, we analyze the consensus among winners returned by different voting rules.
After that in \Cref{sub:rankcons}, we analyze the relationship between the rankings returned by the different voting rules.

\subsubsection{Winner Consensus} \label{sub:winnercons}
\begin{figure*}
	\centering
	\begin{subfigure}[t]{0.24\textwidth}
		\centering
		\includegraphics[width=\textwidth]{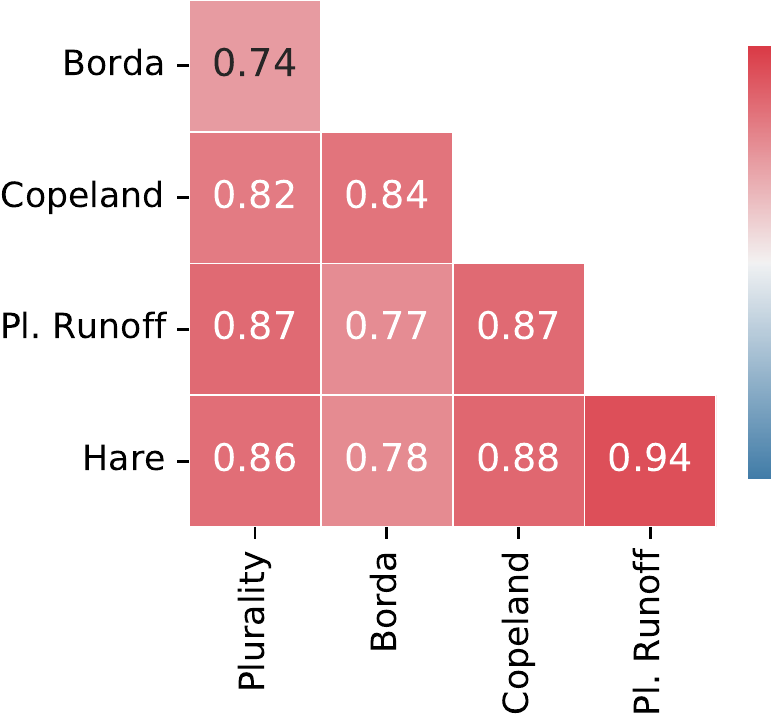}
		\caption{All elections.} \label{first:total}
	\end{subfigure}
	\hfill
	\begin{subfigure}[t]{0.24\textwidth}
		\centering
		\includegraphics[width=\textwidth]{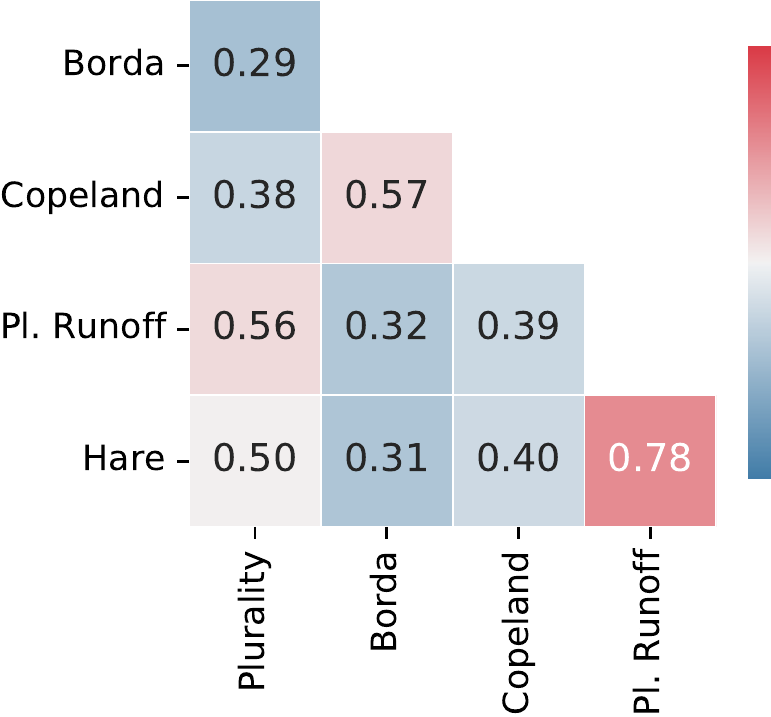}
		\caption{All elections without a strong Condorcet winner.}\label{first:non_cond}
	\end{subfigure}
	\hfill
	\begin{subfigure}[t]{0.24\textwidth}
		\centering
		\includegraphics[width=\textwidth]{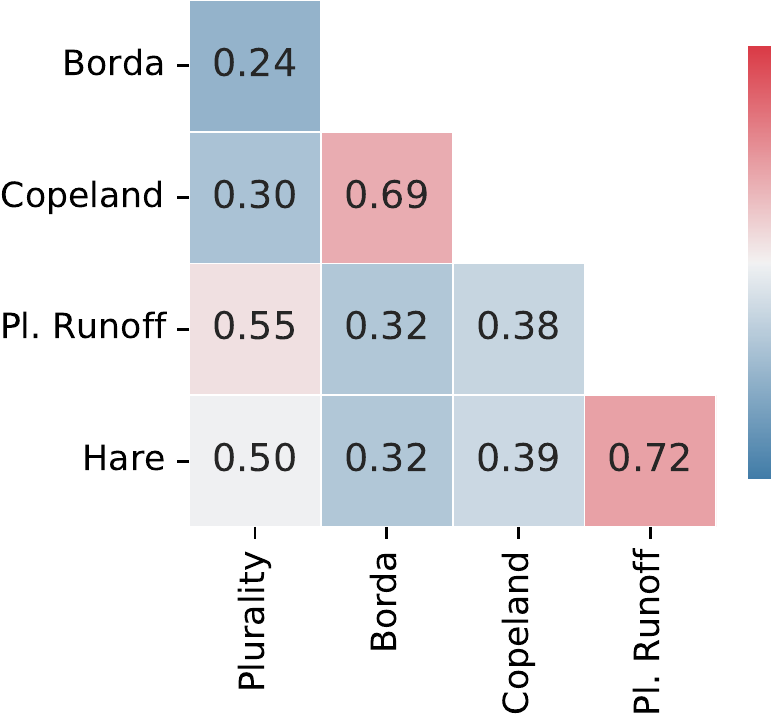}
		\caption{All city ranking elections.}\label{first:city}
	\end{subfigure}
	\hfill
	\begin{subfigure}[t]{0.24\textwidth}
		\centering
		\includegraphics[width=\textwidth]{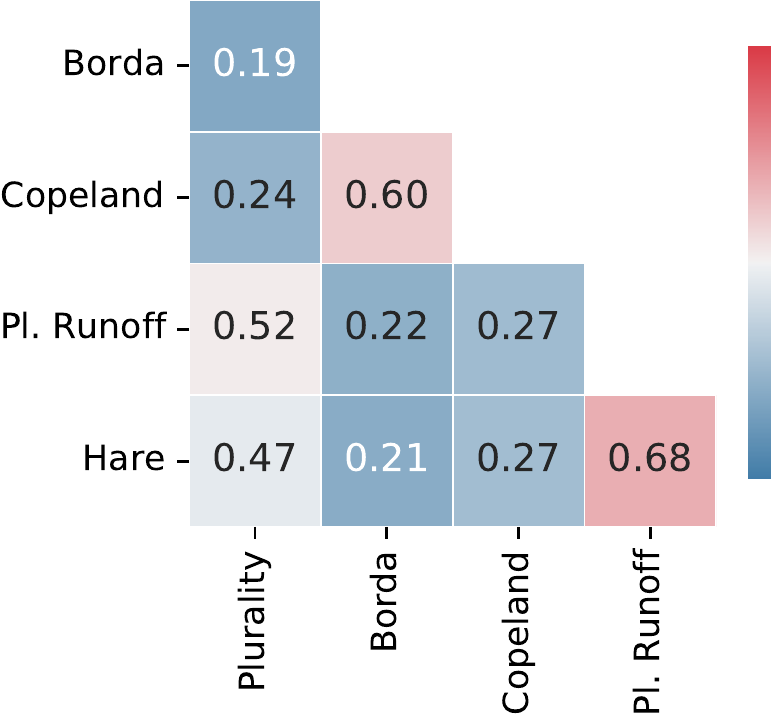}
		\caption{All city ranking elections without a strong Condorcet winner.}\label{first:city_non_cond}
	\end{subfigure}
	\caption{For pairs of voting rules, fraction of elections where rules return same winner after lexicographic tie-breaking.}
	\label{fig:winner-rule-overlap}
\end{figure*}

In \Cref{first:total}, we depict the average lexicographic agreement of each pair of rules. 
The average lexicographic agreement of some pair of rules is the fraction of all elections where the winner returned by the two rules is the same if we apply lexicographic tie-breaking during the execution of both rules.\footnote{In \Cref{app:ties} we also consider two alternative similarity measures for the returned winners. We also observe that voting rules return tied winners in around $5\%$ of elections but that this fraction is much higher for elections without strong Condorcet winners.} 
In general, the consensus among the different voting rules is quite high, ranging from $0.96$ for the only two iterative rules, Hare and Plurality with runoff, to $0.74$ for Borda and Plurality. 
However, the reason for this generally high agreement between voting rules might be connected to our observation from \Cref{sub:cond} that most of our elections have a strong Condorcet winner and that in case a strong Condorcet winner exists, most of  the time rules return it as a winner. 
To verify this, in \Cref{first:non_cond}, we depict the average lexicographic agreement of pairs of voting rules on all elections without a strong Condorcet winner. 
Indeed, the consensus among voting rules is significantly lower in this case: For all pairs of rules except for Hare and Plurality with runoff, whose average lexicographic agreement is still $0.82$, the average lexicographic agreement drops by between $0.27$ and $0.48$ when moving from the full election dataset to elections without strong Condorcet winner. 
\Cref{first:non_cond} further suggests that there exist two groups of voting rules: Plurality, Plurality with Runoff, and Hare on the one hand, and Borda and Copeland on the other hand. 
This partition is also quite intuitive, as all rules from the first group use Plurality scores in some way or the other, while Copeland and Borda in some sense always take into account the full election. 
Overall, our results indicate that a main reason why voting rules seem to typically exhibit a high consensus is because they all favor strong Condorcet winners which often exist. 
This could also explain why previous research \cite{chamberlin1984social,felsenthal1993empirical,regenwetter2006behavioral,regenwetter2007unexpected,darmann2019evaluative,popov2014consensus}
has found a higher consensus among rules than what we have observed: On their data strong Condorcet winners exist more often than on ours.

On the dataset level, results are again very different and correlate with our grouping of the datasets: On the one hand, on close to identity datasets the consensus of voting rules is very high, while, on the other hand, on city rankings it is lowest.
In \Cref{first:city,first:city_non_cond}, we display the average lexicographic agreement on all city ranking elections and on all city ranking elections without strong Condorcet winners.
Both \Cref{first:city,first:city_non_cond} look quite similar (as many city ranking elections do not admit a strong Condorcet winner). 
Again, we can find the already observed partitioning of the rules into groups. 
Here, both the consensus between Plurality with Runoff and Hare and the consensus between Borda and Copeland is particularly high.

\subsubsection{Ranking Consensus} \label{sub:rankcons}
In this section, we analyze the relationship between the rankings returned by the different rules: For Borda, Copeland, and Plurality, candidates are ranked according to their score. 
For Plurality with Runoff, eliminated (remaining) candidates are ranked according to their score in the first (second) round. 
In Hare, the elimination order defines the final ranking. 
We also consider the Kemeny consensus ranking. 
For each pair of voting rules, as already done in previous works \cite{DBLP:conf/aldt/Mattei11,felsenthal1993empirical,darmann2019evaluative}, we compute the Spearman correlation coefficient for rank variables, which is based on the difference of the position of candidates in the two rankings. 
As for the Pearson correlation coefficient, $1$ means a perfect positive correlation, $0$ means no correlation, and $-1$ means a negative correlation.  

\begin{figure*}
	\centering
	\begin{subfigure}[t]{0.24\textwidth}
		\centering
		\includegraphics[width=\textwidth]{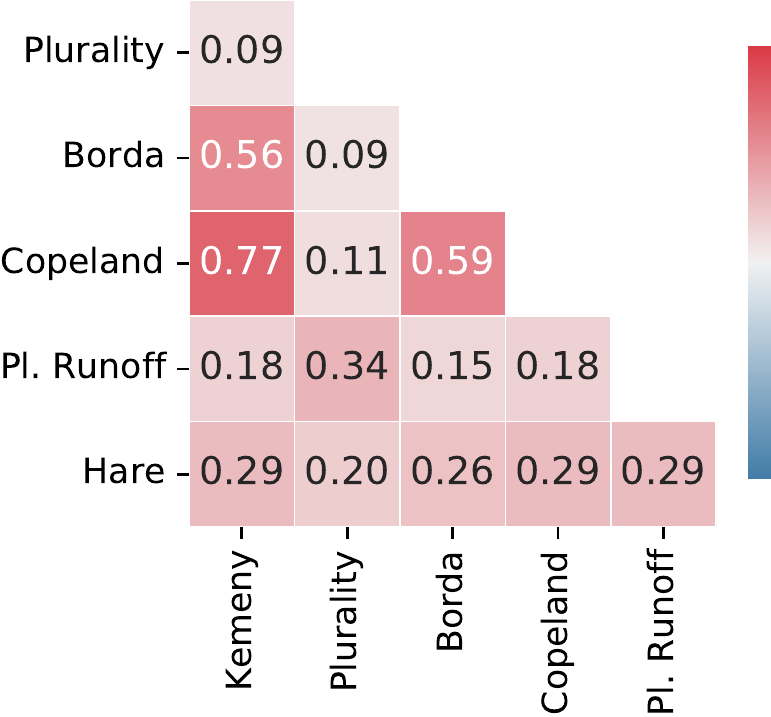}
		\caption{All elections.} \label{fig:spear-all-total}
	\end{subfigure}
	\hfill
	\begin{subfigure}[t]{0.24\textwidth}
		\centering
		\includegraphics[width=\textwidth]{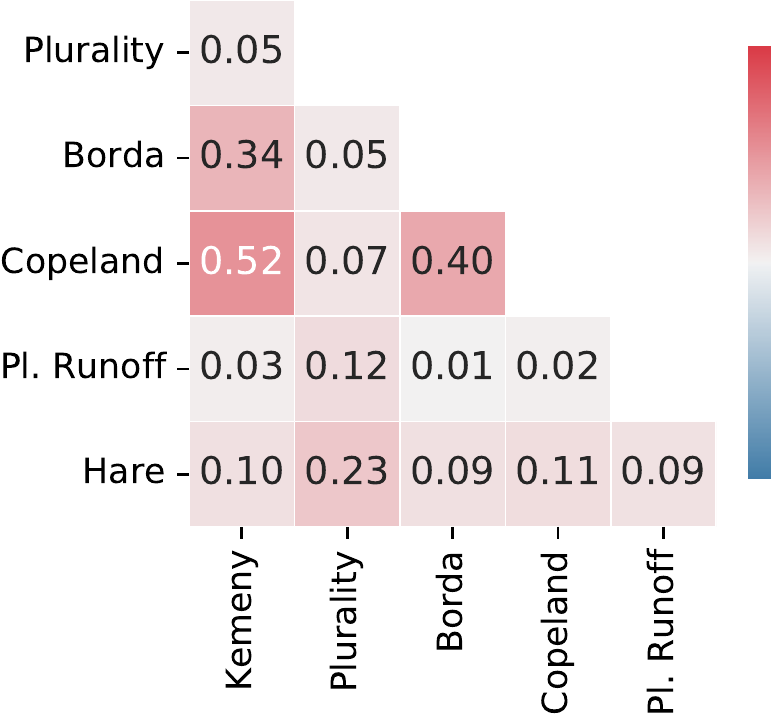}
		\caption{All elections not admitting a Condorcet winner.}\label{fig:spear-non_cond-total}
	\end{subfigure}
	\hfill
	\begin{subfigure}[t]{0.24\textwidth}
		\centering
		\includegraphics[width=\textwidth]{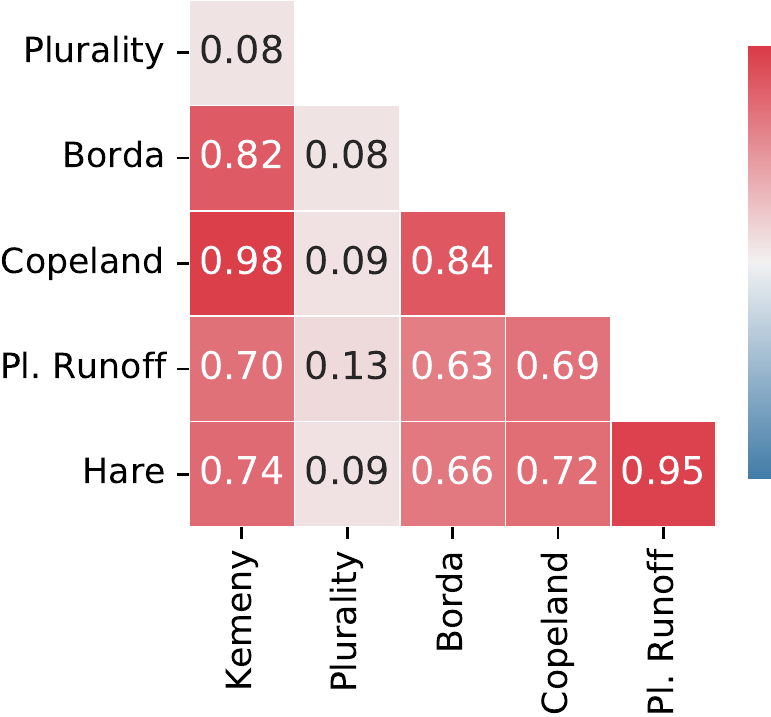}
		\caption{All boxing top 16 elections.}  \label{fig:spear-all-boxing}
	\end{subfigure}
	\hfill
	\begin{subfigure}[t]{0.24\textwidth}
		\centering
		\includegraphics[width=\textwidth]{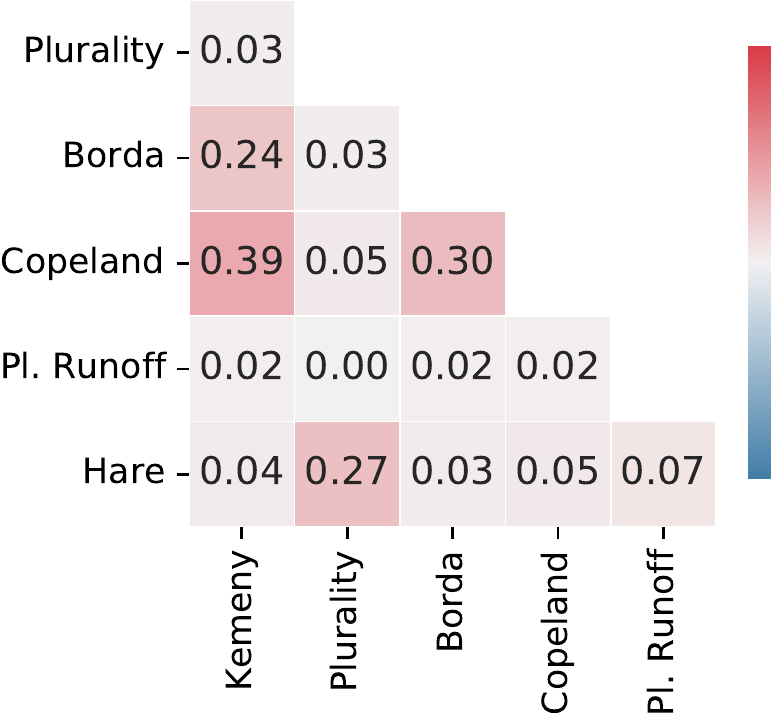}
		\caption{All city rankings elections.}   \label{fig:spear-all-city}
	\end{subfigure}
	\caption{Average Spearman's rank correlation coefficient of different pairs of voting rules on different datasets.}
	\label{fig:spear}
\end{figure*}

In \Cref{fig:spear-all-total}, we include the Spearman correlation coefficient for all pairs of voting rules averaged over all elections. 
Comparing this with \Cref{first:total}, the consensus among rules concerning the full ranking of candidates is considerably lower than concerning the election winner. 
Only Kemeny, Borda, and Copeland exhibit an average correlation over $0.5$.
\Cref{fig:spear-all-total} clearly differs from previous research that reported  high Spearman correlation coefficients \cite{darmann2019evaluative,DBLP:conf/aldt/Mattei11,felsenthal1993empirical}. 
However, a partial explanation for this is that in these works elections with fewer candidates and (much) more voters have been considered. 
Especially rules like Plurality and Plurality with runoff have difficulties to differentiate the strength of candidates if there are few voters per candidate.
Considering elections without a Condorcet winner (see \Cref{fig:spear-non_cond-total}), the correlation between the different rules is even lower.

On the dataset level, results again vary substantially. 
In \Cref{fig:spear-all-boxing}, we show the average Spearman coefficient for boxing top 16 elections where the consensus is the highest, while in \Cref{fig:spear-all-city} we depict the average Spearman coefficient for city rankings where the consensus is the lowest.  
In the boxing top 16 dataset, all rules (except for Plurality which has a very low correlation with all other rules) have a pairwise average Spearman coefficient of at least $0.63$, especially Copeland and Kemeny (with $0.98$) and Hare and Plurality with Runoff ($0.95$) are highly correlated. 
However, given that Plurality and Plurality with runoff work quite similar, their lack of correlation is quite surprising. 
Turning to city ranking elections, most of the rules produce completely uncorrelated rankings. Only Copeland, Borda and Kemeny together with Plurality and Hare still have some noticeable, yet still quite low, correlation.

\section{Conclusion}
We have collected, classified, analyzed, and used a diverse collection of real-world elections and provided various evidence hinting at their usefulness for experimental research. 
To the best of our knowledge, this is the first work that systematically compares elections from numerous different sources.

For future work, it would be interesting to analyze the relationship of the collected elections to elections drawn from various statistical cultures. 
Moreover, also performing our experiments on such synthetic elections could be useful to get a better understanding of their properties. 
In addition, examining the collected elections (even) more carefully would be of great use: While we have been able to provide intuitive explanations for some phenomena we observed, the reasons for others remain unclear.
Furthermore, as we have found only little evidence to support the large-scale practical applicability of already developed parameterized algorithms, identifying new properties that are shared by many elections and that allow for the development of tractable algorithms would be extremely valuable. 
Finally, the main purpose of this project is to provide a helpful source of real-world election datasets, so we hope that others will find our data useful in bridging the gap between theory and practice in computational social choice.

\subsection*{Acknowledgments}
NB was supported by the DFG project MaMu (NI 369/19) and by the DFG project ComSoc-MPMS (NI
369/22). NS was supported by the DFG project MaMu (NI 369/19). The authors would like to thank  Piotr Faliszewski for constructive criticism of the manuscript. 
 
\clearpage

\bibliographystyle{abbrvnat}

%%%%%%%%%%%%%%%%%%%%%%%%%%%%%%%%%%%%%%%%%%%%%%%%%%%%%%%%%%%%%%%%%%%%%%%%
\clearpage 
\appendix

\begin{table}
	\begin{center}
		 \begin{tabular}{ccccccccc}
				\toprule
				name & 
				type& 
				\multicolumn{3}{c}{raw} & \phantom{a} &
				\multicolumn{3}{c}{relevant complete}\\ \cmidrule{3-5}
				\cmidrule{7-9}
				&& \#Elec. & \makecell{Avg.\\ \#Voters} & \makecell{Avg.\\ \#Cand.}   & & \makecell{\#Elec.} & \makecell{Avg.\\ \#Voters} &  \makecell{Avg.\\ \#Cand. }    \\ \midrule \midrule
				%------
				board games & time & 1 & 130 & 5904 && 1 & 130 & 885 \\
				table tennis top & time & 38 & 11.32 & 1429.92 && 36 & 11.5 & 670.47 \\
				Giro d'Italia & time & 99 & 19.27 & 102.2 && 23 & 16.83 & 130.13 \\
				marblelympics & time & 5 & 13.4 & 19.4 && 4 & 12.25 & 15.75\\
				mylaps & time & 635 & 28.67 & 25.87 && 394 & 25.33 & 29.59\\
				football combined & crit. & 23 & 1466.26 & 228.48 && 23 & 506.39 & 217.87\\
				basketball season & time & 1095 & 13.54 & 325.06 && 966 & 14.19 & 336.42\\
				basketball week & crit. & 382 & 38.82 & 342.38 && 378 & 35.8 & 337.34\\
				basketball combined & crit. & 20 & 741.5 & 341.65 && 20 & 671.4 & 315.65\\
				baseball season & time & 174 & 12.74 & 217.99 && 127 & 12.24 & 263.98\\
				baseball week & crit. & 159 & 14.3 & 299.47 && 158 & 10.34 & 192.44\\
				baseball combined & crit. & 10 & 221.6 & 299.8 && 10 & 134.8 & 249.6\\
				\bottomrule
		\end{tabular}
		\caption{Information about not-used election datasets.} \label{tab:basicStats2}
	\end{center}
\end{table}

\section{Additional Material for Section \ref{sec:collect}}
\subsection{Further Datasets} \label{se:moredata}
In the following, we present the additional datasets that we created. Notably, we could have also included all of them in our analysis from the main body but decided against it for the sake of conciseness.
In \Cref{tab:basicStats2}, we present basic information about our further non-used datasets analgous to \Cref{tab:basicStats}. 

\paragraph{Board Games.} 
The board games data (collected by \citet{boardgames}) contains a weekly ranking of the 2000 most popular board games on boardgamegeek.com between October 2018 and December 2021. We created a single \emph{board games} election where each game is a candidate and each vote corresponds to the ranking of the games in one week.

\paragraph{Table Tennis (World) Rankings.} The table tennis data (collected by \citet{tabletennis}) contains the monthly ITTF ranking of the top 500-1500 male and female table tennis players between 2001 and 2020.   
For each year, we created a \emph{table tennis} election where each player is a candidate and each vote corresponds to the ranking of the players in one month.

\paragraph{Giro d'Italia.} For each edition of the Giro d'Italia between 1910 and 2020, the data contains the completion times of all riders for each stage of the edition. The dataset was crawled by us from the website \url{procyclingstats.com}. For each edition, we created one \emph{Giro d'Italia} election in which the riders are the candidates and each vote corresponds to a stage and ranks the riders by their completion time.

\paragraph{Marblelympics.} The marblelympics are a competition of marbles in multiple races broadcasted on \url{youtube.com}.  
For each edition of the marblelympics between 2016 and 2020, the data contains the completion times of all marbles for each stage of the edition. The dataset was crawled by us from the website \url{jellesmarbleruns.fandom.com}. For each edition, we created one \emph{marblelympics} election in which the marbles are the candidates and each vote corresponds to a race and ranks the marbles by their completion time.

\paragraph{Mylaps.} 
The mylaps data contains the completion time of athletes in each lap of a multi-lap competition (specifically, speed skating and cycling competitions). 
The dataset was crawled by us from the website \url{http://results.sporthive.com/}. For each race, we created one \emph{mylaps} election in which the athletes are the candidates and each vote corresponds to one lap and ranks the athletes by their completion time.

\paragraph{College Sports.} Apart from the American football data presented and used in the main body, the college sports data (collected by \citet{college}) also contains weekly power rankings of college basketball teams (between 2001 and 2021) and college baseball teams (between 2010 and 2021) from different media outlets and ranking systems.
For all three data sources, we created three different types of elections with teams as candidates:
First, for each season and each ranking system, we created a \emph{season} election where each vote corresponds to the power ranking of the teams in one week according to the ranking system.
Second, for each week in one of the seasons, we created a \emph{week} election where each vote corresponds to the power ranking of the teams in this week according to one of the ranking systems.
Third, for each season, we created a \emph{combined} election where each vote corresponds to the power ranking of the teams in one of the weeks according to one of the ranking systems.

\section{Additional Material for Section \ref{sec:rest}}

\subsection{Properties of Elections From or Close to a Restricted Domain}\label{sub:restricteddomain-deg}
\label{app:restricteddomain-deg}
\begin{itemize}
	\item Elections from a restricted domain are typically quite degenerate. 
	\item In single-peaked elections the top-choices of all voters typically come from the same part of the societal order.
	\item Single-peaked or single-crossing elections that arise from deleting some candidates or voters from our election are more diverse than initially single-peaked or single-crossing elections.
\end{itemize}

Given that all but two of our elections that are part of a restricted domain come from the boxing top 16 dataset which contains elections where votes are typically very similar to each other, in this section we want to analyze the structural properties of elections that are part of or close to a restricted domain.  
In particular, we are interested in whether these elections are simply ``degenerate`` (with all voters having more or less the same preferences, thereby being part of a restricted domain more ``by accident'' than ``by design'') or ``exploit'' the full space of the domain (recall that, for instance, an election where half of the voters rank the candidates in one order and the other rank them in the opposite order is single-peaked, single-crossing, and group-separable).

\subsubsection{Elections from a Restricted Domain} 
\begin{figure}[t!]
	\centering
	\resizebox{0.4\textwidth}{!}{\input{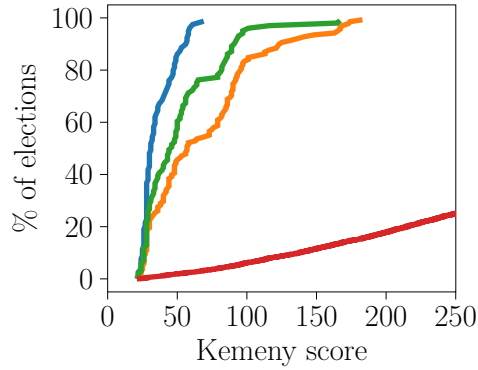}}
	\caption{Fraction of single-peaked (blue), single-crossing (orange), group-separable (green), and all elections (red) whose Kemeny score is at most the value displayed on the $x$-axis.}
	\label{fig:Kemeny-deg}
\end{figure}

We first consider elections that are part of a restricted domain and show in \Cref{fig:Kemeny-deg} a cumulative plot of their Kemeny score: 
A large majority of elections from a restricted domain have a Kemeny score below $100$, while only around $6\%$ of all elections have a Kemeny score below $100$.
In single-peaked elections, votes are especially similar ($86\%$ have a Kemeny score below $50$), whereas for single-crossingness there are at least $22$ elections ($16\%$) with a Kemeny score above $100$. 
Nevertheless, it seems that in real-world elections from a restricted domain voters have very similar preferences and that these elections only cover a very small part of the space of all elections from this restricted domain.

\begin{figure}[t!]
	\centering
	\begin{minipage}[t]{.47\textwidth}
		\centering
		\resizebox{0.8\textwidth}{!}{\begin{tikzpicture}
\definecolor{color0}{rgb}{0.12156862745098,0.466666666666667,0.705882352941177}
\definecolor{color1}{rgb}{1,0.498039215686275,0.0549019607843137}
\definecolor{color2}{rgb}{0.172549019607843,0.627450980392157,0.172549019607843}
\definecolor{color3}{rgb}{0.83921568627451,0.152941176470588,0.156862745098039}

\begin{axis}[
ylabel near ticks, label style={font=\large},
xlabel=candidate deletion distance,
x grid style={white!69.0196078431373!black},
xmin=-1, xmax=14,
xtick style={color=black},
xtick={0,2,4,6,8,10,12},
xticklabels={0,2,4,6,8,10,12},
y grid style={white!69.0196078431373!black},
ymin=-1, ymax=600,
ytick style={color=black},ylabel shift = 0pt,
ylabel=average Kemeny score,
every tick label/.append style={font=\LARGE}, 
label style={font=\LARGE}
]
\addplot[color0,line width=3pt] coordinates {(0,35.6025641025641) (1,59.32183908045977) (2,82.60084033613445) (3,97.45398773006134) (4,121.50319829424308) (5,146.73125) (6,174.4709026128266) (7,201.83815851922165) (8,221.81094736842104) (9,255.35650224215246) (10,340.8034339846063) (11,502.342782468838) (12,593.6039156626506) (13,595.2705450908485)};
\addplot[color1,line width=3pt] coordinates {(0,68.56204379562044) (1,77.3731778425656) (2,85.36117381489842) (3,92.7639405204461) (4,112.02309782608695) (5,135.459273797841) (6,168.03939184519695) (7,197.938117524701) (8,217.67527839643654) (9,253.99309021113243) (10,376.5056928630936) (11,558.7086870146288) (12,594.6692256341789) (13,595.2705450908485)};
\addplot[color2,line width=3pt] coordinates {(0,52.15) (1,66.5239852398524) (2,70.63450292397661) (3,84.73651452282158) (4,102.75075987841946) (5,119.71011235955056) (6,145.20064205457464) (7,175.88028169014083) (8,205.22368421052633) (9,239.65252525252527) (10,373.73205596107056) (11,546.8068464333525) (12,594.9628023352794) (13,595.2705450908485)};

\end{axis}
% \begin{axis}[
% color=blue,  label style={font=\Large}, ylabel shift = 0pt,
% separate axis lines,
% ylabel=fraction of changing pairs,
% axis y line*=left,
% axis x line=none,
% ymin=-0.1, ymax=1.1,
% xmin=-1, xmax=14,every axis 
% plot/.append style={thick},
% every tick label/.append style={font=\Huge}, 
% label style={font=\Huge}
% ]
% \addplot[blue!100,mark=*,line width=3pt] coordinates{
% 	(0,0.0657694962042787) (1,0.08289768483943222) (2,0.0976923076923076) (3,0.13046251993620409) (4,0.1461891643709825) (5,0.17110325873576737) (6,0.2078790238836967) (7,0.2421968787515006) (8,0.29591836734693844) (9,0.44) (10,0.6832329317269076) (11,0.8748278236914603) (12,0.9824175824175823) (13,1.0)
% };
% %\addplot+[lightgray,forget plot,thick] coordinates{(0.248,0) (0.248,250)};
% \end{axis}

\end{tikzpicture}}
		\caption{Average Kemeny score of elections within a given candidate deletion distance to single-peakedness (blue), single-crossingness (orange), or group-separability (green).}
		\label{fig:Kemeny-dist} 
	\end{minipage}\hfill
	\begin{minipage}[t]{.47\textwidth}
		\centering
		\resizebox{0.8\textwidth}{!}{% This file was created with tikzplotlib v0.9.17.
\begin{tikzpicture}

\definecolor{color0}{rgb}{0.12156862745098,0.466666666666667,0.705882352941177}

\begin{axis}[
tick align=outside,
tick pos=left,
xtick={1,2,3,4,5,6,7,8,9,10,11,12},
xticklabels={1,2,3,4,5,6,7,8,9,10,11,12},
x grid style={white!69.0196078431373!black},
xlabel={\#different top choices},
xmin=0, xmax=13,
xtick style={color=black},
y grid style={white!69.0196078431373!black},
ylabel={\#elections},
ymin=0, ymax=1251.6,
ytick style={color=black},
every tick label/.append style={font=\Large}, 
label style={font=\Large}
]
% \draw[draw=none,fill=color0] (axis cs:0.5,0) rectangle (axis cs:1.5,1192);
% \draw[draw=none,fill=color0] (axis cs:1.5,0) rectangle (axis cs:2.5,836);
% \draw[draw=none,fill=color0] (axis cs:2.5,0) rectangle (axis cs:3.5,719);
% \draw[draw=none,fill=color0] (axis cs:3.5,0) rectangle (axis cs:4.5,679);
% \draw[draw=none,fill=color0] (axis cs:4.5,0) rectangle (axis cs:5.5,735);
% \draw[draw=none,fill=color0] (axis cs:5.5,0) rectangle (axis cs:6.5,685);
% \draw[draw=none,fill=color0] (axis cs:6.5,0) rectangle (axis cs:7.5,587);
% \draw[draw=none,fill=color0] (axis cs:7.5,0) rectangle (axis cs:8.5,352);
% \draw[draw=none,fill=color0] (axis cs:8.5,0) rectangle (axis cs:9.5,139);
% \draw[draw=none,fill=color0] (axis cs:9.5,0) rectangle (axis cs:10.5,53);
% \draw[draw=none,fill=color0] (axis cs:10.5,0) rectangle (axis cs:11.5,18);
% \draw[draw=none,fill=color0] (axis cs:11.5,0) rectangle (axis cs:12.5,5);
% \draw[draw=none,fill=color0] (axis cs:12.5,0) rectangle (axis cs:13.5,0);
% \draw[draw=none,fill=color0] (axis cs:13.5,0) rectangle (axis cs:14.5,0);

\addplot[mark=*,line width=3pt] coordinates{
	(1,1192) (2,836) (3,719) (4,679) (5,735) (6,685) (7,587) (8,352) (9,139) (10,53) (11,18) (12,5)
};

\end{axis}

\end{tikzpicture}}
		\caption{Number of elections (y-axis) with a given number of top-choices (x-axis).}
		\label{fig:top-choice-dist}   
	\end{minipage}
\end{figure}

Considering all elections, there is a clear connection between the Kemeny score of an election and its distance to a restricted domain:
In \Cref{fig:Kemeny-dist}, for each restricted domain, we depict the average Kemeny score of elections within some candidate deletion distance. 
The plot clearly shows that the further an election is away from a restricted domain, the higher, on average, is its Kemeny score. 
While for smaller candidate deletion distances, the average Kemeny score roughly increases by $30$ if we increase the distance by one, there is a rapid increase by roughly $200$ when moving from distance $10$ to $11$ (notably, for each restricted domain, roughly $30\%$ of all elections are at candidate deletion distance $11$). 
For the voter deletion distance the picture is similar.
Overall, these findings indicate that in elections from or close to a restricted domain votes are similar to each other.

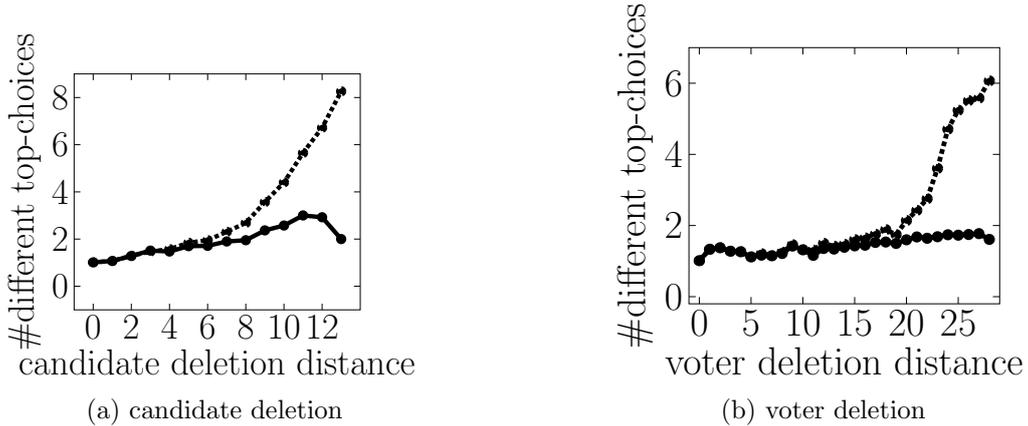
\begin{figure}[t!] 
	\centering
	\begin{subfigure}[b]{0.35\textwidth}
		\centering
		\resizebox{\textwidth}{!}{\begin{tikzpicture}

\begin{axis}[
ylabel near ticks, label style={font=\Huge},
xlabel=candidate deletion distance,
x grid style={white!69.0196078431373!black},
xmin=-1, xmax=14,
xtick style={color=black},
xtick={0,2,4,6,8,10,12},
xticklabels={0,2,4,6,8,10,12},
y grid style={white!69.0196078431373!black},
ymin=-1, ymax=9,
ytick style={color=black},ylabel shift = 0pt,
ylabel=\#different top-choices,
every tick label/.append style={font=\Huge}, 
label style={font=\Huge}
]
\addplot[black!100,mark=*,dotted,line width=3pt] coordinates {(0,1.0126582278481013) (1,1.0710382513661203) (2,1.2930232558139534) (3,1.4715909090909092) (4,1.5734265734265733) (5,1.845029239766082) (6,1.952970297029703) (7,2.307328605200946) (8,2.6865671641791047) (9,3.568106312292359) (10,4.401709401709402) (11,5.652882205513785) (12,6.717564870259481) (13,8.26086956521739)};
\addplot[black!100,mark=*,line width=3pt] coordinates{
	(0,1.0126582278481013) (1,1.0710382513661203) (2,1.283720930232558) (3,1.5113636363636365) (4,1.472027972027972) (5,1.6988304093567252) (6,1.7128712871287128) (7,1.8983451536643026) (8,1.9514925373134329) (9,2.3687707641196014) (10,2.574074074074074) (11,3.00125313283208) (12,2.9251497005988023) (13,2.0)
};
\end{axis}
% \begin{axis}[
% color=blue,  label style={font=\Large}, ylabel shift = 0pt,
% separate axis lines,
% ylabel=fraction of changing pairs,
% axis y line*=left,
% axis x line=none,
% ymin=-0.1, ymax=1.1,
% xmin=-1, xmax=14,every axis 
% plot/.append style={thick},
% every tick label/.append style={font=\Huge}, 
% label style={font=\Huge}
% ]
% \addplot[blue!100,mark=*,line width=3pt] coordinates{
% 	(0,0.0657694962042787) (1,0.08289768483943222) (2,0.0976923076923076) (3,0.13046251993620409) (4,0.1461891643709825) (5,0.17110325873576737) (6,0.2078790238836967) (7,0.2421968787515006) (8,0.29591836734693844) (9,0.44) (10,0.6832329317269076) (11,0.8748278236914603) (12,0.9824175824175823) (13,1.0)
% };
% %\addplot+[lightgray,forget plot,thick] coordinates{(0.248,0) (0.248,250)};
% \end{axis}

\end{tikzpicture}}
		\caption{candidate deletion}\label{fig:structural-props-deg-cand-sp}
	\end{subfigure}
	\qquad \qquad \qquad
	\begin{subfigure}[b]{0.35\textwidth}
		\centering 
		\resizebox{\textwidth}{!}{\begin{tikzpicture}

\begin{axis}[
ylabel near ticks, label style={font=\Huge},
xlabel=voter deletion distance,
x grid style={white!69.0196078431373!black},
xmin=-1, xmax=29,
xtick style={color=black},
y grid style={white!69.0196078431373!black},
ymin=-0.2, ymax=7,
ytick style={color=black},ylabel shift = 0pt,
ylabel=\#different top-choices,
every tick label/.append style={font=\Huge}, 
label style={font=\Huge}
]
\addplot[black!100,mark=*,dotted,line width=3pt] coordinates {(0,1.0126582278481013) (1,1.3333333333333333) (2,1.375) (3,1.2777777777777777) (4,1.263157894736842) (5,1.1176470588235294) (6,1.2083333333333333) (7,1.15) (8,1.25) (9,1.4615384615384615) (10,1.3170731707317074) (11,1.3064516129032258) (12,1.5) (13,1.4193548387096775) (14,1.4722222222222223) (15,1.6043956043956045) (16,1.6504854368932038) (17,1.7542372881355932) (18,1.8840579710144927) (19,1.7450980392156863) (20,2.140096618357488) (21,2.425373134328358) (22,2.7533783783783785) (23,3.605769230769231) (24,4.706645056726094) (25,5.239308462238399) (26,5.510362694300518) (27,5.5788690476190474) (28,6.064935064935065)};
\addplot[black!100,mark=*,line width=3pt] coordinates{
	(0,1.0126582278481013) (1,1.3333333333333333) (2,1.375) (3,1.2777777777777777) (4,1.263157894736842) (5,1.1176470588235294) (6,1.1666666666666667) (7,1.15) (8,1.2142857142857142) (9,1.4230769230769231) (10,1.3170731707317074) (11,1.1612903225806452) (12,1.3548387096774193) (13,1.3387096774193548) (14,1.3888888888888888) (15,1.4285714285714286) (16,1.4466019417475728) (17,1.5254237288135593) (18,1.536231884057971) (19,1.4967320261437909) (20,1.5990338164251208) (21,1.6791044776119404) (22,1.6418918918918919) (23,1.6850961538461537) (24,1.7406807131280388) (25,1.7288444040036397) (26,1.7478411053540588) (27,1.7723214285714286) (28,1.6103896103896105)
};
\end{axis}

\end{tikzpicture}}
		\caption{voter deletion}\label{fig:structural-props-deg-voter-sp}
	\end{subfigure}
	\caption{In solid (dotted), number of different top-choices of voters for elections at a certain distance from single-peakedness after (before) deleting the respective candidates or voters.}
	\label{fig:structural-props-deg-sp}
\end{figure}

\begin{figure*}[t]
	\centering
	\begin{subfigure}[t]{0.3\textwidth}
		\centering
		\resizebox{\textwidth}{!}{% This file was created with tikzplotlib v0.9.17.
\begin{tikzpicture}

\definecolor{color0}{rgb}{0.12156862745098,0.466666666666667,0.705882352941177}

\begin{axis}[
tick align=outside,
tick pos=left,
x grid style={white!69.0196078431373!black},
xlabel={top-choice rank in societal ordering},
xmin=1, xmax=15,
xtick style={color=black},
y grid style={white!69.0196078431373!black},
ylabel={\#votes},
ymin=0, ymax=118.65,
every tick label/.append style={font=\Huge}, 
label style={font=\Huge},
ytick style={color=black}
]
% \draw[draw=none,fill=color0] (axis cs:-0.5,0) rectangle (axis cs:0.5,0);
% \draw[draw=none,fill=color0] (axis cs:0.5,0) rectangle (axis cs:1.5,0);
% \draw[draw=none,fill=color0] (axis cs:1.5,0) rectangle (axis cs:2.5,0);
% \draw[draw=none,fill=color0] (axis cs:2.5,0) rectangle (axis cs:3.5,0);
% \draw[draw=none,fill=color0] (axis cs:3.5,0) rectangle (axis cs:4.5,0);
% \draw[draw=none,fill=color0] (axis cs:4.5,0) rectangle (axis cs:5.5,0);
% \draw[draw=none,fill=color0] (axis cs:5.5,0) rectangle (axis cs:6.5,0);
% \draw[draw=none,fill=color0] (axis cs:6.5,0) rectangle (axis cs:7.5,0);
% \draw[draw=none,fill=color0] (axis cs:7.5,0) rectangle (axis cs:8.5,27);
% \draw[draw=none,fill=color0] (axis cs:8.5,0) rectangle (axis cs:9.5,26);
% \draw[draw=none,fill=color0] (axis cs:9.5,0) rectangle (axis cs:10.5,49);
% \draw[draw=none,fill=color0] (axis cs:10.5,0) rectangle (axis cs:11.5,113);
% \draw[draw=none,fill=color0] (axis cs:11.5,0) rectangle (axis cs:12.5,46);
% \draw[draw=none,fill=color0] (axis cs:12.5,0) rectangle (axis cs:13.5,18);
% \draw[draw=none,fill=color0] (axis cs:13.5,0) rectangle (axis cs:14.5,0);

\addplot[mark=*,line width=3pt] coordinates{
	(1,0) (2,0) (3,0) (4,0) (5,0) (6,0) (7,0) (8,0) (9,27) (10,26) (11,49) (12,113) (13,46) (14,18) (15,0)
};
\end{axis}

\end{tikzpicture}}
		\caption{All of our elections which are single-peaked.} \label{fig:peakposCD0}
	\end{subfigure}       
	\hfill
	\begin{subfigure}[t]{0.3\textwidth}
		\centering
		\resizebox{\textwidth}{!}{% This file was created with tikzplotlib v0.9.17.
\begin{tikzpicture}

\definecolor{color0}{rgb}{0.12156862745098,0.466666666666667,0.705882352941177}

\begin{axis}[
tick align=outside,
tick pos=left,
x grid style={white!69.0196078431373!black},
xlabel={top-choice rank in societal ordering},
xmin=1, xmax=15,
xtick style={color=black},
y grid style={white!69.0196078431373!black},
ylabel={\#votes},
ymin=0, ymax=72.45,
every tick label/.append style={font=\Huge}, 
label style={font=\Huge},
ytick style={color=black}
]
% \draw[draw=none,fill=color0] (axis cs:-0.5,0) rectangle (axis cs:0.5,0);
% \draw[draw=none,fill=color0] (axis cs:0.5,0) rectangle (axis cs:1.5,0);
% \draw[draw=none,fill=color0] (axis cs:1.5,0) rectangle (axis cs:2.5,2);
% \draw[draw=none,fill=color0] (axis cs:2.5,0) rectangle (axis cs:3.5,4);
% \draw[draw=none,fill=color0] (axis cs:3.5,0) rectangle (axis cs:4.5,20);
% \draw[draw=none,fill=color0] (axis cs:4.5,0) rectangle (axis cs:5.5,31);
% \draw[draw=none,fill=color0] (axis cs:5.5,0) rectangle (axis cs:6.5,48);
% \draw[draw=none,fill=color0] (axis cs:6.5,0) rectangle (axis cs:7.5,69);
% \draw[draw=none,fill=color0] (axis cs:7.5,0) rectangle (axis cs:8.5,69);
% \draw[draw=none,fill=color0] (axis cs:8.5,0) rectangle (axis cs:9.5,22);
% \draw[draw=none,fill=color0] (axis cs:9.5,0) rectangle (axis cs:10.5,10);
% \draw[draw=none,fill=color0] (axis cs:10.5,0) rectangle (axis cs:11.5,4);
% \draw[draw=none,fill=color0] (axis cs:11.5,0) rectangle (axis cs:12.5,0);
% \draw[draw=none,fill=color0] (axis cs:12.5,0) rectangle (axis cs:13.5,0);
% \draw[draw=none,fill=color0] (axis cs:13.5,0) rectangle (axis cs:14.5,0);

\addplot[mark=*,line width=3pt] coordinates{
	(1,0) (2,0) (3,2) (4,4) (5,20) (6,31) (7,48) (8,69) (9,69) (10,22) (11,10) (12,4) (13,0) (14,0) (15,0)
};
\end{axis}

\end{tikzpicture}}
		\caption{All of our elections which are single-peaked after applying a random tie-breaking when computing a compatible societal order.} \label{fig:peakposCD0random}
	\end{subfigure}       
	\hfill
	\begin{subfigure}[t]{0.3\textwidth}
		\centering
		\resizebox{\textwidth}{!}{% This file was created with tikzplotlib v0.9.17.
\begin{tikzpicture}

\definecolor{color0}{rgb}{0.12156862745098,0.466666666666667,0.705882352941177}

\begin{axis}[
tick align=outside,
tick pos=left,
x grid style={white!69.0196078431373!black},
xlabel={top-choice rank in societal ordering},
xmin=1, xmax=12,
xtick style={color=black},
y grid style={white!69.0196078431373!black},
ylabel={\#votes},
ymin=0, ymax=903,
every tick label/.append style={font=\Huge}, 
label style={font=\Huge},
ytick style={color=black}
]
% \draw[draw=none,fill=color0] (axis cs:-0.5,0) rectangle (axis cs:0.5,0);
% \draw[draw=none,fill=color0] (axis cs:0.5,0) rectangle (axis cs:1.5,0);
% \draw[draw=none,fill=color0] (axis cs:1.5,0) rectangle (axis cs:2.5,0);
% \draw[draw=none,fill=color0] (axis cs:2.5,0) rectangle (axis cs:3.5,0);
% \draw[draw=none,fill=color0] (axis cs:3.5,0) rectangle (axis cs:4.5,150);
% \draw[draw=none,fill=color0] (axis cs:4.5,0) rectangle (axis cs:5.5,586);
% \draw[draw=none,fill=color0] (axis cs:5.5,0) rectangle (axis cs:6.5,860);
% \draw[draw=none,fill=color0] (axis cs:6.5,0) rectangle (axis cs:7.5,531);
% \draw[draw=none,fill=color0] (axis cs:7.5,0) rectangle (axis cs:8.5,115);
% \draw[draw=none,fill=color0] (axis cs:8.5,0) rectangle (axis cs:9.5,1);
% \draw[draw=none,fill=color0] (axis cs:9.5,0) rectangle (axis cs:10.5,0);
% \draw[draw=none,fill=color0] (axis cs:10.5,0) rectangle (axis cs:11.5,0);
% \draw[draw=none,fill=color0] (axis cs:11.5,0) rectangle (axis cs:12.5,0);
% \draw[draw=none,fill=color0] (axis cs:12.5,0) rectangle (axis cs:13.5,0);
% \draw[draw=none,fill=color0] (axis cs:13.5,0) rectangle (axis cs:14.5,0);

\addplot[mark=*,line width=3pt] coordinates{
	(1,0) (2,0) (3,0) (4,0) (5,150) (6,586) (7,860) (8,531) (9,115) (10,1) (11,0) (12,0)
};
\end{axis}

\end{tikzpicture}}
		\caption{Single-peaked elections obtained from deleting $3$ candidates from elections at candidate deletion distance~$3$.}\label{fig:peakposCD3}
	\end{subfigure}
	\hfill
	\begin{subfigure}[t]{0.3\textwidth}
		\centering
		\resizebox{\textwidth}{!}{% This file was created with tikzplotlib v0.9.17.
\begin{tikzpicture}

\definecolor{color0}{rgb}{0.12156862745098,0.466666666666667,0.705882352941177}

\begin{axis}[
tick align=outside,
tick pos=left,
x grid style={white!69.0196078431373!black},
xlabel={top-choice rank in societal ordering},
xmin=1, xmax=15,
xtick style={color=black},
y grid style={white!69.0196078431373!black},
ylabel={\#votes},
ymin=0, ymax=135.45,
every tick label/.append style={font=\Huge}, 
label style={font=\Huge},
ytick style={color=black}
]
% \draw[draw=none,fill=color0] (axis cs:-0.5,0) rectangle (axis cs:0.5,0);
% \draw[draw=none,fill=color0] (axis cs:0.5,0) rectangle (axis cs:1.5,0);
% \draw[draw=none,fill=color0] (axis cs:1.5,0) rectangle (axis cs:2.5,0);
% \draw[draw=none,fill=color0] (axis cs:2.5,0) rectangle (axis cs:3.5,0);
% \draw[draw=none,fill=color0] (axis cs:3.5,0) rectangle (axis cs:4.5,0);
% \draw[draw=none,fill=color0] (axis cs:4.5,0) rectangle (axis cs:5.5,0);
% \draw[draw=none,fill=color0] (axis cs:5.5,0) rectangle (axis cs:6.5,21);
% \draw[draw=none,fill=color0] (axis cs:6.5,0) rectangle (axis cs:7.5,91);
% \draw[draw=none,fill=color0] (axis cs:7.5,0) rectangle (axis cs:8.5,129);
% \draw[draw=none,fill=color0] (axis cs:8.5,0) rectangle (axis cs:9.5,81);
% \draw[draw=none,fill=color0] (axis cs:9.5,0) rectangle (axis cs:10.5,29);
% \draw[draw=none,fill=color0] (axis cs:10.5,0) rectangle (axis cs:11.5,7);
% \draw[draw=none,fill=color0] (axis cs:11.5,0) rectangle (axis cs:12.5,9);
% \draw[draw=none,fill=color0] (axis cs:12.5,0) rectangle (axis cs:13.5,8);
% \draw[draw=none,fill=color0] (axis cs:13.5,0) rectangle (axis cs:14.5,0);

\addplot[mark=*,line width=3pt] coordinates{
	(1,0) (2,0) (3,0) (4,0) (5,0) (6,0) (7,21) (8,91) (9,129) (10,81) (11,29) (12,7) (13,9) (14,8) (15,0)
};
\end{axis}

\end{tikzpicture}}
		\caption{Single-peaked elections obtained from deleting $12$ voters from elections at voter deletion distance $12$.}\label{fig:peakposVD12}
	\end{subfigure}
	\qquad \qquad \qquad 
	\begin{subfigure}[t]{0.3\textwidth}
		\centering  
		\resizebox{\textwidth}{!}{% This file was created with tikzplotlib v0.9.17.
\begin{tikzpicture}

\definecolor{color0}{rgb}{0.12156862745098,0.466666666666667,0.705882352941177}

\begin{axis}[
tick align=outside,
tick pos=left,
x grid style={white!69.0196078431373!black},
xlabel={top-choice rank in societal ordering},
xmin=1, xmax=15,
xtick style={color=black},
y grid style={white!69.0196078431373!black},
ylabel={\#votes},
ymin=0, ymax=806.4,
every tick label/.append style={font=\Huge}, 
label style={font=\Huge},
ytick style={color=black}
]
% \draw[draw=none,fill=color0] (axis cs:-0.5,0) rectangle (axis cs:0.5,0);
% \draw[draw=none,fill=color0] (axis cs:0.5,0) rectangle (axis cs:1.5,0);
% \draw[draw=none,fill=color0] (axis cs:1.5,0) rectangle (axis cs:2.5,0);
% \draw[draw=none,fill=color0] (axis cs:2.5,0) rectangle (axis cs:3.5,0);
% \draw[draw=none,fill=color0] (axis cs:3.5,0) rectangle (axis cs:4.5,3);
% \draw[draw=none,fill=color0] (axis cs:4.5,0) rectangle (axis cs:5.5,15);
% \draw[draw=none,fill=color0] (axis cs:5.5,0) rectangle (axis cs:6.5,70);
% \draw[draw=none,fill=color0] (axis cs:6.5,0) rectangle (axis cs:7.5,136);
% \draw[draw=none,fill=color0] (axis cs:7.5,0) rectangle (axis cs:8.5,215);
% \draw[draw=none,fill=color0] (axis cs:8.5,0) rectangle (axis cs:9.5,160);
% \draw[draw=none,fill=color0] (axis cs:9.5,0) rectangle (axis cs:10.5,140);
% \draw[draw=none,fill=color0] (axis cs:10.5,0) rectangle (axis cs:11.5,71);
% \draw[draw=none,fill=color0] (axis cs:11.5,0) rectangle (axis cs:12.5,22);
% \draw[draw=none,fill=color0] (axis cs:12.5,0) rectangle (axis cs:13.5,9);
% \draw[draw=none,fill=color0] (axis cs:13.5,0) rectangle (axis cs:14.5,768);

\addplot[mark=*,line width=3pt] coordinates{
	(1,0) (2,0) (3,0) (4,0) (5,3) (6,15) (7,70) (8,136) (9,215) (10,160) (11,140) (12,71) (13,22) (14,9) (15,768)
};
\end{axis}

\end{tikzpicture}}
		\caption{Single-peaked elections obtained from deleting $25$ voters from elections at voter deletion distance $25$.}\label{fig:peakposVD25}
	\end{subfigure}
	\caption{Each single-peaked election comes with a societal order. The diagrams depict where the top-choices of voters lie on this axis. Specifically, for each $i\in [15]$, we depict the number of voters that have the $i$th candidate from the societal order as their top-choice for different types of single-peaked elections.}
	\label{fig:peakpos}
\end{figure*}
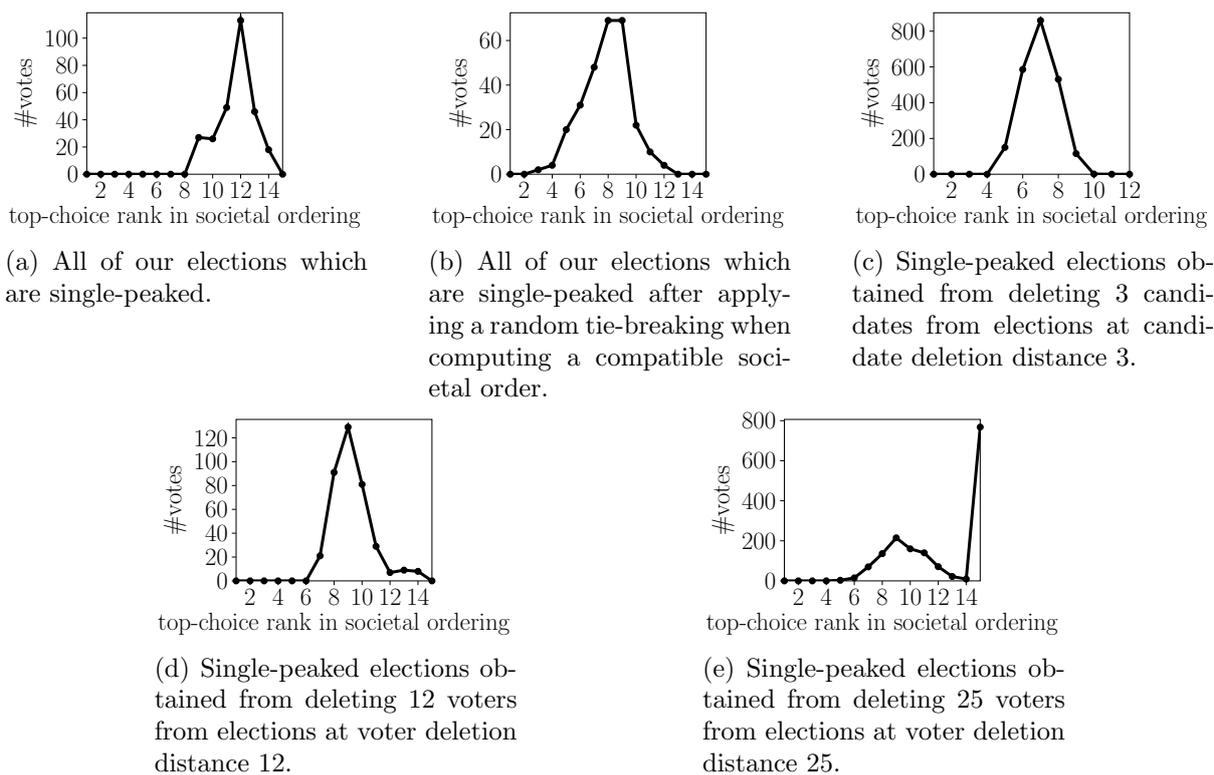

\subsubsection{(Close to) Single-Peakedness} For single-peaked elections a very simple  measure for how degenerate elections are is the number of different top-choices of voters, as the top-choice of a voter in a single-peaked election fundamentally influences the vote and in principle all top-choices are possible.
To get a feeling for how many different top-choices an election typically has, in \Cref{fig:top-choice-dist}, we depict the distribution of the number of different top-choices in an election. 
On average, our elections have only $4.081$ different top-choices.

\paragraph{Number of Different Top-Choices.} Out of our $79$ single-peaked elections, in $78$ of them, all voters have the same top-choice and in only one of them, voters have two different ones. This indicates again the degenerateness of the found single-peaked elections. 
Moreover, in \Cref{fig:structural-props-deg-sp}, in dotted lines we depict the average number of different top-choices in elections at a given voter and candidate deletion distance.
Remarkably, for elections up to candidate deletion distance $6$ and up to voter deletion distance $19$ the average number of top-choices is smaller than $2$ and only moving (even) further away from single-peakedness leads to a substantial increase in the number of top-choices. 
This suggests that also elections close to single-peakedness do not capture the full space of these elections.
However, it is also possible to take a different view here and to consider the elections at some distance to single-peakedness after the deletion of candidates or voters (so the elections that remain and that are actually single-peaked).
The results are depicted in the solid lines in  \Cref{fig:structural-props-deg-sp}.
Interestingly, the elections created this way seem to be a bit more diverse than originally single-peaked elections:  
Until candidate deletion distance $11$ 
the
average 
number of top-choices
constantly
increases 
from around $1.5$ top-choices at distance $3$ to around $3$ top-choices at distance $11$.
For the voter deletion distance we have a more or less constant increase from around $1$ for distance $0$ to around $1.7$ for distance $27$. 
This steady increase is even more remarkable considering that by deleting voters or candidates we decrease the maximum number of possible different top-choices.

Interestingly, the results from the previous paragraph suggest that the Walsh model \cite{DBLP:journals/corr/Walsh15} to sample single-peaked preferences might be more realistic than the Conitzer model \cite{DBLP:journals/jair/Conitzer09}.
In the Walsh model, one samples an election from the space of single-peaked elections uniformly at random and with a high-probability the top choice of all voters is one of the middle (two) candidates from the societal order.
In contrast to this, in the Conitzer model, top-choices of voters are drawn uniformly at random.
This hypothesis gets confirmed in the next paragraph.

\paragraph{Position of Top-Choices in the Societal Order.} Next, we analyze where top-choices of voters in single-peaked elections lie on the societal order. 
Without loss of generality assume that the societal order is $c_1\succ \dots \succ c_{15}$ (if this is not the case we can relabel the candidates). 
Thus, $c_1$ and $c_{15}$ are the extreme candidates in this election (e.g., in case that an election models a political election these could be the left extremist and right extremists candidates), whereas $c_7$ and $c_8$ are in some sense the center candidates. 
We are now interested in the rank of the top-choice of each voter in the societal order, which corresponds to the candidate's index for the societal order  $c_1\succ \dots \succ c_{15}$. 
For this, we look at the single-peaked elections that are obtained from our initial elections by deleting a minimum number of candidates or voters (i.e., the single-peaked ``cores'' of our elections obtained when computing the respective distances to single-peakedness).
We depict the frequency of  top-choice ranks in the societal order in \Cref{fig:peakpos}.
In \Cref{fig:peakposCD0}, we consider all elections from our dataset which are single-peaked. 
The results are remarkable in two ways.

First, the top-choices of voters do not lie in the middle of the societal order (like one would probably intuitively expect it), but are heavily skewered to one side. 
One explanation for this could be the single-peaked detection algorithm by \citet{DBLP:conf/ecai/EscoffierLO08} which we use. 
This algorithm constructs the societal order in a greedy fashion by successively inserting candidates ranked last by some voter at the extremes of the societal order. 
Ties are here broken always by inserting candidates on the top free position in the societal order, which could result in top-choices of voters typically being inserted rather in the bottom of the axis. 
To check this hypothesis, we reran the experiments with an adapted variant of the single-peaked detection algorithm, where we break ties independent uniformly at random and depict the results in \Cref{fig:peakposCD0random}.
Indeed, the top-choices of voters now lie in the middle of the axis. 
Consequently, the voters top-choices are not ``truly'' skewered to one side. 
Nevertheless, these observations still allow for some non-obvious conclusions concerning our data: 
Single-peaked elections are typically consistent with many potentially quite different axes, indicating that the final axis carries only little meaning and in single-peaked elections votes are quite similar.

Second, the top-choices of voters come all more or less from the same part of the societal order. 
Remarkably, there is not a single vote in all single-peaked elections where one of the first eight candidates from the societal order is ranked in the first position. 
In principle, any distribution is possible. 
For instance, the synthetic Conitzer model assumes a uniform distribution of top-choice ranks. 
This seems to be far away from reality.
However, this might be also due to to fact that all but two of our single-peaked elections come from the boxing top 16 dataset in which votes are anyway quite similar. 
Thus, to test this observation we now also look at the single-peaked elections that result from deleting a minimum number of voters or candidates from close to single-peaked elections. 

We start by looking at the single-peaked elections that result from deleting $i$ candidates from elections that are at candidate deletion distance $i$ from being single-peaked. 
We depict the corresponding plot for $i=3$ in \Cref{fig:peakposCD3}. 
Independent of the value $i$, the plots always look similar to a Gaussian distribution, confirming our earlier observation that all voters typically have top-choices from the same part of the societal order. 
However, for increasing $i$ the peak of the distribution moves towards the center of the societal order. For instance, after removing for instance $3$ candidates, the most frequent top-choice rank is $7$. 
This indicates that the generated elections are no longer compatible with a lot of quite different societal orders. 

If we instead consider single-peaked elections that result from deleting $i$ voters from elections that are at voter deletion distance $i$ from being single-peaked, the picture is slightly different: 
Here, for $i\in [0,22]$, the distribution is again similar to a Gaussian distribution with the peak slowly shifting towards the center. 
We depict the distribution for $i=12$ in \Cref{fig:peakposVD12}.
However, for $i\geq 23$, more and more votes start to have the last candidate from the societal order as their top choice (see \Cref{fig:peakposVD25} for $i=25$ for an example). 
Note that if a voter has the last candidate from the axis as its top-choice, then its vote is already fully determined by this (as the reversed societal order). 
This indicates that in elections at a large voter deletion distance the single-peaked core simply consists of identical votes.

\subsubsection{(Close to) Single-Crossingness} Recall that for single-crossingness we require that if we move along the central order of voters the ordering of each pair of candidates changes at most once.
So the extremely diverse case would be that each pair of candidates changes ordering exactly once. 
This is why we consider the fraction of pairs of candidates that change their ordering as a measure for the degenerateness of the election. 
Here for candidate/voter deletion distance larger than zero, we analyze only the properties of the elections that remain after the deletion of candidates or voters (as for the original elections there is no ordering of voters).

\begin{figure}  
	\centering
	\begin{subfigure}[b]{0.35\textwidth}
		\centering
		\resizebox{\textwidth}{!}{\begin{tikzpicture}

% \begin{axis}[
% ylabel near ticks, label style={font=\Large},
% xlabel=candidate deletion distance,
% x grid style={white!69.0196078431373!black},
% xmin=-1, xmax=14,
% xtick style={color=black},
% xtick={0,2,4,6,8,10,12},
% xticklabels={0,2,4,6,8,10,12},
% y grid style={white!69.0196078431373!black},
% ymin=-1, ymax=9,
% ytick style={color=black},ylabel shift = 0pt,
% ylabel=\#different top-choices,
% every tick label/.append style={font=\Huge}, 
% label style={font=\Huge}
% ]
% \addplot[black!100,mark=*,dotted,line width=3pt] coordinates {(0,1.0126582278481013) (1,1.0710382513661203) (2,1.2930232558139534) (3,1.4715909090909092) (4,1.5734265734265733) (5,1.845029239766082) (6,1.952970297029703) (7,2.307328605200946) (8,2.6865671641791047) (9,3.568106312292359) (10,4.401709401709402) (11,5.652882205513785) (12,6.717564870259481) (13,8.26086956521739)};
% \addplot[black!100,mark=*,line width=3pt] coordinates{
% 	(0,1.0126582278481013) (1,1.0710382513661203) (2,1.283720930232558) (3,1.5113636363636365) (4,1.472027972027972) (5,1.6988304093567252) (6,1.7128712871287128) (7,1.8983451536643026) (8,1.9514925373134329) (9,2.3687707641196014) (10,2.574074074074074) (11,3.00125313283208) (12,2.9251497005988023) (13,2.0)
% };
% \end{axis}
\begin{axis}[
ylabel near ticks,
xlabel=candidate deletion distance,
xtick={0,2,4,6,8,10,12},
xticklabels={0,2,4,6,8,10,12},
x grid style={white!69.0196078431373!black},
label style={font=\Huge}, ylabel shift = 0pt,
ylabel=fraction of changing pairs,
ymin=-0.1, ymax=1.1,
xmin=-1, xmax=14,every axis 
plot/.append style={thick},
every tick label/.append style={font=\Huge}, 
label style={font=\Huge}
]
\addplot[mark=*,line width=3pt] coordinates{
	(0,0.0657694962042787) (1,0.08289768483943222) (2,0.0976923076923076) (3,0.13046251993620409) (4,0.1461891643709825) (5,0.17110325873576737) (6,0.2078790238836967) (7,0.2421968787515006) (8,0.29591836734693844) (9,0.44) (10,0.6832329317269076) (11,0.8748278236914603) (12,0.9824175824175823) (13,1.0)
};
%\addplot+[lightgray,forget plot,thick] coordinates{(0.248,0) (0.248,250)};
\end{axis}

\end{tikzpicture}}
		\caption{candidate deletion}\label{fig:structural-props-deg-cand-sc}
	\end{subfigure}
	\qquad \qquad \qquad 
	\begin{subfigure}[b]{0.35\textwidth}
		\centering 
		\resizebox{\textwidth}{!}{\begin{tikzpicture}

\begin{axis}[
ylabel near ticks,label style={font=\Large}, ylabel shift = 0pt,
separate axis lines,
xlabel=voter deletion distance,
ylabel=fraction of changing pairs,
ymin=0.04, ymax=0.55,
xmin=-1, xmax=27,every axis 
plot/.append style={thick},
every tick label/.append style={font=\Huge}, 
label style={font=\Huge}
]
\addplot[mark=*,line width=3pt] coordinates{
	(0,0.0657694962042787) (1,0.1019047619047619) (2,0.09720853858784893) (3,0.07573696145124717) (4,0.09175377468060393) (5,0.10476190476190479) (6,0.10158730158730159) (7,0.11017316017316015) (8,0.1276595744680851) (9,0.11371428571428574) (10,0.14637681159420293) (11,0.15222978080120936) (12,0.1896825396825397) (13,0.1473495058400719) (14,0.25659551176792555) (15,0.24371953505811778) (16,0.29598506069094305) (17,0.3051467051467051) (18,0.3582084582084582) (19,0.4253221288515405) (20,0.46358065706902896) (21,0.4934697179842809) (22,0.47532975631567187) (23,0.4672007120605251) (24,0.46129464285714283) (25,0.4682027649769587) (26,0.4562443845462713)
};
%\addplot+[lightgray,forget plot,thick] coordinates{(0.248,0) (0.248,250)};
\end{axis}

\end{tikzpicture}}
		\caption{voter deletion}\label{fig:structural-props-deg-voter-sc}
	\end{subfigure}
	\caption{Fraction of candidate pairs who are not ranked in the same order by all voters for elections at a certain distance from single-crossingness after deleting the respective candidates or voters.}
	\label{fig:structural-props-deg-sc}
\end{figure}
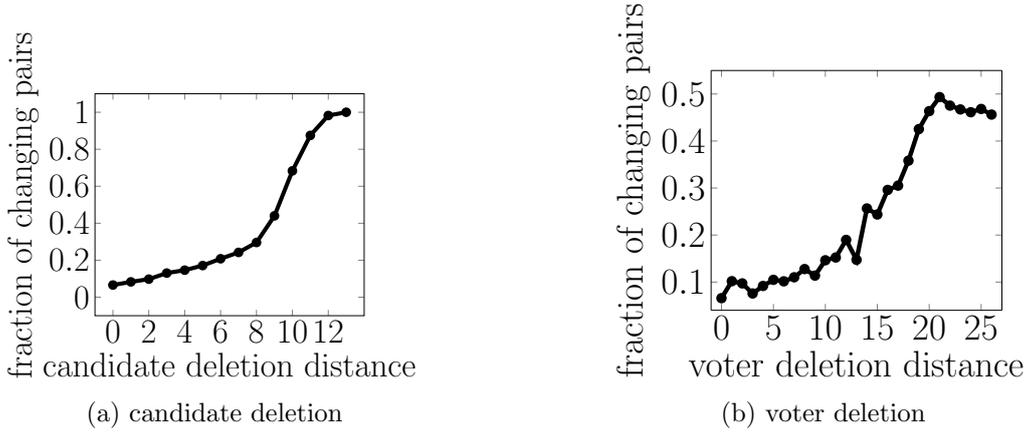

In \Cref{fig:structural-props-deg-sc}, we show the results.
While for single-crossing elections only on average $6.5\%$ of candidate pairs swap their ordering when moving along the central order, for increasing candidate deletion distance this value constantly increases until almost all of the remaining candidate pairs switch their ordering. 
For increasing voter deletion distance it increases until half of all candidate pairs switch their ordering. 
The difference between the behavior for the candidate and voter deletion distance here is probably due to the fact that with fewer candidates remaining it becomes ``easier'' to swap the ordering of all candidate pairs, while with fewer voters remaining it becomes in some sense ``harder''. 
Nevertheless, the results also suggest here that the single-crossing ``core'' of non-single-crossing elections is much more diverse than full single-crossing elections from our dataset.

\subsection{General Preference Restrictions} \label{sub:general-restrictions}
All three considered restricted domains can be characterized via two forbidden configurations (meaning that elections are part of the domain if they do not contain both forbidden configurations).
The most important configurations are: 
\begin{description}
	\item[$\boldsymbol{\alpha}$-configuration]: Two voters $v$ and $v'$ and four candidates $a,b,c$, and $d$ with: 
	$$a\succ b\succ c  \wedge  d \succ b \text{ and } c \succ' b \succ' a \wedge  d \succ' b.$$
	\item[$\boldsymbol{\beta}$-configuration]: Two voters $v$ and $v'$ and four candidates $a,b,c,$ and $d$ with: 
	$$a\succ b\succ c \succ d \text{ and } b \succ' d \succ' a \succ' c.$$
	\item[$\boldsymbol{\gamma}$-configuration]: Three voters $v$, $v'$, and $v''$ and six (possibly identical) candidates $a,b,c,d,e,$ and $f$ with: 
	\begin{align*}
	& b\succ a \wedge c\succ d \wedge e\succ f \text{ and }\\
	& a\succ'b \wedge d \succ' c \wedge e\succ' f \text{ and }\\
	& a\succ'' b \wedge c\succ'' d \wedge f \succ'' e.
	\end{align*}
	\item[$\boldsymbol{\delta}$-configuration]: Four voters $v$, $v'$, $v''$, and $v'''$ and four (possibly identical) candidates $a,b,c,$ and $d$ with: 
	\begin{align*}
	& a\succ b\wedge c \succ d \text{ and }\\
	&  a \succ' b \wedge d \succ' c\text{ and }\\
	& b\succ'' a \wedge c \succ'' d \text{ and }\\
	& b\succ''' a \wedge d \succ''' c.
	\end{align*}
	\item[best-configuration] Three voters $v$, $v'$, and $v''$ and three candidates $a,b,$ and $c$ with: $$ a\succ b \wedge a\succ c \text{ and } b\succ' a \wedge b\succ'c \text{ and } c\succ'' a \wedge c \succ'' b.$$
	\item[worst-configuration]
	Three voters $v$, $v'$, and $v''$ and three candidates $a,b,$ and $c$ with: $$ a\succ c \wedge b\succ c \text{ and } a\succ' b \wedge c\succ' b \text{ and } b\succ'' a \wedge c \succ'' a.$$
	\item[medium-configuration] Three voters $v$, $v'$, and $v''$ and three candidates $a,b,$ and $c$ with: 
	\begin{align*}
	& b\succ a \succ c \vee c\succ a \succ b \text{ and }\\
	& a\succ' b \succ' c \vee c\succ' b \succ' a \text{ and }\\
	& a\succ'' c \succ'' b \vee b \succ'' c \succ'' a.
	\end{align*}
	\item[value-configuration] Three voters $v$, $v'$, and $v''$ and three candidates $a,b,$ and $c$ with:
	$$a\succ b \succ c \text{ and } b\succ' c\succ' a \text{ and } c\succ'' a \succ'' b.$$
\end{description}
The domains of best/worst/medium/value-restricted elections are characterized by the absence of best/worst/medium/value-configurations. 
Elections without a best-, worst-, or medium-configuration also do not admit a value-configuration.

Single-peaked elections are characterized by the absence of $\alpha$- and worst-configurations. 
Single-crossing elections are characterized by the absence of $\gamma$- and $\delta$-configurations.  
Group-separable elections are characterized by the absence of $\beta$- and medium-configurations.

\begin{figure*}
	\centering
	\includegraphics[width=\textwidth]{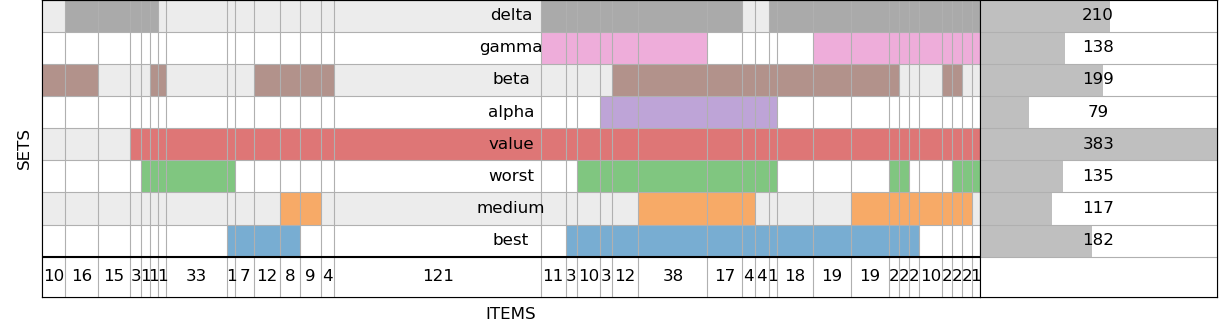}
	\caption{Diagram visualization the number of elections which do not admit a forbidden configuration. Each line corresponds to one configuration. For each configuration, in the right most column, the number of elections without the respective configuration is shown. In the other columns elections that do not admit one or multiple configurations are represented: Their number can be found at the bottom and the configurations they do not contain are highlighted. For instance, the second column says that there are $16$ elections that do not admit a $\beta$- and $\delta$- configuration.}
	\label{fig:subervenn}
\end{figure*}

In \Cref{fig:subervenn}, we visualize the number and relationship of elections that do not admit one (or multiple) of the forbidden configurations.
We find that for single-peaked elections, which are characterized by the absence of $\alpha$- and worst-configurations, worst-configurations are practically irrelevant as all elections that do not admit a $\alpha$-configuration also do not admit a worst-configuration. 
Similarly, for single-crossing elections (characterized by the absence of $\gamma$- and $\delta$-configurations) $\delta$-configurations are fully redundant. 
For group-separable elections (characterized by the absence of $\beta$- and medium-configurations) $\beta$-configurations are redundant in most cases. 
Lastly, there is one configuration which in an overwhelming majority of elections does not appear if any other configuration does not appear (some of these relationships are theoretically guaranteed): value-configurations.
Among the $424$ elections not admitting one of the forbidden configurations, $383$ do not admit a value-configuration.
This hints at the practical relevance of value-restricted elections \cite{sen1966possibility} as a (more) frequently occurring restricted domain. 

\begin{figure}
	\centering
	\begin{subfigure}[b]{0.35\textwidth}
		\centering
		\resizebox{\textwidth}{!}{% This file was created with tikzplotlib v0.9.17.
\begin{tikzpicture}

\definecolor{color0}{rgb}{1,0.647058823529412,0}

\begin{axis}[
legend cell align={left},
legend style={
  fill opacity=0.8,
  draw opacity=1,
  text opacity=1,
  at={(0.03,0.97)},
  anchor=north west,
  draw=white!80!black
},
tick align=outside,
tick pos=left,
x grid style={white!69.0196078431373!black},
xlabel={deleted voters/cand.},
xmin=-1.45, xmax=30.45,
xtick style={color=black},
y grid style={white!69.0196078431373!black},
ylabel={\% of elections},
every tick label/.append style={font=\Huge}, 
label style={font=\Huge},
ymin=-4.9996, ymax=104.9916,
ytick style={color=black}
]
\addplot [line width=3pt, blue]
table {%
0 0
0 1.29999995231628
1 1.31666672229767
1 4.34999990463257
2 4.36666679382324
2 7.93333339691162
3 7.94999980926514
3 10.8666667938232
4 10.8833332061768
4 15.6333332061768
5 15.6499996185303
5 21.3333339691162
6 21.3500003814697
6 28.0666675567627
7 28.0833339691162
7 35.1166648864746
8 35.1333351135254
8 39.5833320617676
9 39.5999984741211
9 44.5999984741211
10 44.6166648864746
10 56.2999992370605
11 56.3166656494141
11 82.9000015258789
12 82.9166641235352
12 99.5999984741211
13 99.6166687011719
13 99.9833297729492
};
\addplot [line width=3pt, blue, dashed]
table {%
0 0
0 1.29999995231628
1 1.31666672229767
1 1.35000002384186
2 1.36666667461395
2 1.48333334922791
3 1.5
3 1.78333330154419
4 1.79999995231628
4 2.09999990463257
5 2.11666655540466
5 2.38333344459534
6 2.40000009536743
6 2.78333330154419
7 2.79999995231628
7 3.45000004768372
8 3.46666669845581
8 3.91666674613953
9 3.93333339691162
9 4.78333330154419
10 4.80000019073486
10 5.46666669845581
11 5.48333311080933
11 6.5
12 6.51666688919067
12 7.53333330154419
13 7.55000019073486
13 8.56666660308838
14 8.58333301544189
14 9.76666641235352
15 9.78333377838135
15 11.2833337783813
16 11.3000001907349
16 13
17 13.0166664123535
17 14.9666662216187
18 14.9833335876465
18 17.2666664123535
19 17.283332824707
19 19.8166675567627
20 19.8333339691162
20 23.2666664123535
21 23.283332824707
21 27.7333335876465
22 27.75
22 32.6666679382324
23 32.6833343505859
23 39.5999984741211
24 39.6166648864746
24 49.8833351135254
25 49.9000015258789
25 68.1999969482422
26 68.216667175293
26 87.5
27 87.5166702270508
27 98.6999969482422
28 98.716667175293
28 99.9833297729492
};
\addplot [line width=3pt, color0]
table {%
0 0
0 2.28333330154419
1 2.29999995231628
1 5.71666669845581
2 5.73333311080933
2 7.38333320617676
3 7.40000009536743
3 8.96666622161865
4 8.98333358764648
4 12.2666664123535
5 12.2833337783813
5 16.9833335876465
6 17
6 24.1166667938232
7 24.1333332061768
7 32.0499992370605
8 32.0666656494141
8 37.4166679382324
9 37.4333343505859
9 43.4166679382324
10 43.4333343505859
10 60.0166664123535
11 60.033332824707
11 92.283332824707
12 92.3000030517578
12 99.8666687011719
13 99.8833312988281
13 99.9833297729492
};
\addplot [line width=3pt, color0, dashed]
table {%
0 0
0 2.28333330154419
1 2.29999995231628
1 2.45000004768372
2 2.46666669845581
2 2.93333339691162
3 2.95000004768372
3 3.28333330154419
4 3.29999995231628
4 3.96666669845581
5 3.98333334922791
5 4.44999980926514
6 4.46666669845581
6 5.05000019073486
7 5.06666660308838
7 5.78333330154419
8 5.80000019073486
8 6.56666660308838
9 6.58333349227905
9 7.40000009536743
10 7.41666650772095
10 8.16666698455811
11 8.18333339691162
11 9.21666622161865
12 9.23333358764648
12 9.81666660308838
13 9.83333301544189
13 10.6999998092651
14 10.7166662216187
14 12.1499996185303
15 12.1666669845581
15 14.2666664123535
16 14.2833337783813
16 16.8166675567627
17 16.8333339691162
17 20.1166667938232
18 20.1333332061768
18 24.8833332061768
19 24.8999996185303
19 31.966667175293
20 31.9833335876465
20 42.716667175293
21 42.7333335876465
21 56.4500007629395
22 56.466667175293
22 70.6500015258789
23 70.6666641235352
23 81.3499984741211
24 81.3666687011719
24 92.0166702270508
25 92.033332824707
25 98.216667175293
26 98.2333297729492
26 99.9833297729492
};
\addplot [line width=3pt, green!50.1960784313725!black]
table {%
0 0
0 1.66666662693024
1 1.68333327770233
1 4.51666688919067
2 4.53333330154419
2 5.69999980926514
3 5.71666669845581
3 8.03333377838135
4 8.05000019073486
4 10.9666662216187
5 10.9833335876465
5 14.8333330154419
6 14.8500003814697
6 20.7666664123535
7 20.783332824707
7 28.3999996185303
8 28.4166660308838
8 35.466667175293
9 35.4833335876465
9 41.25
10 41.2666664123535
10 54.7999992370605
11 54.8166656494141
11 87.1500015258789
12 87.1666641235352
12 99.9166641235352
13 99.9333343505859
13 99.9833297729492
};
\addplot [line width=3pt, green!50.1960784313725!black, dashed]
table {%
0 0
0 1.66666662693024
1 1.68333327770233
1 1.70000004768372
2 1.71666669845581
2 1.89999997615814
3 1.91666662693024
3 2
4 2.01666665077209
4 2.28333330154419
5 2.29999995231628
5 2.58333325386047
6 2.59999990463257
6 3.16666674613953
7 3.18333339691162
7 3.81666660308838
8 3.83333325386047
8 4.53333330154419
9 4.55000019073486
9 5.28333330154419
10 5.30000019073486
10 5.71666669845581
11 5.73333311080933
11 6.21666669845581
12 6.23333311080933
12 6.78333330154419
13 6.80000019073486
13 7.61666679382324
14 7.63333320617676
14 8.5
15 8.51666641235352
15 9.33333301544189
16 9.35000038146973
16 10.5166664123535
17 10.5333337783813
17 11.9166669845581
18 11.9333333969116
18 14
19 14.0166664123535
19 16.6166667938232
20 16.6333332061768
20 20.1666660308838
21 20.1833324432373
21 24.8333339691162
22 24.8500003814697
22 31.6833324432373
23 31.7000007629395
23 41.9333343505859
24 41.9500007629395
24 54.9833335876465
25 55
25 74.1500015258789
26 74.1666641235352
26 92.3333358764648
27 92.3499984741211
27 99.6166687011719
28 99.6333312988281
28 99.9833297729492
};
\addplot [line width=3pt, black]
table {%
0 0
0 6.36666679382324
1 6.38333320617676
1 10.6166667938232
2 10.6333332061768
2 15.4666662216187
3 15.4833335876465
3 22.1666660308838
4 22.1833324432373
4 28.966667175293
5 28.9833335876465
5 34.7999992370605
6 34.8166656494141
6 39.1333351135254
7 39.1500015258789
7 42.9000015258789
8 42.9166679382324
8 50.9166679382324
9 50.9333343505859
9 66
10 66.0166702270508
10 87.466667175293
11 87.4833297729492
11 98.5833358764648
12 98.5999984741211
12 99.9833297729492
};
\addplot [line width=3pt, black, dashed]
table {%
0 0
0 6.36666679382324
1 6.38333320617676
1 6.88333320617676
2 6.90000009536743
2 7.96666669845581
3 7.98333311080933
3 9.36666679382324
4 9.38333320617676
4 10.6000003814697
5 10.6166667938232
5 12.1666669845581
6 12.1833333969116
6 13.9499998092651
7 13.9666662216187
7 16.1499996185303
8 16.1666660308838
8 18.3999996185303
9 18.4166660308838
9 20.4500007629395
10 20.466667175293
10 22.8333339691162
11 22.8500003814697
11 25.3999996185303
12 25.4166660308838
12 27.6499996185303
13 27.6666660308838
13 30.3500003814697
14 30.3666667938232
14 33.3166656494141
15 33.3333320617676
15 36.6166648864746
16 36.6333351135254
16 40.0833320617676
17 40.0999984741211
17 45.3499984741211
18 45.3666648864746
18 52.7000007629395
19 52.716667175293
19 62.4166679382324
20 62.4333343505859
20 73.1500015258789
21 73.1666641235352
21 84.1333312988281
22 84.1500015258789
22 93.1166687011719
23 93.1333312988281
23 97.9833297729492
24 98
24 99.8333358764648
25 99.8499984741211
25 99.9833297729492
};
\end{axis}

\end{tikzpicture}}
		\caption{aggregated}\label{fig:distanceToRestricted-ag-val-all}
	\end{subfigure}
	\qquad \qquad \qquad 
	\begin{subfigure}[b]{0.35\textwidth}
		\centering 
		\resizebox{\textwidth}{!}{% This file was created with tikzplotlib v0.9.17.
\begin{tikzpicture}

\definecolor{color0}{rgb}{1,0.647058823529412,0}

\begin{axis}[
legend cell align={left},
legend style={
  fill opacity=0.8,
  draw opacity=1,
  text opacity=1,
  at={(0.97,0.03)},
  anchor=south east,
  draw=white!80!black
},
tick align=outside,
tick pos=left,
every tick label/.append style={font=\Huge}, 
label style={font=\Huge},
x grid style={white!69.0196078431373!black},
xlabel={deleted voters/cand.},
xmin=-1.3, xmax=27.3,
xtick style={color=black},
y grid style={white!69.0196078431373!black},
ylabel={\% of elections},
ymin=-4.99, ymax=104.79,
ytick style={color=black}
]
\addplot [line width=3pt, blue]
table {%
0 0
1 0.200000047683716
1 1.20000004768372
2 1.39999997615814
2 9.39999961853027
3 9.60000038146973
3 25.6000003814697
4 25.7999992370605
4 51.7999992370605
5 52
5 73.5999984741211
6 73.8000030517578
6 89.8000030517578
7 90
7 98
8 98.1999969482422
8 99.4000015258789
9 99.5999984741211
9 99.8000030517578
};
\addplot [line width=3pt, blue, dashed]
table {%
0 0
3 0.600000023841858
3 1
4 1.20000004768372
4 1.39999997615814
5 1.60000002384186
5 1.79999995231628
6 2
6 2.20000004768372
7 2.40000009536743
7 3
8 3.20000004768372
8 4
9 4.19999980926514
9 6
10 6.19999980926514
10 7.40000009536743
11 7.59999990463257
11 10
12 10.1999998092651
12 13.8000001907349
13 14
13 16.3999996185303
14 16.6000003814697
14 20.2000007629395
15 20.3999996185303
15 27.6000003814697
16 27.7999992370605
16 32.5999984741211
17 32.7999992370605
17 40.7999992370605
18 41
18 48.5999984741211
19 48.7999992370605
19 57.4000015258789
20 57.5999984741211
20 68.4000015258789
21 68.5999984741211
21 77
22 77.1999969482422
22 84
23 84.1999969482422
23 92.8000030517578
24 93
24 95.8000030517578
25 96
25 99.4000015258789
26 99.5999984741211
26 99.8000030517578
};
\addplot [line width=3pt, color0]
table {%
2 0
2 0.600000023841858
3 0.799999952316284
3 6.59999990463257
4 6.80000019073486
4 27
5 27.2000007629395
5 54.5999984741211
6 54.7999992370605
6 80.5999984741211
7 80.8000030517578
7 95.1999969482422
8 95.4000015258789
8 99.5999984741211
9 99.8000030517578
};
\addplot [line width=3pt, color0, dashed]
table {%
6 0
6 0.200000047683716
7 0.399999976158142
7 0.600000023841858
8 0.799999952316284
8 1
9 1.20000004768372
9 2
10 2.20000004768372
10 4.40000009536743
11 4.59999990463257
11 8.60000038146973
12 8.80000019073486
12 10
13 10.1999998092651
13 13.1999998092651
14 13.3999996185303
14 17.7999992370605
15 18
15 26.7999992370605
16 27
16 34.7999992370605
17 35
17 45.4000015258789
18 45.5999984741211
18 55.2000007629395
19 55.4000015258789
19 68
20 68.1999969482422
20 81.8000030517578
21 82
21 90.5999984741211
22 90.8000030517578
22 96.4000015258789
23 96.5999984741211
23 99.1999969482422
24 99.4000015258789
24 99.8000030517578
};
\addplot [line width=3pt, green!50.1960784313725!black]
table {%
2 0
2 0.399999976158142
3 0.600000023841858
3 6.40000009536743
4 6.59999990463257
4 21.3999996185303
5 21.6000003814697
5 48.2000007629395
6 48.4000015258789
6 74.5999984741211
7 74.8000030517578
7 90.5999984741211
8 90.8000030517578
8 98
9 98.1999969482422
9 99.8000030517578
};
\addplot [line width=3pt, green!50.1960784313725!black, dashed]
table {%
7 0
7 0.399999976158142
8 0.600000023841858
8 0.799999952316284
10 1.20000004768372
10 1.79999995231628
11 2
11 2.40000009536743
12 2.59999990463257
12 3.59999990463257
13 3.79999995231628
13 5.40000009536743
14 5.59999990463257
14 8.19999980926514
15 8.39999961853027
15 10.1999998092651
16 10.3999996185303
16 15.6000003814697
17 15.8000001907349
17 22.2000007629395
18 22.3999996185303
18 29.7999992370605
19 30
19 40
20 40.2000007629395
20 50.5999984741211
21 50.7999992370605
21 63.2000007629395
22 63.4000015258789
22 77.4000015258789
23 77.5999984741211
23 90.1999969482422
24 90.4000015258789
24 97.1999969482422
25 97.4000015258789
25 99.5999984741211
26 99.8000030517578
};
\addplot [line width=3pt, black]
table {%
0 0
0 2.79999995231628
1 3
1 15.3999996185303
2 15.6000003814697
2 42.2000007629395
3 42.4000015258789
3 74.4000015258789
4 74.5999984741211
4 89.8000030517578
5 90
5 97.4000015258789
6 97.5999984741211
6 99.5999984741211
7 99.8000030517578
};
\addplot [line width=3pt, black, dashed]
table {%
0 0
0 2.79999995231628
1 3
1 5
2 5.19999980926514
2 9
3 9.19999980926514
3 15
4 15.1999998092651
4 20.6000003814697
5 20.7999992370605
5 26.7999992370605
6 27
6 36.2000007629395
7 36.4000015258789
7 45.5999984741211
8 45.7999992370605
8 54.2000007629395
9 54.4000015258789
9 62.4000015258789
10 62.5999984741211
10 70
11 70.1999969482422
11 76.4000015258789
12 76.5999984741211
12 80.8000030517578
13 81
13 86.1999969482422
14 86.4000015258789
14 90.5999984741211
15 90.8000030517578
15 94.5999984741211
16 94.8000030517578
16 97
17 97.1999969482422
17 98.8000030517578
18 99
18 99.4000015258789
19 99.5999984741211
19 99.8000030517578
};
\end{axis}

\end{tikzpicture}}
		\caption{spotify month}\label{fig:distanceToRestricted-ag-val-spot}
	\end{subfigure}
	\caption{Fraction of elections within a given candidate deletion (solid) or voter deletion distance (dashed) from single-peakedness (blue), single-crossingness (orange), group-separability~(green), and value-restricted (black).}
	\label{fig:distanceToRestricted-ag-val}
\end{figure}
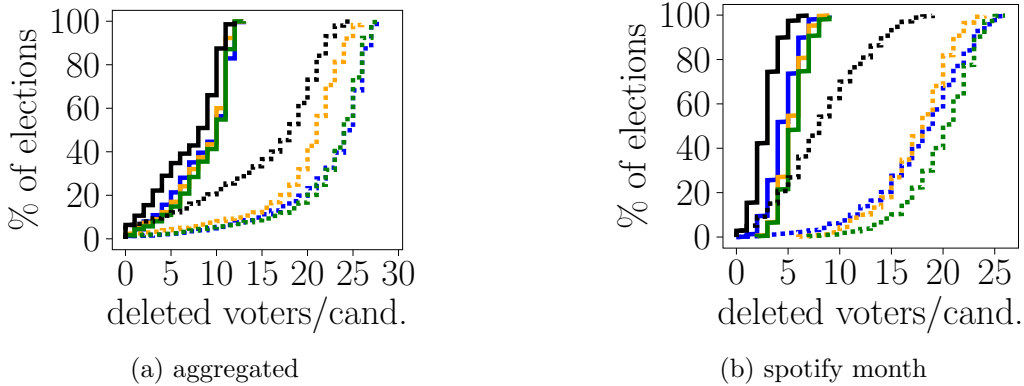

Motivated by this observation, we also checked the voter deletion and candidate deletion of our elections from being value-restricted using the algorithmic ideas of \citet{DBLP:conf/aaai/ElkindL14}. 
In \Cref{fig:distanceToRestricted-ag-val} we depict a cumulative distribution function visualization the fraction of elections that are within a certain candidate/voter deletion distance from single-peaked/single-crossing/group-seperable/value-restricted (note that this figure is the same as \Cref{fig:distanceToRestricted} but with added black lines for value-restricted). 
As visible in \Cref{fig:distanceToRestricted-ag-val-all}, while elections are in fact closer to being value-restricted than to any of the other three restricted domains, there are still only few elections within a small voter/candidate deletion distance.
The difference between value-restrictedness and the other three restricted domains is most pronounced for the voter deletion distance for spotify month elections (\Cref{fig:distanceToRestricted-ag-val-all}): 
Here, usually around $10$ less voters need to be deleted to make an election value restricted than to make it fall into one of the other three restricted domains.
The other datasets from our first group close to identity exhibit a similar behavior.

\section{Additional Material for Section \ref{sec:util}}

\subsection{Number of Ties and Impact on Winner Consensus}\label{app:ties}\label{sub:winner}

\begin{figure}[t]
	\begin{subfigure}[t]{0.35\textwidth}
		\centering
		\includegraphics[width=\textwidth]{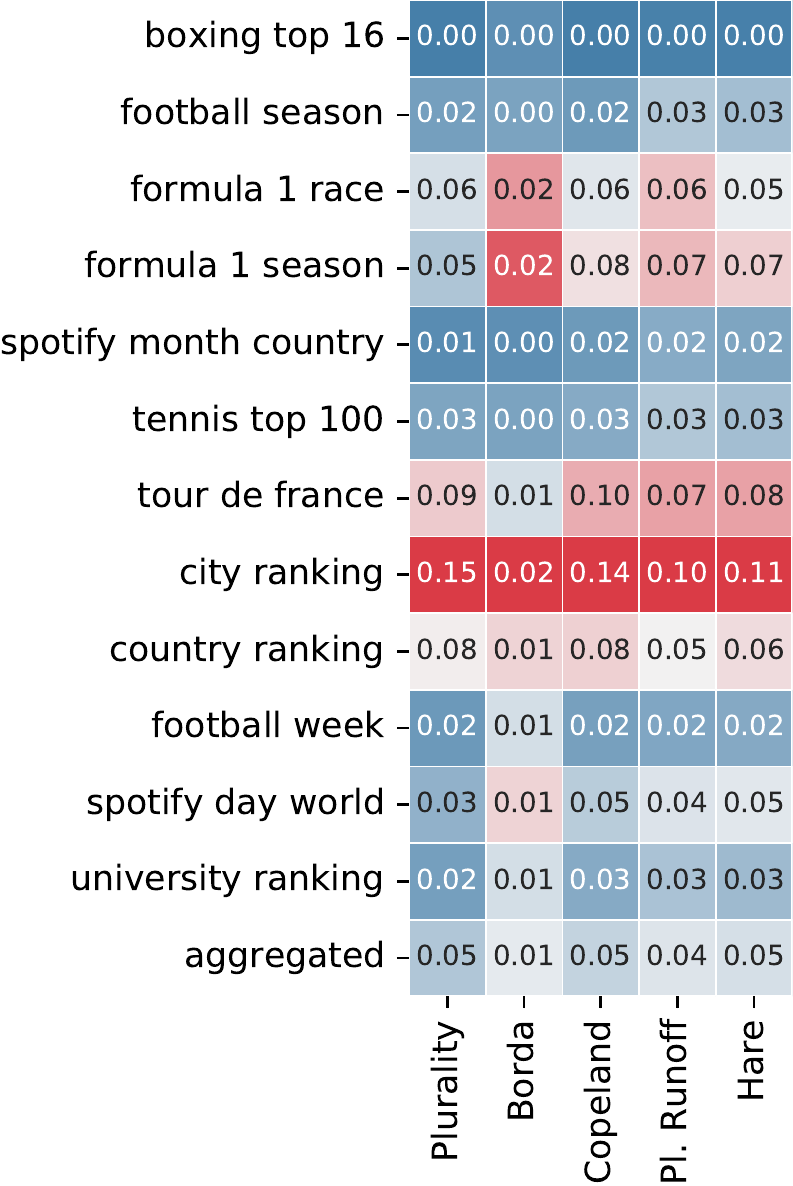}
		\caption{All elections.}\label{fig:ties-full}
	\end{subfigure}
	\qquad \qquad \qquad \qquad
	\begin{subfigure}[t]{0.35\textwidth}
		\centering
		\includegraphics[width=\textwidth]{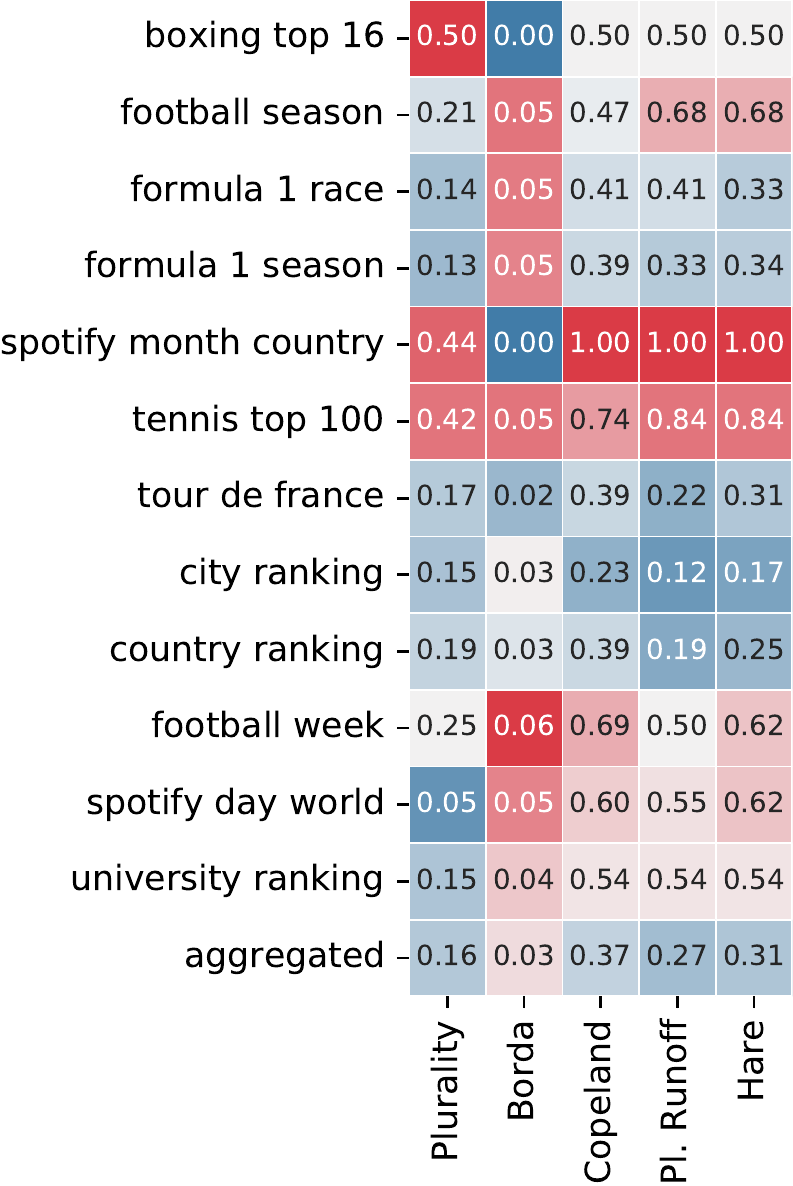}
		\caption{Elections without strong Condorcet winner}\label{fig:ties-Cond}
	\end{subfigure}
	\caption{Fraction of elections where voting rules returned multiple winners. Colors express normalized values which are normalized with respect to the maximum in each column.}
	\label{fig:ties}
\end{figure}

\begin{figure}[t]
	\centering
	\includegraphics[width=0.25\textwidth]{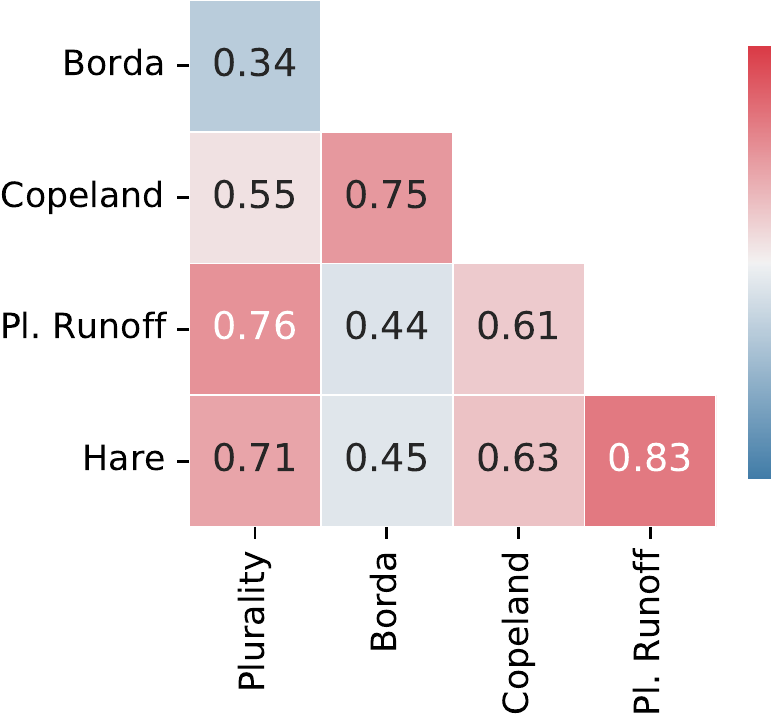}
	\caption{Non-empty overlap on all elections without Condorcet winner.}\label{overlap:non_cond}
\end{figure}

We computed three different ways of evaluating the similarity of winners returned by different voting rules:
\begin{description}
	\item[Lexicographic agreement] This measure is one if in case that we apply lexicographic tie-breaking, the returned winner is the same and zero otherwise. 
	\item[Non-empty overlap] This measure is one if the winner sets overlap and zero otherwise. 
	\item[Normalized overlap] For two winner sets $W_1$ and $W_2$ of two voting rules the overlap is $\nicefrac{|W_1\cap W_2|}{|W_1\cup W_2|}$.
\end{description}
Note that these three measures coincide if both voting rules return a unique winner. 
So let us focus for a moment on how often the considered voting rules select more than one winner.
As shown in \Cref{fig:ties-full}, all our voting rules expect Borda return a tied winner on only $5\%$ of all elections; for Borda it is only $1\%$ of elections. 
One possible explanation for this could be that most of our elections admit a strong Condorcet winner and that in this case rules are likely to return it as the unique winner.
Examining  \Cref{fig:ties-Cond}, where we depict the fraction of elections without a strong Condorcet winner that are tied, this intuition gets confirmed: 
Under all voting rules, the fraction of all elections without a strong Condorcet winner that are tied is at least three times higher than for all elections. 
That is, of all elections without a strong Condorcet winner $16\%$, $3\%$, $37\%$, $27\%$, and $31\%$ are tied under Plurality, Borda, Copeland, Plurality with Runoff and Hare, respectively. 
In particular, for Copeland, Plurality with Runoff and Hare are these values remarkably high. 
This might be due to the fact that all these three voting rules might boil down to a pairwise comparison between two candidates: 
While a strong Condorcet winner always wins such a comparison, without one ties can arise.
Lastly, observe that as the fraction of elections with a strong Condorcet winner varies significantly among our datasets, it is quite intuitive that the fraction of tied elections also depends on the dataset.

Coming back to our original question of comparing our three measures for the consensus of voting rules, lexicographic agreement and normalized overlap produce mostly very similar results.
For lexicographic agreement and non-empty overlap, results are similar for the aggregate dataset but differ slightly more on elections without strong Condorcet winners (which is also quite intuitive because we previously observed that these elections are more likely to be tied). 
In \Cref{overlap:non_cond}, we display the average non-empty overlap for elections without a strong Condorcet winner (for the aggregate dataset the picture for non-empty overlap is very similar to the one for lexicographic agreement).
As the non-empty overlap measure is always one if the lexicographic agreement is one,  comparing \Cref{overlap:non_cond} to the analgous plot (\Cref{first:non_cond}) for the lexicographic agreement, we observe that the values for the non-empty overlap are higher. 
However, given that we previously observed that voting rules return tied winners only on a minority of elections, the extent (e.g., for Hare and Plurality the lexicographic agreement is $0.5$ and the non-empty overlap is $0.71$) is partly surprising.  
Moreover, the results for the non-empty overlap confirm our proposed splitting of voting rules into two groups. 

\end{document}